\documentclass{aa}  

\usepackage{graphicx}
\usepackage{txfonts}
\usepackage{url}
\usepackage{color}
\usepackage{comment}
\def\Lya{Ly$\alpha$}

\def\Hb{H$\beta$}

\def\HeII{He\,{\sc ii}}
\def\OII{[O\,{\sc ii}]}
\def\OIII{O\,{\sc iii}]}

\def\NV{N\,{\sc v}}
\def\CIII{C{\sc iii}]}
\def\CIV{C{\sc iv}}

\def\Oabundance{$12+\log ({\rm O/H})$}
\def\HI{H\,{\sc i}}
\def\HII{H\,{\sc ii}}

\def\Msun{M$_{\odot}$}
\def\nH{$n_{\rm H}$}
\def\xiion{$\xi_{\rm ion}$}
\def\ebv{$E_s(B-V)$}

\def\aj{AJ}
\def\apj{ApJ}
\def\apjs{ApJS}
\def\apjl{ApJL}
\def\aap{A\&A}  
\def\mnras{MNRAS}
\def\pasj{PASJ}
\def\pasp{PASP}

\def\msun{\ifmmode M_{\odot} \else M$_{\odot}$\fi}
\def\msunyr{\ifmmode M_{\odot} {\rm yr}^{-1} \else M$_{\odot}$ yr$^{-1}$\fi}
\def\zsun{\ifmmode Z_{\odot} \else Z$_{\odot}$\fi}
\def\cloudy{\textsc{Cloudy}}
\def\ergsHz{erg$^{-1}$\,Hz}

\begin{document} 

%%%%%%%%%%%%%%%%%%%%%%%%%%%%%%%%%%%%%%%%%%%%%%%%%%%%
%%%%%%%%%%%%%%%%%%%%%%%%%%%%%%%%%%%%%%%%%%%%%%%%%%%%

% Title
\title{%
        The VIMOS Ultra Deep Survey: 
        Nature, ISM properties, and ionizing spectra of \CIII$\lambda 1909$ emitters at $z=2-4$
        \thanks{Based on data obtained with the European Southern Observatory
        Very Large Telescope, Paranal, Chile, under Large Program 185.A-0791.}
}       
\titlerunning{%
        UV spectroscopic properties of VUDS \CIII\ emitters
}

%%%%%%%%%%%%%%%%%%%%%%%%%%%%%%%%%%%%%%%%%%%%%%%%%%%%
                
\author{
        K. Nakajima\inst{1,2}
        \thanks{JSPS Overseas Research Fellow}
        \and
        D. Schaerer\inst{1,3}
        \and
        O. Le F\`{e}vre\inst{4}
        \and
        R. Amor\'{i}n\inst{5,6,7}
        \and
        M. Talia\inst{8,9}
        \and
        B. C. Lemaux\inst{4,10}
        \and 
        L. A. M. Tasca\inst{4}
        \and 
        E. Vanzella\inst{9}
        \and
        G. Zamorani\inst{9}
        \and
        S. Bardelli\inst{9}
        \and 
        A. Grazian\inst{5}
        \and 
        L. Guaita\inst{5,11}
        \and
        N. P. Hathi\inst{12}
        \and
        L. Pentericci\inst{5}
        \and
        E. Zucca\inst{9}
}
\authorrunning{%
        Nakajima, Schaerer, Le F\`{e}vre et al.
}

\institute{
        Observatoire de Gen\`{e}ve, Universit\'{e} de Gen\`{e}ve, 
        51 Ch. des Maillettes, 1290 Versoix, Switzerland
        \and
        European Southern Observatory, 
        Karl-Schwarzschild-Str. 2, D-85748, Garching bei M\"{u}nchen, Germany\\
        \email{knakajim@eso.org}
        \and
        CNRS, IRAP, 14 Avenue E. Belin, 31400 Toulouse, France
        \and
        Aix Marseille Universit\'{e}, CNRS, 
        LAM (Laboratoire d'Astrophysique de Marseille) 
        UMR 7326, 13388 Marseille, France
        \and
        INAF-- Osservatorio Astronomico di Roma, via di Frascati 33, I-00078, Monte Porzio Catone, Italy
        \and
        Cavendish Laboratory, University of Cambridge, 19 JJ Thomson Avenue, Cambridge, CB3 0HE, UK 
        \and
        Kavli Institute for Cosmology, University of Cambridge, Madingley Road, Cambridge CB3 0HA, UK
        \and
        Dipartimento di Fisica e Astronomia, Universit\`{a} di Bologna, Via Gobetti 93/2, I-40129, Bologna, Italy 
        \and
        INAF-- Osservatorio Astronomico di Bologna, Via Gobetti 93/3, I-40129, Bologna, Italy
        \and 
        Department of Physics, University of California, Davis, One Shields Ave., Davis, CA 95616, USA
        \and 
        N\'ucleo de Astronom\'ia, Facultad de Ingenier\'ia, Universidad Diego Portales, Av. Ej\'ercito 441, Santiago, Chile
        \and
        Space Telescope Science Institute, 3700 San Martin Drive, Baltimore, MD 21218, USA
}

%%%%%%%%%%%%%%%%%%%%%%%%%%%%%%%%%%%%%%%%%%%%%%%%%%%%

\date{Received 12 September 2017; Accepted 19 December 2017}

% \abstract{}{}{}{}{} 
% 5 {} token are mandatory

%%%%%%%%%%%%%%%%%%%%%%%%%%%%%%%%%%%%%%%%%%%%%%%%%%%%

\abstract
   {Ultraviolet (UV) emission-line spectra are used to 
   spectroscopically confirm high-$z$ galaxies and increasingly also to
   determine their physical properties.}
   {We construct photoionization models to interpret the observed 
   UV spectra of distant galaxies in terms of the dominant radiation field
   and the physical condition of the interstellar medium (ISM).
   These models are applied to new spectroscopic observations
   from the VIMOS Ultra Deep Survey (VUDS).}
   {We construct a large grid of photoionization models, which use
   several incident radiation fields (stellar populations, 
   active galactic nuclei (AGNs), mix of stars and AGNs, blackbodies, and others),
   and cover a wide range of metallicities and ionization parameters. 
   From these models we derive new spectral UV line diagnostics 
   using equivalent widths (EWs) of 
   \CIII$\lambda 1909$ doublet, 
   \CIV$\lambda 1549$ doublet
   and the line ratios of \CIII, \CIV, and \HeII$\lambda 1640$ recombination lines.
   We apply these diagnostics to a sample of $450$ \CIII-emitting galaxies at
   redshifts $z=2-4$ previously identified in VUDS.}
   {We demonstrate that our photoionization models successfully reproduce 
   observations of nearby and high-redshift sources with known radiation field 
   and/or metallicity.
   For star-forming galaxies our models predict that \CIII\ EW peaks at 
   sub-solar metallicities, whereas \CIV\ EW peaks at even lower metallicity.
   Using the UV diagnostics, 
   we show that the average star-forming galaxy 
   (EW(\CIII) $\sim 2$\,\AA)
   based on the composite of the $450$ UV-selected galaxies' spectra
   is well described by stellar photoionization from single and binary stars.
   The inferred metallicity and ionization parameter is typically 
   $Z =0.3$ -- $0.5\,Z_{\odot}$ and $\log U = -2.7$ to $-3$, 
   in agreement with earlier works at similar redshifts.
   The models also indicate an average age of $50$ -- $200$\,Myr
   since the beginning of the current star-formation, 
   and an ionizing photon production rate, \xiion,
   of $\log$\,\xiion$/$\ergsHz\ $= 25.3$ -- $25.4$.
   Among the sources with EW(\CIII) $>=10$\,\AA, 
   approximately $30$\,\%\ are likely dominated by AGNs.
   The metallicity derived for galaxies with EW(CIII) $=10-20$\,\AA\
   is low, $Z=0.02-0.2\,Z_{\odot}$, and the ionization parameter 
   higher ($\log U\sim -1.7$) 
   than the average star-forming galaxy. 
   To explain the average UV observations of the strongest but 
   rarest \CIII\ emitters (EW(\CIII) $>20$\,\AA), we find that stellar 
   photoionization is clearly insufficient. A radiation field consisting 
   of a mix of a young stellar population ($\log$\,\xiion$/$\ergsHz\ $\sim 25.7$) 
   plus an AGN component is required. 
   Furthermore an enhanced C$/$O abundance ratio (up to the solar value) 
   is needed for metallicities 
    $Z =0.1$ -- $0.2\,Z_{\odot}$ and $\log U = -1.7$ to $-1.5$.
   }
   {A large grid of photoionization models has allowed us to propose new 
   diagnostic diagrams to classify the nature of the ionizing radiation field 
   (star formation or AGN) of distant galaxies using UV emission lines, 
   and to constrain their ISM properties.
   We have applied this grid to a sample of \CIII-emitting galaxies at $z=2-4$ detected 
   in VUDS, finding a range of physical properties and clear evidence for significant 
   AGN contribution in rare sources with very strong \CIII\ emission.
   The UV diagnostics we propose should also serve as an important basis
   for the interpretation of upcoming observations of high-redshift galaxies.}


%%%%%%%%%%%%%%%%%%%%%%%%%%%%%%%%%%%%%%%%%%%%%%%%%%%%

\maketitle

%%%%%%%%%%%%%%%%%%%%%%%%%%%%%%%%%%%%%%%%%%%%%%%%%%%%
%%%%%%%%%%%%%%%%%%%%%%%%%%%%%%%%%%%%%%%%%%%%%%%%%%%%
%%%%%%%%%%%%%%%%%%%%%%%%%%%%%%%%%%%%%%%%%%%%%%%%%%%%
%%%%%%%%%%%%%%%%%%%%%%%%%%%%%%%%%%%%%%%%%%%%%%%%%%%%

%%%%%%%%%%%%%%%%%%%%%%%%%%%%%%%%%%%%%%%%%%%%%%%%%%%%
%%%%%%%%%%%%%%%%%%%%%%%%%%%%%%%%%%%%%%%%%%%%%%%%%%%%
\section{Introduction}
\label{sec:introduction}
%%%%%%%%%%%%%%%%%%%%%%%%%%%%%%%%%%%%%%%%%%%%%%%%%%%%
%%%%%%%%%%%%%%%%%%%%%%%%%%%%%%%%%%%%%%%%%%%%%%%%%%%%

Rest-frame ultraviolet (UV) spectra provide a powerful tool, 
enabling us to examine galaxies at $z>2$
that are redshifted and observable with the ground-based 
optical spectrographs
(e.g., \citealt{shapley2003}). 
Historically, \Lya\ in emission at $\lambda 1215.7$\,\AA\ 
has been used for identifying high-$z$ galaxies. 
This is because 
i) the \Lya\ line is theoretically the strongest feature from star-forming galaxies 
that is observable with a ground-based telescope
in the rest-frame UV/optical, 
ii) it shows a unique distinguishable asymmetric line profile, and
iii) galaxies that present prominent \Lya\ emission are more 
abundant at higher-$z$ at least up to $z\sim 6$
(e.g., \citealt{schaerer2003,vanzella2009,stark2011,cassata2015}).
However, since \Lya\ is resonantly scattered by neutral hydrogen, 
it becomes unlikely to be the best probe of galaxies in the reionization
epoch at $z>6$, when the intergalactic medium (IGM)
neutral fraction is significantly larger than zero
(e.g., \citealt{fan2006}).
Indeed, several studies have pointed out a decline of the fraction 
of \Lya\ emitters with respect to all star-forming galaxies at $z\gtrsim 7$
(e.g., \citealt{pentericci2014,schenker2014}).
A similar drop in the abundance of \Lya\ emitters is also reported 
in the luminosity function study of \citet{konno2014}.
We thus need to explore alternative probes of galaxies in the early universe,
in order to characterize as well as spectroscopically confirm them.

An advantage of other UV lines is that they are observable
from galaxies at $z>6-7$ with near-infrared spectrographs 
currently available at ground-based observatories and the 
Hubble Space telescope (HST). 
Indeed, recent deep spectroscopic surveys of $z>6-7$
identify the following rest-UV emission lines; 
[C{\sc iii}]$\lambda 1907$, C{\sc iii}]$\lambda 1909$ , 
(hereafter their sum is referred to as \CIII$\lambda 1909$),
\CIV$\lambda\lambda 1548,1550$ (hereafter \CIV$\lambda 1549$),
\HeII$\lambda 1640$, and/or \NV$\lambda 1240$
\citep{stark2015_c3,stark2015_c4,stark2017,mainali2017,%
schmidt2017,laporte2017}. 
The next question pertains to whether or not we can use these UV emission lines
to infer the detailed properties of stars and gas.

In order to gain insight into the UV spectroscopic properties, 
previous studies have worked on galaxies at intermediate redshifts
($z=1.5 - 3.5$; \citealt{shapley2003,fosbury2003,erb2010,
christensen2012a,stark2014,bayliss2014,debarros2016,amorin2017,
stroe2017})
with ground-based instruments, 
as well as in the nearby universe \citep{garnett1995,berg2016,senchyna2017} 
using the HST.
At these redshifts of $z=0 - 3.5$, their rest-frame optical spectra 
are also available,
which are well calibrated to determine the Interstellar medium (ISM) 
properties such as gas-phase metallicity and ionization parameter
using empirical methods 
(e.g., \citealt{pagel1979,storchi-bergmann1994,PP2004})
and photoionization modelling (e.g., \citealt{KD2002}).
\citet{stark2014} have reported \CIII\ detection in $16$ out of $17$ 
strongly-lensed star-forming dwarfs at $z=1.5-3$. 
\CIV\ emission was also detected in the three most extreme \CIII\ emitters. 
Using photoionization models of \citet{gutkin2016}, \citet{stark2014}
have inferred with the UV lines only that the four galaxies in their sample
with the best UV spectra (including the three \CIV\ emitters) are all
characterized by a metal-poor ($Z=0.04$--$0.13\,Z_{\odot}$),
highly ionized ($\log U\sim -2$) gas with a young age stellar population
($\lesssim 50$\,Myr). The gaseous properties are in good 
agreement with those estimated with the rest-frame optical emission
lines.
A similar low-metallicity gas with a high ionization parameter 
has been estimated 
for galaxies with a prominent \CIII\ emission
\citep{erb2010,christensen2012a,berg2016,senchyna2017}. 
These objects all have stellar masses below $10^{10}$\,\Msun.
Since young, metal-poor, low-mass galaxies are considered to be 
more common at higher-$z$, their UV lines, especially the \CIII, are 
likely to be strong enough to be used for their identification.

A positive correlation between equivalent widths (EWs) of \Lya\ and \CIII\ emission
as suggested by some observations (e.g., \citet{shapley2003,stark2014})
also supports the idea of using \CIII\ as an alternative probe of early 
galaxies instead of \Lya. 
It remains unclear, however, whether the positive correlation holds
universally. For example, \citet{rigby2015} suggested that 
the correlation is mostly dominated by the strongest emitters with 
EW(\Lya) $\gtrsim 50$\,\AA\ and EW(\CIII) $\gtrsim 5$\,\AA, 
and that the correlation becomes less significant for weaker emitters.
The correlation needs to be confirmed with larger samples 
and fully understood from a theoretical point of view.

Other UV lines are also important for understanding the nature of 
the ionizing spectrum radiated in high-$z$ objects.
This is because there exist some high ionization lines 
such as \CIV$\lambda 1549$ and the \HeII$\lambda 1640$ recombination line
which require the presence of extremely high-energy photons.
The \HeII\ recombination line is especially sensitive to the abundance 
of ionizing photons with energies higher than $54.4$\,eV,
which can partly be produced by stars, but more abundantly by 
an active galactic nucleus (AGN). It is thus suggested that the emission 
line ratios including \HeII\ can be used as a diagnostic to judge whether 
the ionizing source is powered by stellar population or AGN
\citep{feltre2016,gutkin2016}. 
If only stellar populations are considered, the strength of
the UV lines could be sensitive
to the hardness and the shape of the ionizing spectrum. 
\citet{schaerer2003} demonstrated that the \HeII\ emission strength
provides a diagnostic to examine the ionizing spectrum generated 
by a stellar population that is sensitive to the shape of initial mass function
(IMF), metallicity, age, and so on.
More recently, \citet{JR2016} implied that metal-poor galaxies 
with metallicity below $0.2\,Z_{\odot}$
require 
harder spectral energy distribution (SED) models that include binary stars 
\citep{stanway2015} to 
reproduce large EWs of \CIII\ of $\sim 10$ -- $20$\,\AA.
Some other evidence favoring stellar population models which include 
massive interacting binaries have been pointed out recently 
(e.g., \citealt{steidel2016}).
Photoionization modeling with several input radiation fields 
appears necessary to carefully interpret the observed UV spectra.

More detailed photoionization models are also motivated 
by new data from larger surveys coming on line. 
Using the VIMOS Ultra Deep Survey (VUDS; \citealt{lefevre2015}), 
\citet{lefevre2017} have established a sample of $447$ galaxies 
at $z=2-4$ that present a prominent ($>3\sigma$) \CIII\ emission, 
including $16$ with rest-frame EW(\CIII) larger than $20$\,\AA. 
Since there are other UV emission lines (such as \CIV\ and \HeII)
detected in many of the VUDS spectra, these \CIII\ emitters provide us 
with a unique opportunity to examine ISM properties and 
ionizing spectra of a large sample of high-redshift galaxies.
By comparing the \CIII\ emitters with their parent sample in the 
same redshift interval in the plot between stellar mass and 
star-formation rate (SFR), \citet{lefevre2017} interestingly suggest 
that a significant fraction of galaxies with EW(\CIII) $\gtrsim 10$\,\AA\ 
tend to fall below the relation typically found in the parent sample
of star-forming galaxies
(i.e., below \textit{the star-formation main-sequence}).
This trend is particularly striking for strong \CIII\ emitters with 
EW $>20$\,\AA.
\citet{lefevre2017} therefore suggest that the presence of an AGN
in these stronger \CIII\ emitters may be responsible for the 
star-formation quenching in these galaxies. 
On the other hand, \citet{amorin2017} also identify with the early 
VUDS data ten metal-poor compact star-forming galaxies at 
$z\sim 3$ showing an intermediately strong \CIII\ emission of 
EW(\CIII) $= 5.5$ -- $21.5$\,\AA\ (median value: $11.5$\,\AA).
This earlier VUDS study already suggests that a variety of objects 
may show strong \CIII\ emission. 
In this companion study, we investigate the nature of the 
\CIII\ emitters in VUDS using a large grid of photoionization models.
Particularly in this paper, we exploit UV diagnostic diagrams 
using the three UV lines of \CIII, \CIV, and \HeII.
These are the three most prominent UV emission lines 
after \Lya\ 
observed in the spectra of star-forming galaxies at $z=2-4$
(e.g., in VUDS; \citealt{lefevre2015}), as well as in sources 
at $z>6$
\citep{stark2015_c3,stark2015_c4,stark2017,mainali2017,%
schmidt2017}.
We discuss diagnostics involving the EW of \CIII\ and \CIV,
which complement those presented in the recent studies of 
\citet{feltre2016} and \citet{gutkin2016}. 
The EWs in conjunction with line ratios provide efficient and 
informative tools to investigate the ISM properties as well as 
the nature of the ionizing radiation field of distant galaxies.

This paper is organized as follows. 
We describe our photoionization modeling in Sect.\ \ref{sec:modelling},
and present the photoionization model predictions of the UV line fluxes 
and EWs in Sect.\ \ref{sec:results_model}.
In Sect.\ \ref{sec:results_UV_diagrams} we develop UV diagrams 
to diagnose the shape of the incident radiation field and the ISM properties.
Using the sample of VUDS \CIII\ emitters described in Sect.\ \ref{sec:c3emitters}
and \citet{lefevre2017}, we derive the physical properties of 
the \CIII\ emitters found by VUDS in Sect.\ \ref{sec:nature_vuds}.
The ten metal-poor VUDS galaxies studied by \citet{amorin2017}
are also re-examined with our UV diagrams. 
In Sect. \ref{sec:discussion}, we discuss the nature of the \CIII\ and \CIV\ 
emitters found by VUDS and detected at even higher-$z$ from the literature,
and investigate the practicality of using emission lines in the rest-frame UV
for identifying high-$z$ sources.
Section\ \ref{sec:conclusions} summarizes and concludes the paper.
Throughout this paper, 
we assume a solar chemical composition following \citet{asplund2009},
where $\log (Z/Z_{\odot})=12+\log({\rm O}/{\rm H})-8.69$.
We adopt a standard $\Lambda$CDM cosmology with 
($\Omega_m$, $\Omega_{\Lambda}$, $H_0$) $=$
(0.3, 0.7, 70\,km\,s$^{-1}$\,Mpc$^{-1}$).
We use a definition of EW with positive values indicating emission. 
All EW measurements are given in the rest frame.

%FFFFFFFFFFFFFFFFFFFFFFFFFFFFFFFFFFFFFFFFFFFFFFFFFFFFFFFFFFFFFFFFFFFFFFFFF%
\begin{figure*}
  \centerline{
    \includegraphics[width=0.88\textwidth]{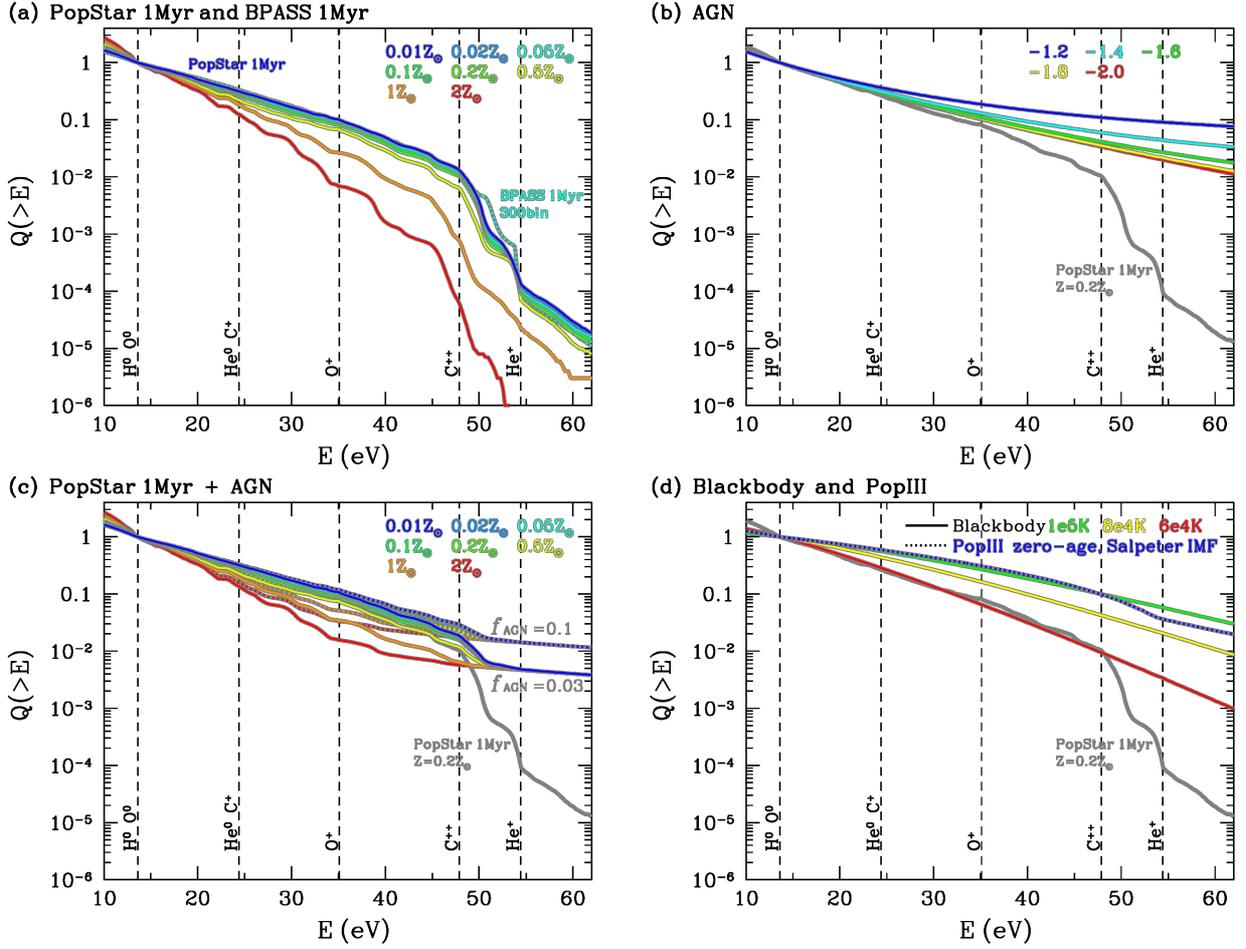}
  }
  \caption{
        The number of photons emitted above the energy $E$ as a 
        function of $E$, for different incident radiation fields. 
        The ionization thresholds of key ions discussed in this paper
        are presented by the vertical dashed lines.
        (a) The \textsc{PopStar} models at an age of $1$\,Myr
        color-coded by stellar metallicity, 
        from $Z=0.01\,Z_{\odot}$ (solid blue) to $2\,Z_{\odot}$ (solid red). 
        In addition, a BPASS 300bin model at an age of $1$\,Myr
        with a metallicity of $Z=0.05\,Z_{\odot}$ is presented 
        (dotted cyan).
        (b) The AGN models color-coded by power-law
        index, from $\alpha=-1.2$ (blue) to $\alpha=-2.0$ (red).
        The \textsc{PopStar} model at an age of $1$\,Myr with
        a metallicity of $Z=0.2\,Z_{\odot}$ is shown by the gray
        curve for comparison (same for the rest of the panels). 
        (c) The \textsc{PopStar} models at an age of $1$\,Myr
        combined with AGN with an $\alpha=-1.6$. 
        Two cases of the contribution of the AGN to the total 
        number of ionizing photons are presented; 
        $f_{\rm AGN}=0.03$ (solid) and $0.1$ (dotted).
        The color-code  is the same as the panel (a).
        (d) The blackbody models with a high effective temperature
        of $T_{\rm eff}=6\times10^{4}$\,K (solid red), 
        $8\times 10^{4}$\,K (solid yellow), and 
        $1\times 10^{5}$\,K (solid green).
        In addition, a PopIII model of \citet{raiter2010} corresponding to 
        zero-age with a \citet{salpeter1955} IMF $1$--$100$\,\Msun\
        is presented (dotted blue).
        }
\label{fig:sed_incident}
\end{figure*}
%FFFFFFFFFFFFFFFFFFFFFFFFFFFFFFFFFFFFFFFFFFFFFFFFFFFFFFFFFFFFFFFFFFFFFFFFF%

%ttttttttttttttttttttttttttttttttttttttttttttttttttttttttttttttttttttttttt%
\begin{table*}
  \centering
  \caption{Main parameters of our \cloudy\ photoionization models explored in this paper.}
  \label{tbl:parameters_cloudy}
  \renewcommand{\arraystretch}{1.4}
  \begin{tabular}{@{}lllllll@{}}
    \hline
    Parameter 
     & \textsc{PopStar} 
     & BPASS 
     & AGN 
     &  \textsc{PopStar}$+$AGN 
     & Blackbody
     & PopIII star 
    \\
    \hline
    SED shape
     & Fig. \ref{fig:sed_incident}(a)$^{(\ddag)}$
     & Fig. \ref{fig:sed_incident}(a)$^{(\ddag)}$
     & Fig. \ref{fig:sed_incident}(b)
     & Fig. \ref{fig:sed_incident}(c)$^{(\ddag)}$
     & Fig. \ref{fig:sed_incident}(d)
     & Fig. \ref{fig:sed_incident}(d)
    \\
    \hline
    $Z\, (Z_{\odot})$ 
     & $0.01,\, 0.02,\, 0.05,$ 
     & $0.05,\, 0.1,\, 0.2,$
     & $0.01,\, 0.02,\, 0.05,$
     & $0.01,\, 0.02,\, 0.05,$
     & $0.01,\, 0.02,\, 0.05,$
     & $10^{-4},\, 10^{-3},\, 0.01,$
    \\
     & $0.1,\, 0.2,\, 0.5,$
     & $0.5,\, 1.0$
     & $0.1,\, 0.2,\, 0.5,$
     & $0.1,\, 0.2,\, 0.5,$
     & $0.1,\, 0.2,\, 0.5,$
     & $0.02,\, 0.05,\, 0.1$
    \\
     & $1.0,\, 2.0$
     & 
     & $1.0,\, 2.0,\, 5.0$
     & $1.0,\, 2.0$
     & $1.0$
     & $1.0$
    \\
    \hline
    $\log U$ 
     & \multicolumn{6}{l}{
        $-3.5,\, -3.0,\, -2.5,\, -2.0,\, -1.5,\, -1.0,\, -0.5$
        }
    \\
    \hline
    \nH\, (cm$^{-3}$)
     & $10,\,10^2\,^{(\dag)},\, 10^3,$
     & $10,\,10^2\,^{(\dag)},\, 10^3,$
     & $10,\,10^2,\, 10^3\,^{(\dag)},$
     & $10,\,10^2\,^{(\dag)},\, 10^3,$
     & $10,\,10^2\,^{(\dag)},\, 10^3,$
     & $10,\,10^2\,^{(\dag)},\, 10^3,$
    \\
     & $10^4,\, 10^5$
     & $10^4,\, 10^5$
     & $10^4,\, 10^5$
     & $10^4,\, 10^5$
     & $10^4,\, 10^5$
     & $10^4,\, 10^5$
    \\
    \hline
    C$/$O
     & \multicolumn{6}{l}{
        C$/$H $= 6.0\times 10^{-5}Z/Z_{\odot} +2.0\times 10^{-4}(Z/Z_{\odot})^2$
        \citep{dopita2006}$^{(\dag)}$, 
        fixed C$/$O ratios up to $1.0$ ($\sim 2\times$ (C$/$O)$_{\odot}$) }
    \\
    \hline
    Other 
     & $\cdots$
     & Upper-mass  
     & $\alpha=-1.2,\, -1,4,$
     & $\alpha=-1.6\,^{(\dag)},\, -1.2$; 
     & $T_{\rm eff}=6\times10^4,$
     & Zero-age, 
    \\
    param.
     & 
     & of $300\,M_{\odot}$
     & $\,-1.6\,^{(\dag)},\, -1.8,$
     & $f_{\rm AGN}=0.03, 0.1$
     & $\,8\times10^4,\, 1\times 10^5$ K
     & Salpeter IMF,  
    \\
     & 
     & 
     & $\,-2.0$
     &
     & 
     & $1$--$100\,M_{\odot}$
    \\
    \hline
  \end{tabular}
  \renewcommand{\arraystretch}{1.0}
  \\ 
  \vspace{-3mm}
  \begin{flushleft}
        \small
        $(\dag)$ Default value adopted for the input SED model.
        $(\ddag)$ Ages of current star-formation of $1$\,Myr and older 
        are considered under the assumption of constant star-formation. 
  \end{flushleft}
\end{table*}
%ttttttttttttttttttttttttttttttttttttttttttttttttttttttttttttttttttttttttt%

%%%%%%%%%%%%%%%%%%%%%%%%%%%%%%%%%%%%%%%%%%%%%%%%%%%%
%%%%%%%%%%%%%%%%%%%%%%%%%%%%%%%%%%%%%%%%%%%%%%%%%%%%
\section{Photoionization models} 
\label{sec:modelling}
%%%%%%%%%%%%%%%%%%%%%%%%%%%%%%%%%%%%%%%%%%%%%%%%%%%%
%%%%%%%%%%%%%%%%%%%%%%%%%%%%%%%%%%%%%%%%%%%%%%%%%%%%

We perform photoionization model calculations with \cloudy\
(version 13.03; \citealt{ferland1998,ferland2013})
to examine the UV spectroscopic properties of distant galaxies, 
with particular attention to \CIII\ emission. 
We use a large range of ionizing fields to try and cover the 
large range of galaxy properties. 
These models are used to interpret the \CIII\ emitters found by
VUDS in Sect. \ref{sec:c3emitters}.
We assume constant-density gas clouds with a plane-parallel
geometry.
We include dust physics and the depletion factors of the various elements 
from the gaseous phase in the same manner as the analyses of 
\citet{dopita2006} and \citet{nagao2011}.
Dust grains in the photoionized regions are considered to play an important role 
in photoelectric heating of the cloud, and to affect the emitted spectrum, 
especially in the UV and optical wavelength regimes (e.g., \citealt{vanhoof2004}).
Depletion factors used in this paper are given by \cloudy\ 
except for nitrogen, for which the value from \citet{dopita2006} is used instead
that is more consistent with the \citet{asplund2009} abundance set.
The choice of the nitrogen depletion factor does not affect
the UV diagnostics we propose in this paper.
We understand that usages of the adopted depletion factors remain debatable
for every model we explore in this paper. 
In non-solar metallicities, we assume that both the dust model and the
depletion factors are unchanged, but the dust abundance is assumed to
scale linearly with the gas metallicity, as adopted in \citet{nagao2011}. 
All elements except nitrogen, carbon, and helium are taken to be 
primary nucleosynthesis elements. 
For carbon and nitrogen, we use forms given by \citet{dopita2006}
and \citet{lopez-sanchez2012} by default, respectively, 
to take account of their secondary nucleosynthesis components
in a high-metallicity range. 
For helium, we use a form in \citet{dopita2006}. 
These prescriptions are empirically derived from observations of 
nearby extragalactic \HII\ regions. 
In this paper, the carbon abundance is particularly important 
since it directly affects the strength of the \CIII\ and \CIV\ emission.
We note that the prescription of the carbon abundance we adopt 
is thought to be valid for $z=2-3$ galaxies on average 
(e.g., \citealt{erb2010,steidel2016,amorin2017}), 
despite their large scatter. 
We thus additionally consider cases where the carbon abundance
does not follow the empirical method (Sect.\ \ref{ssec:results_CO}),
and discuss a possible variation of the carbon abundance 
later in this paper (Sects.\ \ref{sssec:nature_vuds_strongC3_highCO}
and \ref{ssec:discussion_origin_highCO}).
Briefly, models are constructed by varying ISM properties of 
metallicity ($Z$), ionization parameter ($\log U$), and 
electron density (\nH) over a wide parameter space, as indicated in 
Table \ref{tbl:parameters_cloudy}.
For our {\em default models} we assume a gas density of 
$10^2$\,cm$^{-3}$, a typical value that is observationally suggested at 
these redshifts (e.g., \citealt{sanders2016}) except for the AGN SED,
which adopts a higher default gas density of $10^3$\,cm$^{-3}$
as adopted by \citet{kewley2013_theory}
(see also \citealt{dors2014}).
Models with a higher gas density are also presented 
in Sect.\ \ref{ssec:results_single_stellar}.

To address the nature and estimate the physical properties of 
high-redshift galaxies, we have considered a diversity of ionizing radiation
fields including those of star-forming galaxies generated by stellar population 
synthesis codes which include single or binary evolution, AGNs, 
combinations of stars and AGN, blackbodies, and a PopIII stellar population.
The SEDs are illustrated in Fig.\  \ref{fig:sed_incident}.
Such diverse SEDs are in particular needed to explain
and understand the range of observed properties including the strongest 
\CIII\ equivalent widths, which have not been explored in star-forming 
galaxy populations (e.g., \citealt{rigby2015,JR2016}). 
SEDs with different shapes of the ionizing spectrum 
(i.e.,\ wavelengths $\lambda <912$ \AA), 
sometimes also referred to as the hardness of the SED, 
 directly affect the predicted 
 emission line ratios.
For predictions of the  equivalent widths (EWs) of emission lines, that is,\ the ratio of line flux with 
respect to the continuum, the observable UV continuum 
(here $\lambda \sim 1500-2000$ \AA) is also of importance. 
Therefore it is important to also examine models with different ratios of the 
ionizing photon flux to UV luminosity, that is,\ different parameters 
\xiion\ $=Q_{{\rm H}^0}/L_{\rm UV}$ (e.g., \citealt{robertson2013}).
We now explain the different SED models.
The models presented here do not consider the effects of shocks
on the emission lines, which are discussed, for example, by \cite{allen2008}.
In general, the effects of shocks on the integrated UV spectra of
star-forming galaxies are expected to be secondary, and few
predictions exist on this topic.
\citet{JR2016} show how \CIII\ emission is increased by shocks 
with low velocities and strong magnetic fields, leading to spectral 
signatures comparable to those of AGN.

%%%%%%%%%%%%%%%%%%%%%%%%%%%%%%%%%%%%%%%%%%%%%%%%%%%%
\subsection{Star-forming galaxies} \label{ssec:SFGs}
%%%%%%%%%%%%%%%%%%%%%%%%%%%%%%%%%%%%%%%%%%%%%%%%%%%%

As the reference case for models of star-forming galaxies, we adopt
the incident radiation fields from the population
synthesis code \textsc{PopStar} \citep{molla2009} for a 
\citet{chabrier2003} IMF at an age of $1$\,Myr,
which have been studied in detail and compared to data from the Sloan survey by \citet{stasinska2015}.
Such young instantaneous burst models are usually used for
modeling \HII-regions (e.g., \citealt{KD2002})
since the nebular line ratios are sensitive to the youngest 
stellar population; age effects are discussed later 
(Sect.\ \ref{ssec:results_ages}). 
Stellar metallicities are matched to the gas-phase ones.
The shapes of the ionizing photon spectra with eight different
metallicities are shown as solid curves 
in panel (a) of Figure \ref{fig:sed_incident}.
The model calculations are stopped when the electron fraction, 
defined as the ratio of the number density of electrons to that of 
total hydrogen, falls below $10^{-2}$.

The \textsc{PopStar} models are based on a single stellar population. 
In order to include a binary evolution in modeling the stellar 
populations, we additionally consider SEDs provided by 
BPASS (v2; \citealt{stanway2015}) for models of 
star-forming galaxies.
We use publicly available%
\footnote{
\url{http://bpass.auckland.ac.nz/2.html}
} BPASS SEDs 
adopting similar assumptions as in the \textsc{PopStar} models,
assuming an instantaneous starburst at an age of $1$\,Myr, 
a \citet{kroupa2001}-like IMF with the upper-mass cut of $300$\,\Msun, 
and a binary stellar population.
We call the models ``BPASS-300bin'' models.
The stellar metallicities of $Z=0.05$--$1\,Z_{\odot}$ are used for 
the calculation.
For reference the shape of the ionizing spectrum with a metallicity of 
$Z=0.05\,Z_{\odot}$ is shown as a dotted curve in the panel (a) of 
Figure \ref{fig:sed_incident}.
The BPASS-300bin models provide a slightly harder shape of the ionizing 
spectrum than the \textsc{PopStar} models. A notable enhancement is 
found in the high-energy regime of $E\sim 50$\,eV. 
If the models are normalized at a UV wavelength ($\sim 1500$\,\AA) 
instead of at $912$\,\AA, the BPASS models present a larger number of 
ionizing photons than the \textsc{PopStar} models in 
Fig. \ref{fig:sed_incident}(a). 
This means the BPASS-300bin models show a higher \xiion\ parameter
than the single stellar population models for a given metallicity, as 
expected (cf. \citealt{stanway2015,trainor2016}).
Therefore we expect emission line ratios, which are only sensitive to the
shape of the ionizing spectrum, to differ less than line equivalent widths
between models using the \textsc{PopStar} and BPASS SEDs.

%%%%%%%%%%%%%%%%%%%%%%%%%%%%%%%%%%%%%%%%%%%%%%%%%%%%
\subsection{Active galactic Nuclei} \label{ssec:AGNs}
%%%%%%%%%%%%%%%%%%%%%%%%%%%%%%%%%%%%%%%%%%%%%%%%%%%%

We consider a narrow-line region (NLR) surrounding an AGN ionizing radiation field,
which is frequently characterized by a power law.
Our AGN models are generated by the \cloudy\ ``AGN'' continuum command
with the default parameters except for the power-law energy slope between the optical
and X-ray bands, $\alpha_{ox}$ \citep{zamorani1981}. 
The parameter corresponds to the power-law index
$\alpha$, where $f_{\nu}\propto \nu^{\alpha}$, determined in the range of 
a few to a few thousand eV%
\footnote{
E.g., \citet{elvis2002} define the $\alpha_{ox}$ index as 
$\alpha_{ox}=\log(f_x/f_o)/2.605$, where $f_o$ and $f_x$ are the fluxes per unit 
frequency at $2500$\,\AA\ ($\sim4.96$\,eV) and $2$\,keV, respectively.
The value $2.605$ is calculated as $\sim \log(2000/4.96)$.
\citet{elvis2002} derive a typical value of the $\alpha_{ox}$ index 
in the range from $-1.6$ to $-1.4$.
}.
Hereafter, the notation $\alpha$ stands for the $\alpha_{ox}$ index.
Models are constructed with the parameters of 
$\alpha=-1.2,\,-1.4,\,-1.6,\,-1.8,\,-2.0$.
The shapes of the ionizing photon spectra with five different $\alpha$
are shown in Figure \ref{fig:sed_incident}(b).
We note that the shape of the produced AGN continuum in the 
UV and longer-wavelength regimes
is provided based on 
observations of quasars (e.g., \citealt{elvis1994,francis1993}). The 
continuum is considered to be dominated by emission from 
the accretion disk, 
especially in the UV wavelength regime 
we are interested in (e.g., \citealt{elvis1994}).
Since we consider type-II AGNs, 
in which the accretion disk is hidden by torus, 
the UV continuum might be overestimated by our models 
due to a possible attenuation.
The EWs of the AGN models could thus become larger 
than calculated here if the attenuation is taken into account
(see Sect.\ \ref{ssec:results_AGNs}).
The AGN models are truncated at a neutral column density of 
$N$(\HI) $=10^{21}$\,cm$^{-2}$, in accordance with the NLR models 
of \citet{kewley2013_theory}.
For the AGN models, we adopt dust-free gas clouds and assume 
no depletion of elements onto dust grains, since previous studies 
suggest that dust-free models are in better agreement with the 
observations of high-redshift radio galaxies, type-II QSOs, and 
local type-II AGNs than dusty models (e.g., \citealt{nagao2006a}). 
For the AGN models, we additionally consider a further metal-rich 
gas cloud, $Z=5\,Z_{\odot}$.

%%%%%%%%%%%%%%%%%%%%%%%%%%%%%%%%%%%%%%%%%%%%%%%%%%%%
\subsection{Star-forming galaxies with an active galactic nucleus} \label{ssec:SFGs+AGNs}
%%%%%%%%%%%%%%%%%%%%%%%%%%%%%%%%%%%%%%%%%%%%%%%%%%%%

The shape of the highest-energy ionizing photon spectrum at 
$E>40$--$50$\,eV for star-forming galaxies is not yet fully understood.
Current stellar population synthesis codes 
may not properly account for the evolution of very massive stars 
in the low-metallicity regime
and could underestimate the number of high-energy ionizing photons 
(e.g., \citealt{stasinska2015}).
We consider models of star-forming galaxies with 
an additional contribution of an AGN. 
The SEDs of the stellar population are provided by \textsc{PopStar} models 
at an age of $1$\,Myr as described in Sect.\ \ref{ssec:SFGs}.
For the AGN component, we use a power-law spectrum with the index 
$\alpha=-1.6$ over the range of $E=1.0$ -- $100$\,Ryd
(i.e., $E=13.6$\,eV -- $1.36$\,keV) as a default. 
We alternatively use a harder spectrum of $\alpha=-1.2$ 
to check if the hardness affects the resulting visibility of the
emission lines.
We adopt two different values of the contribution of the AGN to the total 
number of ionizing photons, $f_{\rm AGN} = 0.03$\ and $0.1$. 
The contribution of $3$\,\%\ is suggested for explaining the \HeII$/$\Hb\ ratios 
of local blue compact dwarf galaxies with a high ionization parameter
\citep{stasinska2015}, 
and thus could be needed for their high-$z$ analogs of emission-line galaxies.
The combined SEDs are presented in panel (c) of Figure \ref{fig:sed_incident}.
The model calculations are performed in exactly the same manner as 
for the star-forming galaxy models (Sect.\ \ref{ssec:SFGs}).

%FFFFFFFFFFFFFFFFFFFFFFFFFFFFFFFFFFFFFFFFFFFFFFFFFFFFFFFFFFFFFFFFFFFFFFFFF%
\begin{figure*}
  \centerline{
    \includegraphics[width=0.85\textwidth]{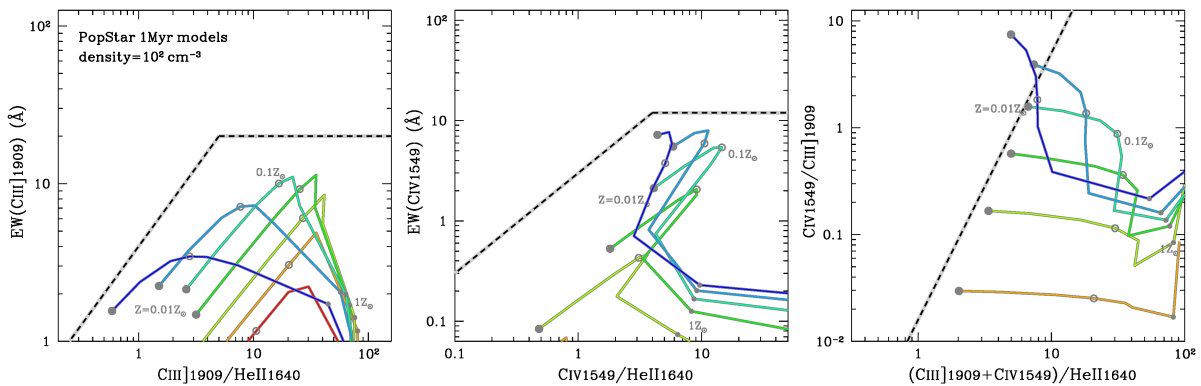}
  }
  \caption{
        UV line diagrams 
        for the SED of \textsc{PopStar} models (Fig. \ref{fig:sed_incident}a).
        A gas density of $10^2\,{\rm cm}^{-3}$ is assumed.
        The curves present our grid of models in the range of 
        $Z=0.01$--$2\,Z_{\odot}$ for $\log\,U = -3.5$ (red), $-3.0$ (orange), 
        $-2.5$ (yellow), $-2.0$ (green), $-1.5$ (cyan), $-1.0$ (skyblue), 
        and $-0.5$ (blue).
        The metallicity of $Z=0.01$, $0.1$, and $1\,Z_{\odot}$
        is denoted with 
        a large filled, medium open, and small  gray-filled circle, 
        respectively, along each of the curves.
        The black dashed curves are the demarcations 
        between a star-forming galaxy and AGN
        that will be defined in Sect.\ \ref{ssec:UV_diagnostics}. 
        }
\label{fig:models_UVlines_popstar}
\end{figure*}
%FFFFFFFFFFFFFFFFFFFFFFFFFFFFFFFFFFFFFFFFFFFFFFFFFFFFFFFFFFFFFFFFFFFFFFFFF%

%FFFFFFFFFFFFFFFFFFFFFFFFFFFFFFFFFFFFFFFFFFFFFFFFFFFFFFFFFFFFFFFFFFFFFFFFF%
\begin{figure*}
  \centerline{
    \includegraphics[width=0.85\textwidth]{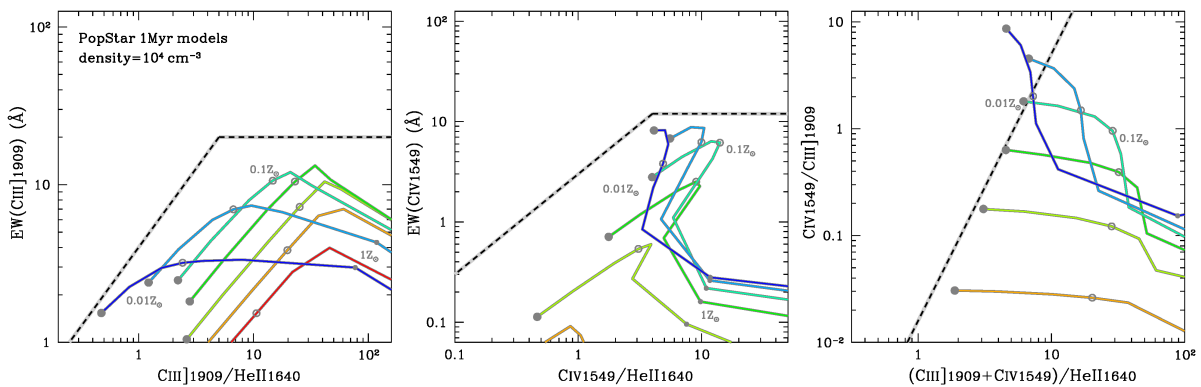}
  }
  \caption{
        As in Fig. \ref{fig:models_UVlines_popstar}, but for models 
        with a higher gas density of $10^4\,{\rm cm}^{-3}$.
        }
\label{fig:models_UVlines_popstar_d4}
\end{figure*}
%FFFFFFFFFFFFFFFFFFFFFFFFFFFFFFFFFFFFFFFFFFFFFFFFFFFFFFFFFFFFFFFFFFFFFFFFF%

%FFFFFFFFFFFFFFFFFFFFFFFFFFFFFFFFFFFFFFFFFFFFFFFFFFFFFFFFFFFFFFFFFFFFFFFFF%
\begin{figure*}
  \centerline{
    \includegraphics[width=0.85\textwidth]{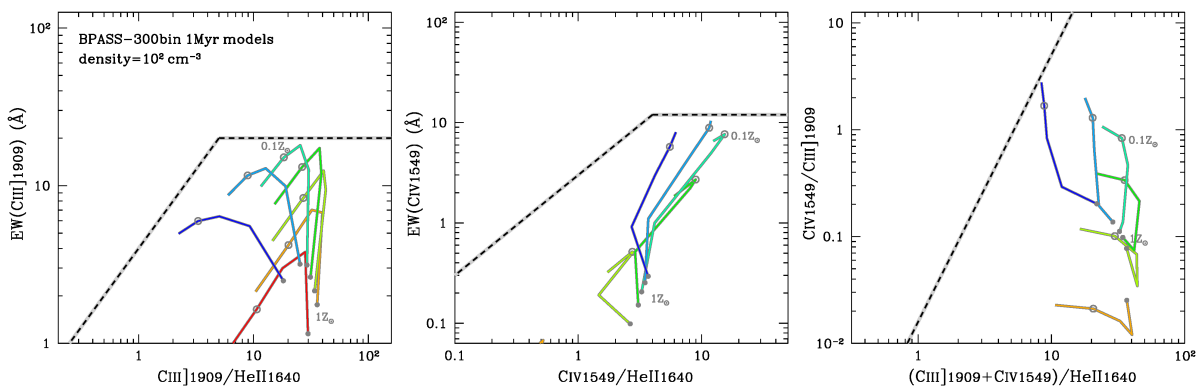}
  }
  \caption{
        As in Fig. \ref{fig:models_UVlines_popstar}, 
        but for the SED of BPASS binary star models, BPASS-300bin
        (cyan dotted curve in Fig. \ref{fig:sed_incident}a). 
        Only the metallicity range of $Z=0.05$--$1\,Z_{\odot}$
        is considered.
        }
\label{fig:models_UVlines_bpass}
\end{figure*}
%FFFFFFFFFFFFFFFFFFFFFFFFFFFFFFFFFFFFFFFFFFFFFFFFFFFFFFFFFFFFFFFFFFFFFFFFF%

%%%%%%%%%%%%%%%%%%%%%%%%%%%%%%%%%%%%%%%%%%%%%%%%%%%%
\subsection{Blackbodies} \label{ssec:BBs}
%%%%%%%%%%%%%%%%%%%%%%%%%%%%%%%%%%%%%%%%%%%%%%%%%%%%

We also consider models with ionizing spectra described by a simple blackbody.
We consider Blackbodies at very high temperature, 
$T_{\rm eff}=60,000$\,K, $80,000$\,K, and $100,000$\,K. 
The SEDs are shown in Figure \ref{fig:sed_incident}(d). 
Motivation for examining blackbodies lies in the fact that 
the shape/hardness of ionizing spectra of 
blackbodies with $T\gtrsim (6-7)\times 10^{4}$\,K resembles 
that of \textsc{PopStar} models at an age of $4$\,Myr, when 
Wolf-Rayet (WR) stars begin to dominate the ionizing spectrum 
\citep{stasinska2015}. 
We aim to examine the UV spectrum if the incident radiation
field is as hard as (or harder than) that during the WR-dominated 
phase.
Furthermore, these high-temperature blackbodies
could be the main ionizing source
if very massive stars, as observed in the central 
region of the star clusters in the nearby universe 
(e.g., \citealt{crowther2016}), dominate the integrated spectrum
of the system.
As we see below, another advantage of blackbody spectra is that
they have a higher ionizing photon flux per unit of UV luminosity,  \xiion,
than more realistic stellar spectra, as they show no discontinuity
at the Lyman edge, in contrast with stellar spectra \citep[see e.g.,\ Fig.\ 15 from][]{raiter2010}.
Although we do not know which astrophysical sources would show this
behavior, it is useful to use these models 
to examine the origin of the largest EWs of \CIII\ 
found in some VUDS spectra.

%%%%%%%%%%%%%%%%%%%%%%%%%%%%%%%%%%%%%%%%%%%%%%%%%%%%
\subsection{PopIII star} \label{ssec:popIII}
%%%%%%%%%%%%%%%%%%%%%%%%%%%%%%%%%%%%%%%%%%%%%%%%%%%%

The final scenario we consider is the PopIII stellar population 
as the main ionizing source.
We take one PopIII SED from \citet{raiter2010} at zero-age 
with a \citet{salpeter1955} IMF and its lower and upper mass 
cut-offs of $1$\,\Msun\ and $100$\,\Msun, respectively.
The SED is presented in panel (d) of Figure \ref{fig:sed_incident} 
(dotted).
For the popIII models, we additionally consider extremely metal-poor 
gas nebulae of $Z=10^{-4}\,Z_{\odot}$ and $10^{-3}\,Z_{\odot}$. 
The model calculations are terminated at the electron fraction of $10^{-2}$
as done for the star-forming galaxy models (Sect. \ref{ssec:SFGs}).

%FFFFFFFFFFFFFFFFFFFFFFFFFFFFFFFFFFFFFFFFFFFFFFFFFFFFFFFFFFFFFFFFFFFFFFFFF%
\begin{figure*}
  \centerline{
    \includegraphics[width=0.85\textwidth]{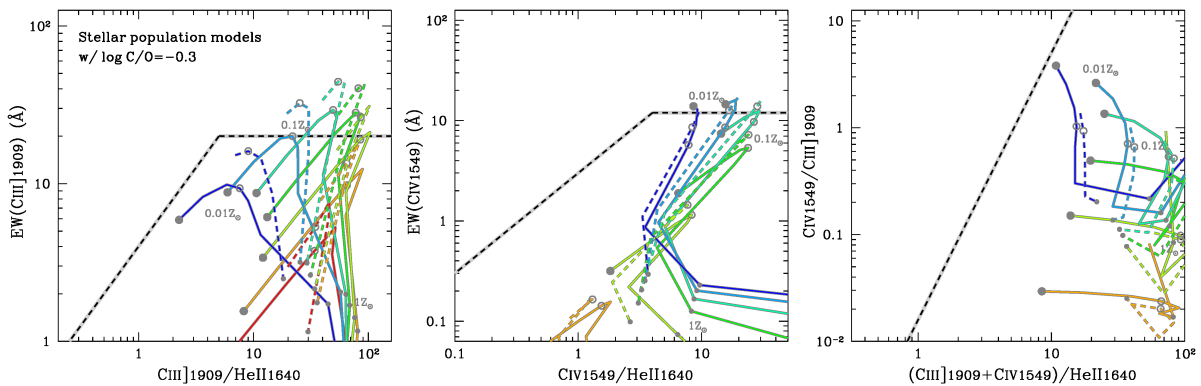}
  }
  \caption{
        As in Figs. \ref{fig:models_UVlines_popstar} and \ref{fig:models_UVlines_bpass}
        but for models with a fixed C$/$O abundance ratio of $\log$\,C$/$O $=-0.3$.
        The solid and dashed curves are the \textsc{PopStar} and 
        the BPASS-300bin models, respectively.
        }
\label{fig:models_UVlines_sfg2_co}
\end{figure*}
%FFFFFFFFFFFFFFFFFFFFFFFFFFFFFFFFFFFFFFFFFFFFFFFFFFFFFFFFFFFFFFFFFFFFFFFFF%

%%%%%%%%%%%%%%%%%%%%%%%%%%%%%%%%%%%%%%%%%%%%%%%%%%%%
%%%%%%%%%%%%%%%%%%%%%%%%%%%%%%%%%%%%%%%%%%%%%%%%%%%%
\section{UV line predictions} \label{sec:results_model}
%%%%%%%%%%%%%%%%%%%%%%%%%%%%%%%%%%%%%%%%%%%%%%%%%%%%
%%%%%%%%%%%%%%%%%%%%%%%%%%%%%%%%%%%%%%%%%%%%%%%%%%%%

In this Section we present the photoionization model predictions 
of the UV line fluxes and EWs that can be directly compared 
with the observations.
Specifically, in this paper we use three diagrams of 
\CIV$/$\CIII\ versus (\CIII$+$\CIV)$/$\HeII, 
EW(\CIII) versus \CIII$/$\HeII, and 
EW(\CIV) versus \CIV$/$\HeII. 
These three lines were chosen as they are the 
three most prominent UV emission lines after \Lya\ 
observed in the spectra of galaxies at $z=2$ -- $4$
(e.g., \citealt{lefevre2015}).
The first line ratios diagram is hereafter referred to as the C4C3--C34 diagram.
This combination is motivated by the optical line ratios diagram of
$[$\OIII$\lambda 5007$$/$\OII$\lambda 3727$ versus 
(\OII$+$$[$\OIII)$/$\Hb\ 
%R23-index 
(e.g., \citealt{NO2014}), which is useful for inferring the metallicity and ionization parameter.
The latter two diagrams use, in part, EWs, which are 
observable even from galaxies at high-$z$ 
(e.g., \citealt{stark2015_c3,stark2015_c4}).
These three diagrams confer the advantage that objects can be 
plotted and diagnosed even if only one of the three lines is detected. 
The main parameters that govern the positions in the diagrams are: 
metallicity,
ionization parameter, 
density, and 
shape of ionizing photon spectrum.
In addition, for star-forming galaxies, the age of the current 
star-formation (or \xiion) is 
also important for the prediction of EWs.
In the following subsections, we explain the behaviors 
of each of the models presented in Sect.\ \ref{sec:modelling}
using the three UV diagrams described above.

%%%%%%%%%%%%%%%%%%%%%%%%%%%%%%%%%%%%%%%%%%%%%%%%%%%%
\subsection{Single stellar population} \label{ssec:results_single_stellar}
%%%%%%%%%%%%%%%%%%%%%%%%%%%%%%%%%%%%%%%%%%%%%%%%%%%%

Figure \ref{fig:models_UVlines_popstar} presents
the three diagrams of C4C3--C34, EW(\CIII)--\CIII$/$\HeII, and EW(\CIV)--\CIV$/$\HeII\
for the conventional star-forming galaxies,
using the $1$\,Myr SED from the \textsc{PopStar} models and $n_{H}=10^2$\,cm$^{-3}$
(hereafter referred to as our ``default model"), 
as a function of metallicity and ionization parameter. 
Variations of gas density and age of SEDs are discussed below.

An interesting feature found in 
Figs. \ref{fig:models_UVlines_popstar} and \ref{fig:models_UVlines_popstar_d4} 
is that EWs of \CIII\ and \CIV\ generated by a single stellar population
have maximum values; EW(\CIII) $\lesssim 12$\,\AA, and EW(\CIV) $\lesssim 9$\,\AA.
Since lower EWs are predicted for older ages (cf.\ below) 
this indicates that objects with an EW of \CIII\ and/or \CIV\ over $\sim 10$\,\AA\ 
are difficult to explain by photoionization of a single stellar population.
The limit for \CIII\ may be slightly relaxed to EW(\CIII) $\lesssim 18$\,\AA\ when
binary stars are included, as discussed in Sect.\ \ref{ssec:results_binary_stellar}.
The EWs of \CIII\ and \CIV\  increase with metallicity (i.e., with carbon abundance) 
in the low-metallicity regime, up to a  turn-over metallicity, above which they
decrease again.
The turn-over metallicity values for the EWs of \CIII\ and \CIV\ are
$Z=0.1-0.2$ and $0.05-0.1\,Z_{\odot}$, respectively.
These turn-overs are mainly due to the fact that the gas temperature decreases 
with increasing metallicity, and the ultraviolet collisional excitation lines  
therefore become weaker.
The EW(\CIII) increases with increasing ionization parameter up to 
$\log U\sim -2$. For even higher ionization parameters \CIV\ becomes dominant and the 
EW(\CIII) decreases.

Figure\ \ref{fig:models_UVlines_popstar} also shows that high ratios 
of \CIII\ and \CIV\ to \HeII\ are predicted, as expected \citep[cf.][]{feltre2016}.
This is due to a weak \HeII\ emission.
Since the \HeII\ is a recombination line in a He$^{++}$ region, 
very high-energy ionizing photons of $E>54.4$\,eV are needed,
which is not expected for normal stellar populations (see Fig.\ \ref{fig:sed_incident}a).
The ratios of \CIII$/$\HeII, \CIV$/$\HeII, and (\CIII$+$\CIV)$/$\HeII\ are predicted
to increase with metallicity (Fig.\ \ref{fig:models_UVlines_popstar}).
This is due to a softening of the radiation field with increasing metallicity,
which causes a decrease of \HeII\ emission.
Furthermore, \CIII\ and \CIV\ emission increases with metallicity, up to
the ``turn-over metallicity''. 
The \CIV$/$\CIII\ ratio is also sensitive to the ionization parameter,
increasing with increasing ionization parameter.

Since \CIII\ is a collisionally excited line, the strength depends on gas density,
as shown in Fig. \ref{fig:models_UVlines_popstar_d4}.
Here we present the models with $n_{H}=10^4$\,cm$^{-3}$,
which is likely to be the highest gas density observed for 
star-forming galaxies at high-$z$ \citep{sanders2016}.
The \CIII\ emission gets stronger if a higher gas density is assumed. 
The trend is visible especially in the high-metallicity regime ($Z\gtrsim 0.5\,Z_{\odot}$), 
where the carbon ions are more abundant. 
However, the change of the peaks of the EW(\CIII) are minimal 
over $2$ orders of magnitude in density. 
Since the maximum value of the EW(\CIII) is obtained if the gas metallicity is 
$Z= 0.1-0.2\,Z_{\odot}$, 
the enhancement of the maximum value is not prominent. 
The critical density for \CIII\ is about $3\times 10^9$\,cm$^{-3}$, and
the collision de-excitation is almost negligible for these models.

%%%%%%%%%%%%%%%%%%%%%%%%%%%%%%%%%%%%%%%%%%%%%%%%%%%%
\subsection{Binary stellar population} \label{ssec:results_binary_stellar}
%%%%%%%%%%%%%%%%%%%%%%%%%%%%%%%%%%%%%%%%%%%%%%%%%%%%

Figure \ref{fig:models_UVlines_bpass} 
shows the UV lines and EWs predicted for the binary stellar population models.
These cases correspond to galaxies whose ionizing spectrum 
is harder than previously suggested by single stellar population models
(e.g., \citealt{steidel2016}).
A feature of the binary models is that \HeII\ emission becomes stronger compared to 
the  \textsc{PopStar} models (see Fig.\ \ref{fig:models_UVlines_popstar}), due to a harder 
ionizing spectrum of the binary models \citep{stanway2015}.

A notable difference between the single and binary star models lies in 
the EWs of \CIII\ and \CIV. The binary models predict larger EWs by 
a factor of $\sim 1.4$, which is primarily due to the increased ionizing photon flux per 
UV continuum luminosity (measured e.g.,\ by \xiion).
If the BPASS SEDs are appropriate, we expect star-forming galaxies 
to show EWs of \CIII\ and \CIV\ as large as  $\sim 20$\,\AA\ and $\sim 12$\,\AA,
respectively.

Predictions concerning emission line ratios are almost the same as seen 
in the single stellar population. 
The \CIV$/$\CIII\ ratio does not change significantly from single to binary stellar populations, 
since the shape/hardness of the ionizing radiation field is almost indistinguishable in the energy 
range below $E\sim 48$\,eV, as seen in Fig. \ref{fig:sed_incident}(a).

%FFFFFFFFFFFFFFFFFFFFFFFFFFFFFFFFFFFFFFFFFFFFFFFFFFFFFFFFFFFFFFFFFFFFFFFFF%
\begin{figure*}[h]
  \centerline{
    \includegraphics[width=0.75\textwidth]{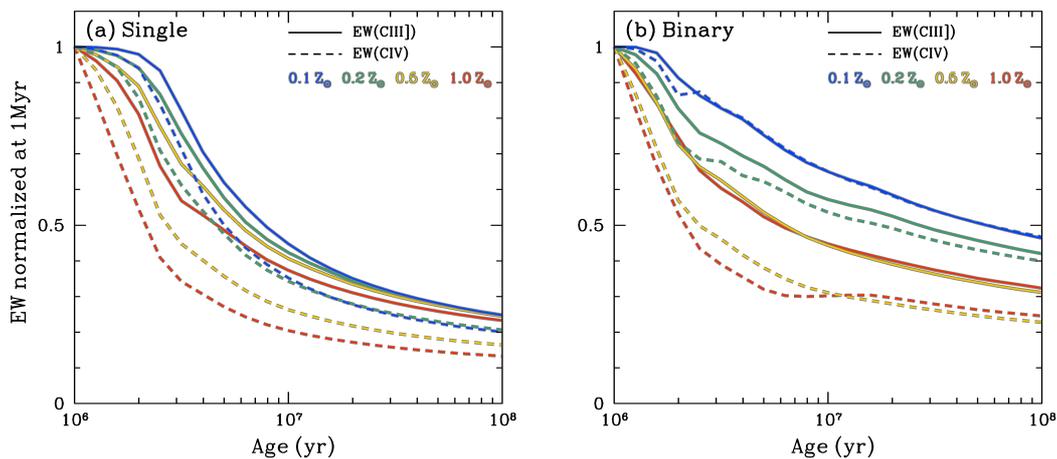}
  }
  \caption{
        EWs of \CIII\ (solid) and \CIV\ (dashed) as a function of age 
        for single- (left) and binary- (right) stellar population
        under the assumption of continuous star-formation.
        The age is defined since the beginning of the 
        current star-formation. 
        They are normalized at $1$\,Myr.
        The decrease of EW of \CIII\ (\CIV) is estimated 
        by counting the number of photons with an energy of $>24.4$\,eV
        ($47.9$\,eV) divided by the UV (stellar$+$nebular) continuum 
        around the \CIII\ (\CIV) emission at a given age. 
        The blue, green, yellow, and red curves denote the age 
        dependencies for a metallicity of $Z=0.1$, $0.2$, $0.5$, and 
        $1.0\,Z_{\odot}$, respectively.
        }
        \label{fig:ewcs_age_rel}
\end{figure*}
%FFFFFFFFFFFFFFFFFFFFFFFFFFFFFFFFFFFFFFFFFFFFFFFFFFFFFFFFFFFFFFFFFFFFFFFFF%

%%%%%%%%%%%%%%%%%%%%%%%%%%%%%%%%%%%%%%%%%%%%%%%%%%%%
\subsection{Variation in C$/$O abundance} \label{ssec:results_CO}
%%%%%%%%%%%%%%%%%%%%%%%%%%%%%%%%%%%%%%%%%%%%%%%%%%%%

As the standard model we assume the formula of \citet{dopita2006}
to give a Carbon abundance at each metallicity.
Although the relation is generally in good agreement with observations
of galaxies at $z=2-3$ (e.g., \citealt{steidel2016}), a non-negligible scatter
toward higher C$/$O is also reported (e.g., \citealt{amorin2017}). 
Since the C$/$O abundance ratio affects the predicted strengths of 
the \CIII\ and \CIV\ lines, a possible variation in C$/$O ratio
needs to be explored. In this subsection, we present models with a 
higher C$/$O ratio, which might be applicable to some strongly \CIII\ emitting galaxies.

Figure \ref{fig:models_UVlines_sfg2_co} shows the \textsc{PopStar} and 
the BPASS-300bin models with a fixed C$/$O ratio of $\log$\,C$/$O $=-0.3$, 
a solar value, irrespective of metallicity.
Compared to the values given by the \citet{dopita2006} formula,
the C$/$O abundance ratio is higher by 
$0.6$, $0.6$, $0.6$, $0.5$, $0.4$, and $0.2$\,dex
for the metallicities $Z=0.01$, $0.02$, $0.05$, $0.1$, $0.2$, and $0.5\,Z_{\odot}$ in our models.
Figure \ref{fig:models_UVlines_sfg2_co} shows that 
a high C$/$O abundance ratio increases carbon-to-helium line ratios 
and the EWs of the carbon emission. 
The EWs and the C34 ratio scale approximately linearly with the C$/$O ratio.
Therefore, the increase with respect to our standard models is more significant 
for low-metallicity models.
The C4C3 ratio is less sensitive to the change of the C$/$O ratio,
as expected.

In some models the EWs of \CIII\ and/or \CIV\ become 
larger than the limits, EW(\CIII) $\simeq 20$\,\AA\ and 
EW(\CIII) $\simeq 12$\,\AA, that we find in the standard models.
These models have a low metallicity ($Z= 0.05-0.2\,Z_{\odot}$
where the \CIII\ and/or \CIV\ emission peaks) but a 
Carbon abundance as high as the solar value in C$/$O, which can
hardly be considered normal (cf. Sect.\ \ref{ssec:discussion_origin_highCO}).
Subsequently we therefore adopt the standard models 
using the prescription of \citet{dopita2006} for the C$/$O ratio,
and refer to models with higher C$/$O ratios only when the standard
models are insufficient.

%%%%%%%%%%%%%%%%%%%%%%%%%%%%%%%%%%%%%%%%%%%%%%%%%%%%
\subsection{Age effects} \label{ssec:results_ages}
%%%%%%%%%%%%%%%%%%%%%%%%%%%%%%%%%%%%%%%%%%%%%%%%%%%%

Although the youngest stellar population of $1$\,Myr is convincingly 
reproducing emission line ratios of star-forming galaxies, 
it is not adequate to discuss their EWs since the continuum level strongly 
depends on the older stellar population.
Here we discuss the age dependencies of the EWs for both 
the single and binary-stellar population models.
Age is defined here as the time since the onset of the current 
star-formation period.
The ages are sensitive to the UV continuum level,
the shape of the ionizing spectrum, and thus to the
\xiion\ parameter.
Figure \ref{fig:ewcs_age_rel} presents the EWs of \CIII\ and \CIV\ 
and their changes as a function of age assuming a 
constant star formation.
They are normalized at the age of 1\,Myr, where the EW is maximal.
To first order, as shown by the solid (dotted) lines, the relative decrease 
of the EW of \CIII\ (\CIV) with age is determined 
by the decrease of the number of ionizing photons with energies 
above the corresponding ionization potentials of 
$E>24.4$\,eV ($E>47.9$\,eV)
and the increase of the non-ionizing UV-continuum flux density 
around $1909$\,\AA\ ($1549$\,\AA).
Moreover, the nebular continuum is added to the UV continuum,
as predicted by  \cloudy. For this Figure we choose a typical ionization parameter
\footnote{
$\log U$=$-2.3$, $-2.5$, $-2.8$, and $-3.0$
for $Z=0.1$, $0.2$, $0.5$, and $1\,Z_{\odot}$
following an empirical relation (e.g., \citealt{kojima2017}).
} and scale the nebular continuum with the number of ionizing photons.
The decreases of EWs with age are in good agreement with
the predictions directly modeled by \citet{JR2016}
for both the single and binary stellar population models.
For example, for the single stellar population with $Z=0.2\,Z_{\odot}$, 
the EW of \CIII\ (\CIV) is suggested to decrease by a factor of about 
$2.4$ ($2.9$) and $4$ ($5$) at the age of $10$ and $100$\,Myr, 
respectively, with respect to the EW at the age of $1$\,Myr. 
On the other hand, if the binary stellar population is assumed,
the EWs of \CIII\ (\CIV) become weaker by a factor of about 
$1.7$ ($1.9$) and $2.4$ ($2.5$) at the age of $10$ and $100$\,Myr, 
respectively, 
even with the same metallicity of $Z=0.2\,Z_{\odot}$.
The age dependence of EW is less significant if the binary stellar 
population is assumed \citep{JR2016}, as the effects of binary 
evolution tend to prolong the epoch over which blue stars dominate 
the spectrum \citep{stanway2017}.

%FFFFFFFFFFFFFFFFFFFFFFFFFFFFFFFFFFFFFFFFFFFFFFFFFFFFFFFFFFFFFFFFFFFFFFFFF%
\begin{figure*}[h]
  \centerline{
    \includegraphics[width=0.85\textwidth]{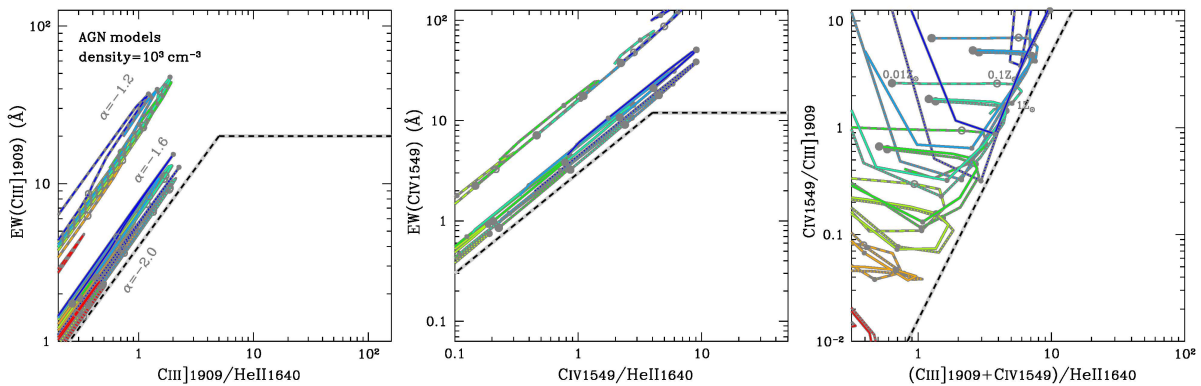}
  }
  \caption{
        As in Fig. \ref{fig:models_UVlines_popstar}, 
        but for the SED of AGN models (Fig. \ref{fig:sed_incident}b).
        Models have power-law indices of  
        $\alpha=-1.2$ (dashed), $-1.6$ (solid), and $-2.0$ (dotted) 
        and the metallicity ranges from $Z=0.01$ to $5\,Z_{\odot}$.
        }
\label{fig:models_UVlines_agn}
\end{figure*}
%FFFFFFFFFFFFFFFFFFFFFFFFFFFFFFFFFFFFFFFFFFFFFFFFFFFFFFFFFFFFFFFFFFFFFFFFF%

%FFFFFFFFFFFFFFFFFFFFFFFFFFFFFFFFFFFFFFFFFFFFFFFFFFFFFFFFFFFFFFFFFFFFFFFFF%
\begin{figure*}[h]
  \centerline{
    \includegraphics[width=0.85\textwidth]{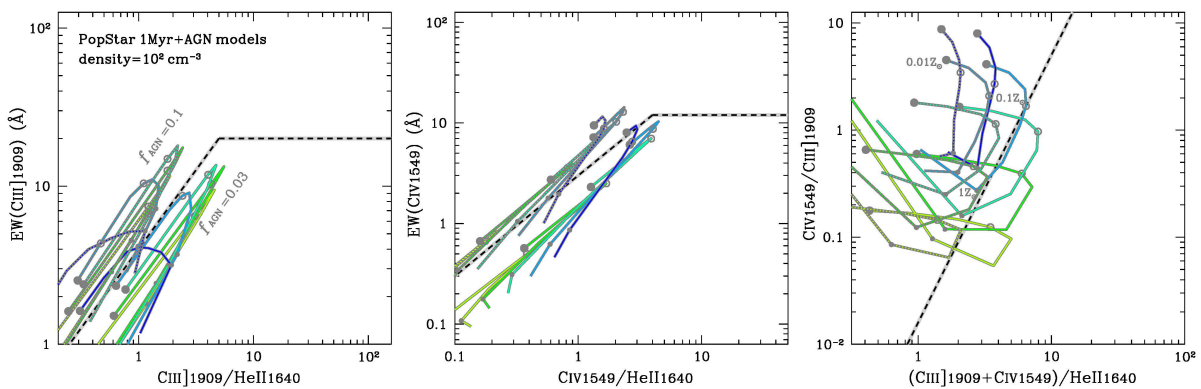}
  }
  \caption{
        As in Fig. \ref{fig:models_UVlines_popstar}, 
        but for the SED of \textsc{PopStar}$+$AGN ($\alpha=-1.6$)
        models with the AGN contribution of 
        $f_{\rm AGN} = 0.03$ (solid; Fig. \ref{fig:sed_incident}c)
        and $0.1$ (dotted; Fig. \ref{fig:sed_incident}c).
        Only the models for the ionization parameters of $\log U > -2.5$ are displayed.
        }
\label{fig:models_UVlines_popstar_agn}
\end{figure*}
%FFFFFFFFFFFFFFFFFFFFFFFFFFFFFFFFFFFFFFFFFFFFFFFFFFFFFFFFFFFFFFFFFFFFFFFFF%

%%%%%%%%%%%%%%%%%%%%%%%%%%%%%%%%%%%%%%%%%%%%%%%%%%%%
\subsection{Active galactic nuclei} \label{ssec:results_AGNs}
%%%%%%%%%%%%%%%%%%%%%%%%%%%%%%%%%%%%%%%%%%%%%%%%%%%%

Figure\ \ref{fig:models_UVlines_agn} shows the grid of AGN models 
with power-law indices of $\alpha=-2.0$, $-1.6$, and $-1.2$.
The UV line diagrams look quite different if the AGN SEDs are adopted 
instead of the SEDs discussed in previous sub-sections.
One notable difference is the lack of high (\CIII$+$\CIV)$/$\HeII\ 
(and \CIII$/$\HeII, \CIV$/$\HeII) ratio.
This is due to the strong \HeII\ emission in the AGN models
(e.g., \citealt{feltre2016}).
The power-law spectra are hard enough to provide 
a large number of ionizing photons with energies above $54.4$\,eV, 
by $2-3$ orders of magnitudes more than pure star-forming galaxies 
(see Fig. \ref{fig:sed_incident}b).
The AGNs' hard spectra also enhance the \CIV\ and \CIII\ emission. 
However, the enhancement of \CIII\ is less significant 
because the AGN models assume dust-free gas clouds and 
lack the enhancement of the line strength due to the photoelectric heating 
in photoionized regions \citep{vanhoof2004}.
If we adopt a harder $\alpha$ index from $\alpha=-2.0$ to $-1.2$, 
the \HeII\ and \CIV\ become stronger when compared to \CIII.
Despite the changes of the line ratios, AGN models are generally
distributed in a well-defined area on the C4C3--C34 diagram, separated from
the star-forming galaxy models, as also shown by the ``dividing'' lines
shown in Fig.\ \ref{fig:models_UVlines_agn} and discussed below.

Adopting a higher gas density ($n_{\rm H}=10^5$\,cm$^{-3}$
instead of $n_{\rm H}=10^3$\,cm$^{-3}$), leads to changes of the EWs and 
line ratios similar to those seen in star-forming galaxies 
(Figs. \ref{fig:models_UVlines_popstar} vs. \ref{fig:models_UVlines_popstar_d4}).
The higher gas density results in a stronger \CIII\ emission, typically by a factor 2.
For AGN SEDs, the maximum EW(\CIII) is obtained for metallicities close to solar.
An interesting possibility to note is that the EWs for the AGN models might be underestimated 
due to a possible attenuation of the continuum emission from 
the central accretion disk for type-II AGNs in the UV wavelength range 
(Sect.\ \ref{ssec:AGNs}).
If this were true, the EWs plotted in Fig. \ref{fig:models_UVlines_agn} would give
lower-limits of EWs for a given ISM condition and $\alpha$ index.
If corrected for this effect, the difference between the star-forming galaxy 
models and the AGNs in the two EW plots would become much more 
significant (see also Sect.\ \ref{sssec:UV_diagnostic_ews}).

%%%%%%%%%%%%%%%%%%%%%%%%%%%%%%%%%%%%%%%%%%%%%%%%%%%%
\subsection{Star-forming galaxies with an active galactic nucleus} \label{ssec:results_SFGs+AGNs}
%%%%%%%%%%%%%%%%%%%%%%%%%%%%%%%%%%%%%%%%%%%%%%%%%%%%

The UV line diagrams for star-forming galaxies with a contribution 
from an AGN are given in 
Fig. \ref{fig:models_UVlines_popstar_agn}. 
Two cases are presented, 
$f_{\rm AGN} =0.03$ and $0.1$, 
with $3$\,\%\ and $10$\,\%\ contributions of AGN to the total number of 
ionizing photons, respectively.
The main difference between pure star-formation and 
mixed star-formation$+$AGN SEDs appears in the high-energy regime of 
$E\gtrsim 50$\,eV (see Fig.\ \ref{fig:sed_incident}).
Therefore, stronger \HeII\ emission is predicted for the 
star-formation$+$AGN models.
Indeed, Fig. \ref{fig:models_UVlines_popstar_agn} shows that 
only a few percent of AGN contribution strongly affects the \HeII-related 
line ratios. If the AGN contribution exceeds $10$\,\%\ the line ratios 
are almost indistinguishable from those produced from pure AGNs.
The EWs of \CIII\ and \CIV, on the other hand, do not change significantly 
even if an AGN component is added to the ionizing radiation. 
The UV continuum level and the \xiion\ parameter are thus considered to 
be particularly important for the UV EWs.
If we adopt a harder $\alpha$ index of $\alpha=-1.2$,
the carbon-to-helium ratios increase by less than 
$\sim1.3$, and the EWs of \CIII\ and \CIV\ 
by $\lesssim1.2$.
Therefore, the hardness of the power-law component has little influence
on the UV diagrams.
Since the hardness of the power-law component and the AGN fraction
are degenerate, we simply fix the $\alpha$ index to $\alpha=-1.6$
and vary the AGN fraction in the following analyses.
We note that even the star-formation$+$AGN models cannot predict high \CIII\ 
equivalent widths EW(\CIII) $\ga 20$ \AA.

%FFFFFFFFFFFFFFFFFFFFFFFFFFFFFFFFFFFFFFFFFFFFFFFFFFFFFFFFFFFFFFFFFFFFFFFFF%
\begin{figure*}[h]
  \centerline{
    \includegraphics[width=0.85\textwidth]{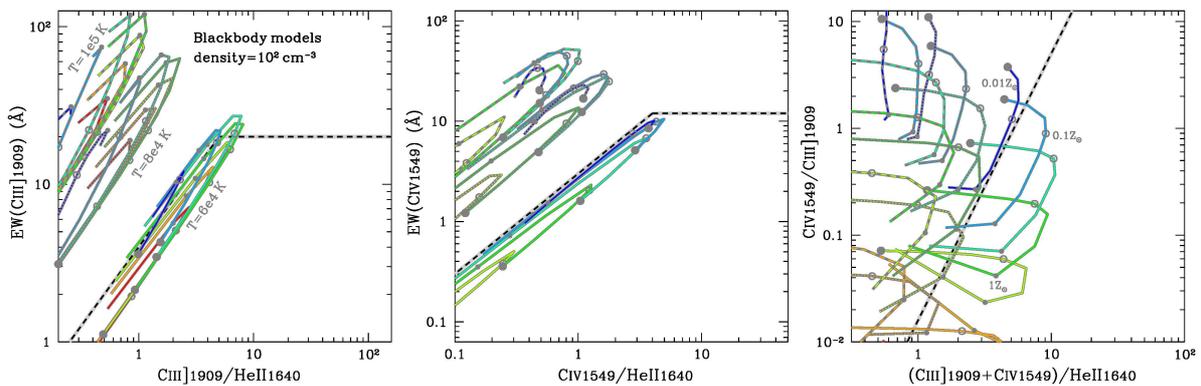}
  }
  \caption{
        As in Fig. \ref{fig:models_UVlines_popstar}, 
        but for the SED of blackbody models with an effective temperature 
        of $T_{\rm eff}=6\times 10^{4}$, $8\times 10^{4}$, and $1\times 10^{5}$\,K
        (Fig. \ref{fig:sed_incident}d).
        }
\label{fig:models_UVlines_BB}
\end{figure*}
%FFFFFFFFFFFFFFFFFFFFFFFFFFFFFFFFFFFFFFFFFFFFFFFFFFFFFFFFFFFFFFFFFFFFFFFFF%

%FFFFFFFFFFFFFFFFFFFFFFFFFFFFFFFFFFFFFFFFFFFFFFFFFFFFFFFFFFFFFFFFFFFFFFFFF%
\begin{figure*}[h]
  \centerline{
    \includegraphics[width=0.85\textwidth]{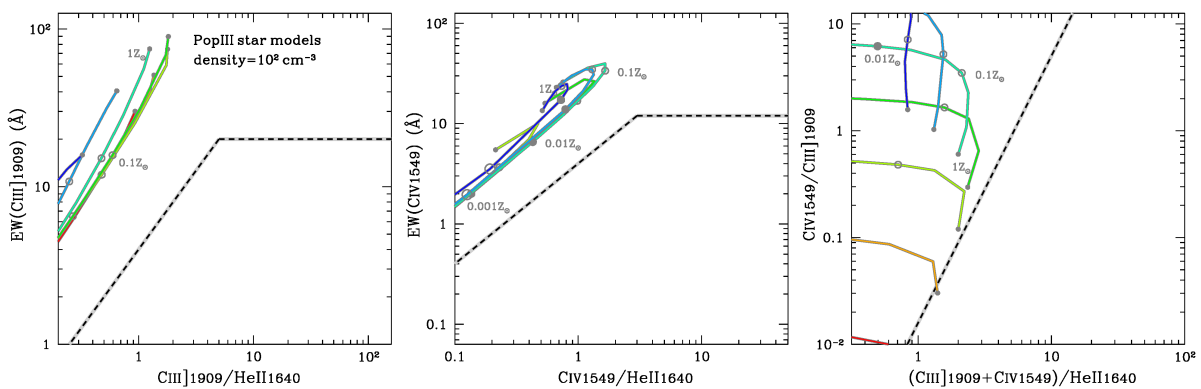}
  }
  \caption{
        As in Fig. \ref{fig:models_UVlines_popstar}, 
        but for the SED of PopIII stars (blue dotted curve in Fig. \ref{fig:sed_incident}d).
        A metallicity for each curve ranges from $Z=10^{-4}$ to $1\,Z_{\odot}$.
        }
\label{fig:models_UVlines_popIII}
\end{figure*}
%FFFFFFFFFFFFFFFFFFFFFFFFFFFFFFFFFFFFFFFFFFFFFFFFFFFFFFFFFFFFFFFFFFFFFFFFF%

%%%%%%%%%%%%%%%%%%%%%%%%%%%%%%%%%%%%%%%%%%%%%%%%%%%%
\subsection{Blackbodies} \label{ssec:results_BBs}
%%%%%%%%%%%%%%%%%%%%%%%%%%%%%%%%%%%%%%%%%%%%%%%%%%%%

Figure \ref{fig:models_UVlines_BB} presents the UV line diagrams for blackbodies 
with high effective temperatures as proxies for a harder ionizing photon spectrum 
by stars.
The shape of the SED of the blackbody with a temperature of 
$T_{\rm eff}=6\times 10^4$\,K appears similar to those provided by \textsc{PopStar} 
and BPASS in the range of $E=13.6$--$\sim 48$\,eV. 
Ionizing photons with higher energies of $E\gtrsim 50$\,eV are, however, 
much more abundant for the blackbody than in stellar population models.
Thus, stronger \HeII\ emission is expected with blackbody spectra.
However, a warning is needed for the EW predictions using blackbody spectra.
In these models, we use a single blackbody SED over the entire energy range.
Since these spectra show no Lyman break -- in contrast to stellar models
\citep[see e.g.,][]{raiter2010} -- the predicted EWs are significantly larger than derived
from more realistic stellar population models.
For example, the maximum EW of \CIII\
reaches $\sim 30$\,\AA\ and $\sim 100$\,\AA\ for blackbodies with  
$T_{\rm eff}=6\times 10^4$\,K  and $10^5$\,K, respectively.
Therefore, these EWs can be regarded as giving a very conservative upper-limit,
which could be achieved by an extremely top heavy IMF at almost zero age
or for spectra dominated by extreme massive stars such as the popIII stars 
(Sect.\ \ref{ssec:results_popIII})
The increase of the \CIV$/$\CIII\ ratio with increasing temperature, 
due to the harder ionizing spectrum, is also noticeable.
If the temperature is higher than $T_{\rm eff}\gtrsim 8\times 10^4$\,K,
the UV line ratios and EWs become similar to those of AGN.

%FFFFFFFFFFFFFFFFFFFFFFFFFFFFFFFFFFFFFFFFFFFFFFFFFFFFFFFFFFFFFFFFFFFFFFFFF%
\begin{figure*}
  \centerline{
    \includegraphics[width=1.0\textwidth]{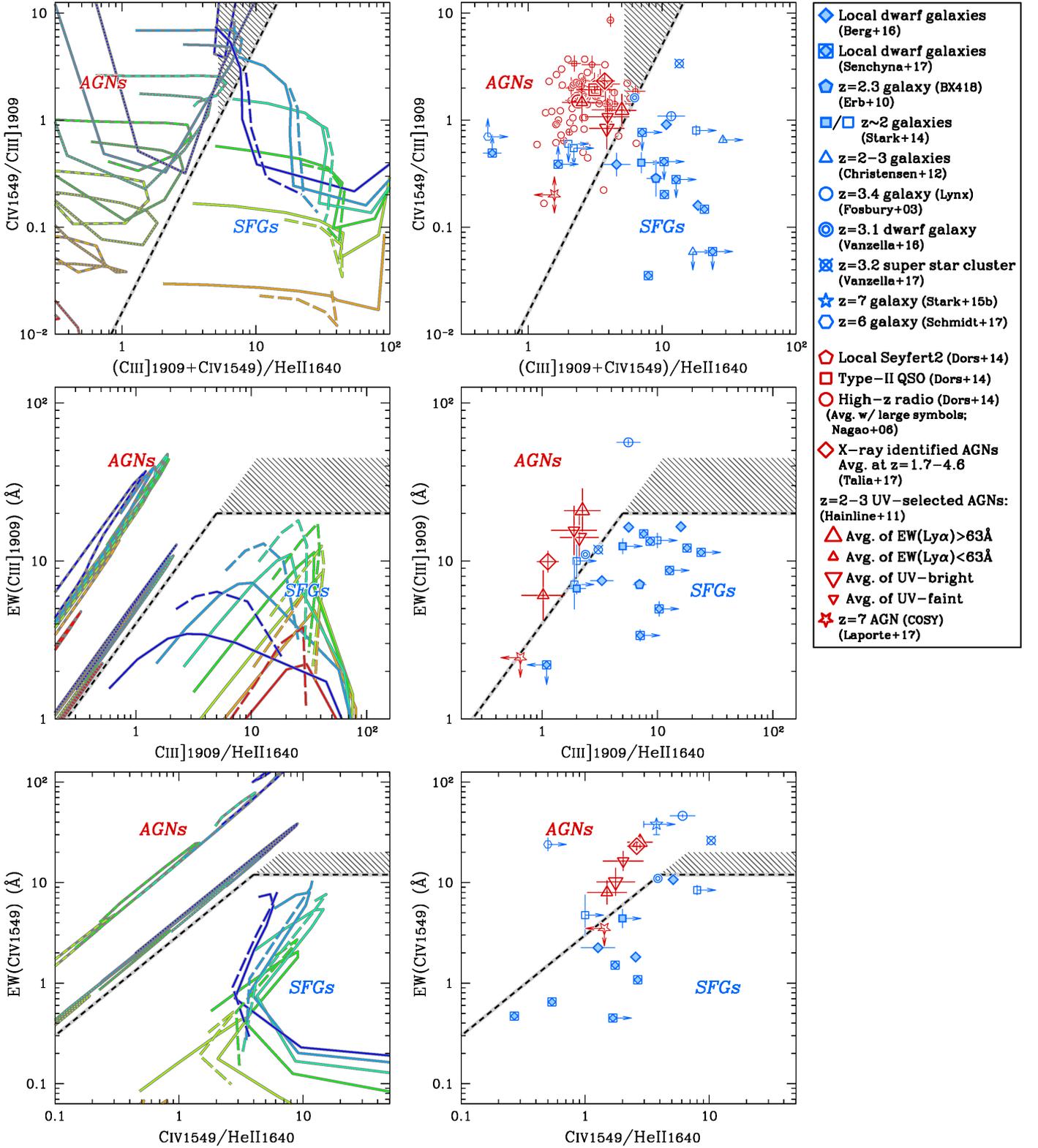}
  }
  \caption{
        UV diagnostics of star formation and AGN 
        backed up by our photoionization models (left) and 
        observational data compiled from the literature (right).
        From top to bottom, the UV lines diagrams of 
        C4C3 versus C34 (top), 
        EW(\CIII) versus \CIII$/$\HeII\ ratio (middle), and 
        EW(\CIV) versus \CIV$/$\HeII\ ratio (bottom) are presented.
        (Left)
        For star-forming galaxy, the SEDs provided by \textsc{PopStar} 
        (solid) and BPASS (long-dashed) are shown. 
        For AGN, two models with different power-law indices of 
        $\alpha=-2.0$ (dotted) and $-1.2$ (dashed) are presented.
        The same color-scheme is adopted as in 
        Fig. \ref{fig:models_UVlines_popstar}.
        (Right)
        Blue and red symbols are star-forming galaxies and AGNs, 
        respectively, compiled from the literature as shown in the legend.
        The filled blue symbols present galaxies that are supported 
        to be star-forming according to the optical BPT-diagnostic.
        The open blue symbols are objects that lack either of the optical
        BPT-lines but show no sign of AGN.
        If the EWs and/or the fluxes are below the $2\sigma$ detections, 
        we apply the $2\sigma$ upper-limits for the quantities and 
        denote them with an arrow.
        The black dashed curves are our proposed separating method
        of a star-forming galaxy from an AGN 
        (Eqs. (\ref{eq:separate_c4c3_c34})--(\ref{eq:separate_ewc4_c4he2})).
        The gray shaded areas show the regions where both AGN
        and SF models could overlap, 
        due to a low metallicity (C4C3--C34, top)
        and a high C$/$O ratio (EW diagrams; middle and bottom).
        }
\label{fig:UV_SFGsAGNs_both}
\end{figure*}
%FFFFFFFFFFFFFFFFFFFFFFFFFFFFFFFFFFFFFFFFFFFFFFFFFFFFFFFFFFFFFFFFFFFFFFFFF%

%%%%%%%%%%%%%%%%%%%%%%%%%%%%%%%%%%%%%%%%%%%%%%%%%%%%
\subsection{PopIII star} \label{ssec:results_popIII}
%%%%%%%%%%%%%%%%%%%%%%%%%%%%%%%%%%%%%%%%%%%%%%%%%%%%

Fig. \ref{fig:models_UVlines_popIII} shows the results if 
the popIII stellar spectrum is used.
The behaviour of the models in the UV line diagrams,
in particular in the C4C3--C34 diagram, 
is quite similar 
to those of the $T_{\rm eff}=10{^5}$\,K blackbody models in 
Fig. \ref{fig:models_UVlines_BB}. 
This is reasonable, since the blackbody with a temperature of 
$T_{\rm eff}\simeq 1\times10{^5}$\,K provides a good approximation
to the popIII SED below $912$\AA\
\citep{raiter2010}.
Since the more realistic popIII SED of \citet{raiter2010} shows a Lyman break 
due to the hydrogen opacity in the atmosphere of the hot stars, 
it possesses a higher level of UV continuum. 
This results in the slightly decreased EWs of \CIII\ and \CIV\ 
(by a factor of $\lesssim 2$)
compared to the $T_{\rm eff}=10{^5}$\,K blackbody models.
Overall, the UV line ratios and the EWs of the popIII star models
resemble those of AGN.

We should note that if we assume a very metal-poor gas of 
$Z<10^{-3}\,Z_{\odot}$,
the EW(\CIII) becomes smaller than $\sim 1$\,\AA\ and
the C34 index smaller than $\sim 0.1$.
These popIII galaxy models fall out of the ranges presented in
Fig. \ref{fig:models_UVlines_popIII}.
Therefore, the models presented in Fig. \ref{fig:models_UVlines_popIII}
correspond to cases where the interstellar medium is relatively 
evolved (enriched), while the ionizing radiation field is very hot/hard as represented 
by pop III-like stellar spectrum. Whether or not such conditions are realized in nature remains to be determined.

%%%%%%%%%%%%%%%%%%%%%%%%%%%%%%%%%%%%%%%%%%%%%%%%%%%%
%%%%%%%%%%%%%%%%%%%%%%%%%%%%%%%%%%%%%%%%%%%%%%%%%%%%
\section{Applications of the UV diagrams} 
\label{sec:results_UV_diagrams}
%%%%%%%%%%%%%%%%%%%%%%%%%%%%%%%%%%%%%%%%%%%%%%%%%%%%
%%%%%%%%%%%%%%%%%%%%%%%%%%%%%%%%%%%%%%%%%%%%%%%%%%%%

We now show how the UV diagrams allow for distinction between AGN 
and stellar sources, and how they can be used to constrain 
physical properties of the ISM and star-formation ages.

%%%%%%%%%%%%%%%%%%%%%%%%%%%%%%%%%%%%%%%%%%%%%%%%%%%%
\subsection{UV diagnostics to separate AGN from stellar photoionization} 
\label{ssec:UV_diagnostics}
%%%%%%%%%%%%%%%%%%%%%%%%%%%%%%%%%%%%%%%%%%%%%%%%%%%%

The \cloudy\ models presented in Sect.\ \ref{sec:results_model} show
that UV lines are sensitive to the shape of the incident radiation field.
This implies that, in an analogous manner to that for optical lines, the UV lines
can be used to distinguish the nature of the dominant ionizing source,
that is, stellar photoionization of AGN-dominated objects, as already
shown by earlier studies (e.g., \citealt{feltre2016,gutkin2016}).
Here we discuss so-called diagnostic diagrams based on the \CIII, \CIV, 
and \HeII\ lines, the three most prominent/informative UV 
emission lines, 
including for the first time (to the best of our knowledge) 
diagnostics involving the equivalent widths of these UV lines.

We note that we have tested and confirmed that our models of 
star-forming galaxies and AGNs successfully reproduce the optical 
BPT diagram (\citealt{baldwin1981,kewley2001,kauffmann2003_agn}), 
except for the metal-poor AGN models with $Z<0.5\,Z_{\odot}$.
The difficulty in isolating metal-poor AGNs with the BPT diagram has 
already been pointed out by earlier theoretical work
(e.g., \citealt{kewley2013_theory}), suggesting that a different 
diagnostic might be needed for completely distinguishing AGNs from 
star-forming galaxies. 
The UV diagnostic diagrams we present below potentially have the ability
to work for such metal-poor AGNs.

% -% -% -% -% -% -% -% -% -% -% -% -% -% -% -% -% -% -% -% -% -% -% -% -% -% -% -% -% -% -% 
\subsubsection{Diagnostic of C4C3--C34} \label{sssec:UV_diagnostic_c34}
% -% -% -% -% -% -% -% -% -% -% -% -% -% -% -% -% -% -% -% -% -% -% -% -% -% -% -% -% -% -%

The top left panel of Fig. \ref{fig:UV_SFGsAGNs_both} shows 
the line ratios of 
\CIV$/$\CIII\ (C4C3) and (\CIII$+$\CIV)$/$\HeII\ (C34) 
for the models with the SEDs of star-forming galaxies and AGNs.
We display both the single and binary stellar models for 
star-forming galaxies. For the AGN models, we plot those 
for the softest and the hardest spectra (i.e.,\ $\alpha=-2$ and $-1.2$).
A remarkable difference is found in the C34 parameter between 
stellar and AGN models, such that AGNs tend to have a smaller
C34 value for a fixed C4C3. 
The difference is primarily caused by the \HeII\ strength,
which is sensitive to the shape of the ionizing spectrum in the 
very high-energy regime of $E>54.4$\,eV (Sect.\ \ref{sec:results_model}).

The top-right panel of Fig. \ref{fig:UV_SFGsAGNs_both} shows 
the same C4C3--C34 diagram, illustrating now the position of 
observed star-forming galaxies and AGNs collected from the literature.
We  note that the compiled measurements of the \CIV\ and \HeII\ emission
for star-forming galaxies
need to be free of any possible stellar absorption and emission
(see also Sect.\ \ref{sec:c3emitters} for the corrections done for
the VUDS \CIII\ emitters). 
If the \CIV\ strength is measured with the emission line component
alone (without the absorption component taken into account), 
we simply adopt the measured value as the nebular emission.
As for the \HeII\ emission, only a resolved narrow component is used
as the nebular-origin \HeII\ emission
(e.g., \citealt{erb2010,vanzella2016,senchyna2017}).
Although the sample of sources remains quite small, the difference between 
galaxies and AGNs on the C4C3--C34 diagram is significant,
and agrees well with our mode predictions in the top-left panel.
Using the distributions of galaxies and AGNs in conjunction with 
our models, we define a method to distinguish star-forming galaxies 
from AGNs in the C4C3--C34 diagram as follows:
%
% EQUATION
\begin{eqnarray}
        \log {\rm C4C3}
                < 2.5 \log {\rm C34} - 1.8.
        \label{eq:separate_c4c3_c34}
\end{eqnarray}
% EQUATION
We note that Eq. (\ref{eq:separate_c4c3_c34}) is also largely valid 
if galaxies have a higher C$/$O abundance ratio 
(Fig. \ref{fig:models_UVlines_sfg2_co}).

Using the sample of $14$ star-forming galaxies, which are classified by 
the optical BPT-diagnostics \citep{erb2010,stark2014,berg2016,senchyna2017}, 
and $81$ individual AGNs, Eq. (\ref{eq:separate_c4c3_c34}) gives 
clean samples of galaxies and AGNs 
whose  ``classification success rates"
are $92$\,\%\ ($=12/13$)%
\footnote{
We omit one galaxy (ID 110 from \citealt{senchyna2017}) since it cannot 
be classified due to its weak lower limits on the ratios. 
} and $96$\,\%\ ($=78/81$),
respectively. 
We also note that one star-forming galaxy (ID 111 from \citealt{senchyna2017}) falls 
significantly onto the AGN region. This could be because the strong 
\HeII\ emission is not well corrected for the stellar emission.
One caveat is that, 
as seen from the top-left panel of Fig. \ref{fig:UV_SFGsAGNs_both}, 
there is a small region in this diagram 
where both AGN and SF models overlap, that is, our classification may be ambiguous. 
This region, with C4C3 $\ga 1$ and C34 $\simeq 5$ -- $10$ 
which is highlighted with a gray-shaded region, 
corresponds to SF models 
with a very low metallicity ($Z\lesssim 0.01\,Z_{\odot}$) and/or a high ionization 
parameter of $\log U\gtrsim -1$. So far, such a high ionization parameter or 
low metallicities have rarely been reported.
Using larger samples to further test and improve the proposed classification
in the future will certainly be useful. 
For all other regions of this phase space the SF and AGN models are
clearly separated.

%FFFFFFFFFFFFFFFFFFFFFFFFFFFFFFFFFFFFFFFFFFFFFFFFFFFFFFFFFFFFFFFFFFFFFFFFF%
\begin{figure}
  \centerline{
    \includegraphics[width=0.9\columnwidth]{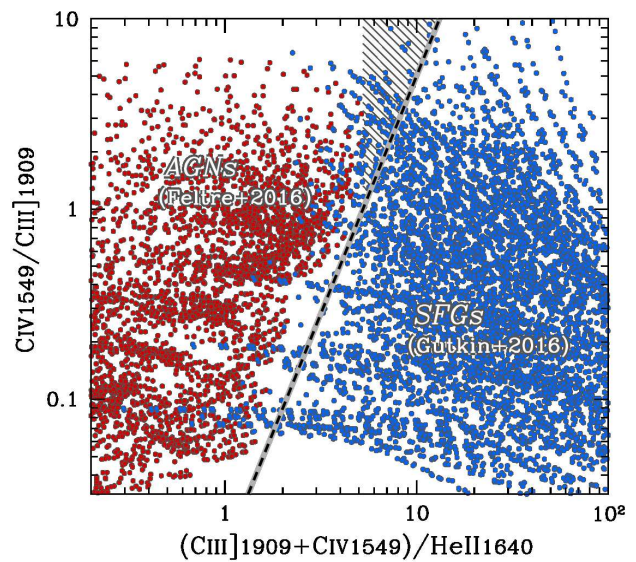}
  }
  \caption{
        Diagram of C4C3 vs. C34 presenting photoionization models of 
        AGNs (red; \citealt{feltre2016}) and 
        star-forming galaxies (SFGs; blue; \citealt{gutkin2016}).
        The separating line of Eq. (\ref{eq:separate_c4c3_c34}) 
        consistently divides the two populations,
        with a small region where metal-poor, highly ionized
        galaxy models could contaminate the AGN area
        (gray shaded) as we also indicate in Fig. \ref{fig:UV_SFGsAGNs_both}.
        }
\label{fig:c4c3_c34_SFGsAGNs_compare}
\end{figure}
%FFFFFFFFFFFFFFFFFFFFFFFFFFFFFFFFFFFFFFFFFFFFFFFFFFFFFFFFFFFFFFFFFFFFFFFFF%

Figure \ref{fig:c4c3_c34_SFGsAGNs_compare} shows 
AGN and star-forming galaxy models from
\citet{feltre2016} and \citet{gutkin2016},
respectively, on the C4C3-C34 diagram. 
These models include more variants than ours, 
such as dust-to-metal mass ratio, C$/$O abundance ratio, 
IMF, and so on, and are thus useful to check our diagnostic. 
Albeit with our simpler modeling, 
it is demonstrated that our models and the separation criteria of 
Eq. (\ref{eq:separate_c4c3_c34}) are nicely consistent with these 
earlier UV modeling studies. 
We note that a small fraction of the modeled star-forming galaxies 
contaminate the AGN regime on the C4C3-C34 diagram. 
They are usually metal-poor galaxy models with $Z\lesssim 0.05\,Z_{\odot}$.
Practically, such a metal-poor gas cloud is thought to be associated 
with a high ionization parameter ($\log U\gtrsim -2.5$;
e.g., \citealt{onodera2016,kojima2017}).
We thus do not worry about the contaminated models with C4C3 $\lesssim 1$.
For the contaminated SF models with C4C3 larger than unity and C34 
$\simeq 5$ -- $10$ (gray-shaded region), 
such metal-poor and highly ionized galaxies could exist 
and could be missed by the current method, as we have found with our 
own models. 
We need more data to calibrate our model predictions.
We emphasize again that these earlier studies only make use of the 
flux ratios. This paper additionally presents the EWs behaviors as a function
of the incident radiation field for the first time, as presented in the following
Section (Sect.\ \ref{sssec:UV_diagnostic_ews}).

%FFFFFFFFFFFFFFFFFFFFFFFFFFFFFFFFFFFFFFFFFFFFFFFFFFFFFFFFFFFFFFFFFFFFFFFFF%
\begin{figure*}
  \centerline{
    \includegraphics[width=0.9\textwidth]{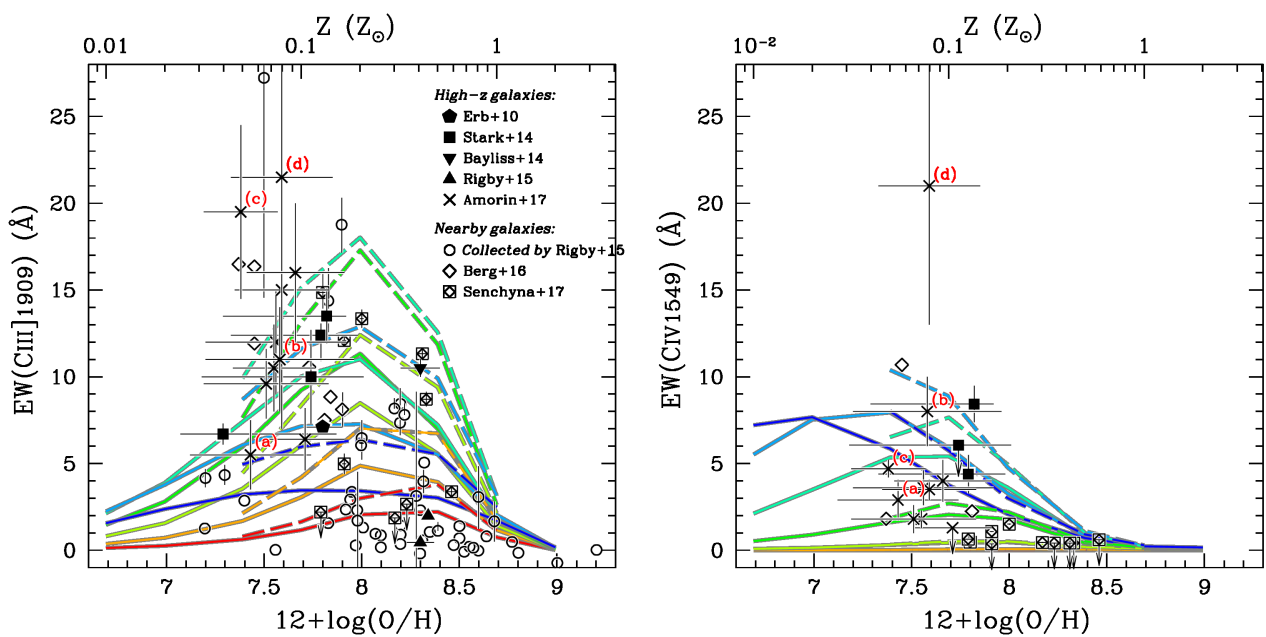}
  }
  \caption{
        EWs of \CIII\ (left) and \CIV\ (right) as a function of metallicity
        comparing observations (black dotted points) with 
        our model predictions (colored curves) for star-forming galaxies.
        The observations consist of nearby and high-$z$ galaxies
        compiled from the literature as shown in the legend of the 
        left panel.
        The models use the single- (solid) and binary- (dashed) stellar 
        populations at the starburst age of $1$\,Myr 
        adopting several ionization parameters with 
        a different color, 
        as shown in the top-left panel of Fig. \ref{fig:UV_SFGsAGNs_both}.
        Four objects from  \citet{amorin2017} that will be discussed 
        in Sect.\ \ref{ssec:nature_vuds_Amorin2017} with 
        IDs: 510583858, 5101421970, 5100998761, and 5100565880
        are labeled as (a), (b), (c), and (d), respectively.
        }
\label{fig:Z_ew}
\end{figure*}
%FFFFFFFFFFFFFFFFFFFFFFFFFFFFFFFFFFFFFFFFFFFFFFFFFFFFFFFFFFFFFFFFFFFFFFFFF%

% -% -% -% -% -% -% -% -% -% -% -% -% -% -% -% -% -% -% -% -% -% -% -% -% -% -% -% -% -% -% 
\subsubsection{Diagnostics using EWs of \CIII\ and \CIV} \label{sssec:UV_diagnostic_ews}
% -% -% -% -% -% -% -% -% -% -% -% -% -% -% -% -% -% -% -% -% -% -% -% -% -% -% -% -% -% -%

We now discuss the use of the UV diagnostics involving line EWs.
The middle and bottom panels of Fig. \ref{fig:UV_SFGsAGNs_both} 
show the UV diagrams of 
EW(\CIII) versus \CIII$/$\HeII\ ratio and EW(\CIV) versus \CIV$/$\HeII\ ratio, respectively. 
The models presented in the left panels show that 
star-forming galaxies and AGNs are distributed differently in these plots, with
AGNs having an EW of \CIII\ (\CIV) larger than that of star-forming galaxies 
at a fixed line ratio of \CIII$/$\HeII\ (\CIV$/$\HeII). 
From this we propose the following demarcation line to distinguish
between star formation and AGN:
%
% EQUATION
\begin{eqnarray}
        \begin{array}{l}
                {\rm EW}({\rm C}\,{\scriptstyle{\rm III}}]) \\
                \\
        \end{array}
        \begin{array}{ll}
                  \ \ < \ \ 4 \times \left({\rm C}\,{\scriptstyle{\rm III}}]/{\rm He}\,{\scriptstyle{\rm II}} \right) 
                        & ({\rm C}\,{\scriptstyle{\rm III}}]/{\rm He}\,{\scriptstyle{\rm II}} < 5) \\
                  \ \ < \ \ 20. 
                        & ({\rm C}\,{\scriptstyle{\rm III}}]/{\rm He}\,{\scriptstyle{\rm II}} \ge 5).
        \end{array}
        \label{eq:separate_ewc3_c3he2}  
\end{eqnarray}
% EQUATION
%
Similarly, for the EW(\CIV) plot:
%
% EQUATION
\begin{eqnarray}
        \begin{array}{l}
                {\rm EW}({\rm C}\,{\scriptstyle{\rm IV}}]) \\
                \\
        \end{array}
        \begin{array}{ll}
                  \ \ < \ \ 3 \times \left({\rm C}\,{\scriptstyle{\rm IV}}]/{\rm He}\,{\scriptstyle{\rm II}} \right) 
                        & ({\rm C}\,{\scriptstyle{\rm IV}}]/{\rm He}\,{\scriptstyle{\rm II}} < 4) \\
                  \ \ < \ \ 12. 
                        & ({\rm C}\,{\scriptstyle{\rm IV}}]/{\rm He}\,{\scriptstyle{\rm II}} \ge 4).
        \end{array}
        \label{eq:separate_ewc4_c4he2}  
\end{eqnarray}
% EQUATION

Comparisons between the observed data points and our models are found 
in the right-hand panels of Fig. \ref{fig:UV_SFGsAGNs_both}.
Again, although the observations are quite sparse, the EW diagnostics 
appear to work exceptionally well.
Only $14$ and $9$ star-forming galaxies are found in the literature
that can be plotted in the EW(\CIII) and EW(\CIV) plots, respectively,
and that are diagnosed to be a star-dominated galaxy
by the optical BPT-diagram. 
All of these objects are in the star-formation regions that we set 
in the EW(\CIII) and EW(\CIV) plots.
Regarding the AGN sample, we use the \citet{talia2017}
and the \citet{hainline2011} composite samples
because most of the individual AGNs shown in the top right
panel have no published EWs. 
\citet{talia2017} use X-ray-identified type-II AGNs at $z=1.7-4.6$,
while \citet{hainline2011} rely on the high-ionization UV lines of 
\NV$\lambda 1240$ and \CIV\ to construct the AGN sample
at $z=2-3$ (cf. \citealt{steidel2002}). 
Figure \ref{fig:UV_SFGsAGNs_both} shows that 
all of their composite AGNs are classified as AGN
in both EW plots. 
In addition, a candidate AGN at $z=7.15$ named COSY is recently
reported by \citet{laporte2017} based on the detection of high-ionization
lines of \NV\ and \HeII\ but the absence of \CIII\ and \CIV.
As shown in Fig. \ref{fig:UV_SFGsAGNs_both} the upper limits on 
C34 and the Carbon equivalent widths are compatible with AGN 
models.

As mentioned in Sect.\ \ref{ssec:results_CO}, the maximum EWs 
would become larger if metal-poor sources with an increased C$/$O ratio existed.
Since the carbon-to-helium ratios also scale linearly with C/O,
a simple extension of the dividing line in the range of 
\CIII$/$\HeII\ $<5$ and \CIV$/$\HeII\ $<4$ in Eq. 
(\ref{eq:separate_ewc3_c3he2}) and (\ref{eq:separate_ewc4_c4he2}),
respectively, would work to find such an unusual population
as illustrated with a gray-shaded area in the middle and bottom
panels of Fig. \ref{fig:UV_SFGsAGNs_both}
We set the maximum EW of \CIII\ as $45$\,\AA\ and that of \CIV\ as $20$\,\AA\ 
when the C$/$O ratio is as high as the solar ratio (Fig. \ref{fig:models_UVlines_sfg2_co}).
Interestingly, a few outliers 
are found in the star-forming galaxy sample
with high \CIII\ or \CIV\ 
equivalent widths in the AGN region.
These are the Lynx arc \citep{fosbury2003}, which could however also be 
AGN-powered \citep{binette2003},
the $z=3.2$ super star-clusters candidate \citep{vanzella2017},
and galaxies at redshifts beyond $6$ \citep{stark2015_c4,schmidt2017},
which are discussed later (Sect.\ \ref{ssec:discussion_comparison}).

In summary, Fig. \ref{fig:UV_SFGsAGNs_both} nicely demonstrates 
that the UV line ratios and EWs of \CIII, \CIV, and \HeII\ can be combined 
in a single diagnostic diagram to identify whether the dominant ionizing 
source is from star formation or AGN, and that our models agree 
very well with the observations.

% -% -% -% -% -% -% -% -% -% -% -% -% -% -% -% -% -% -% -% -% -% -% -% -% -% -% -% -% -% -% 
\subsubsection{Diagnostics for LINERs} \label{sssec:UV_diagnostic_liners}
% -% -% -% -% -% -% -% -% -% -% -% -% -% -% -% -% -% -% -% -% -% -% -% -% -% -% -% -% -% -%

Some of the AGN models presented in Sect. \ref{ssec:AGNs} 
produce optical line ratios typical of 
low-ionization narrow emission-line regions (LINERs)
based on the optical classification (e.g., \citealt{baldwin1981,kewley2006}).
These LINER models are characterized by a high metallicity ($Z\gtrsim 1\,Z_{\odot}$)
and a low ionization parameter ($\log U\lesssim -2.5$),
both of which are consistent with earlier work \citep{kewley2006}. 
They occupy a region on the C4C3--C34 diagram 
with C4C3 $\lesssim 0.1$ and C34 $\lesssim 1$. 
The C4C3--C34 diagram can thus work to discriminate between LINERs 
and other AGNs.
There are, however, several caveats. 
Since the power source of LINERs is still under debate 
(e.g., \citealt{kewley2006,kewley2013_theory} and references therein), 
it remains unclear whether or not the LINERs included in the AGN models 
are complete. 
Moreover, we cannot predict robust EWs for LINERs due to the 
uncertainty of their UV-continuum.
Therefore it is preferable to use emission line ratios 
instead of EW diagrams in this case.

%FFFFFFFFFFFFFFFFFFFFFFFFFFFFFFFFFFFFFFFFFFFFFFFFFFFFFFFFFFFFFFFFFFFFFFFFF%
\begin{figure*}
  \centerline{
    \includegraphics[width=0.92\textwidth]{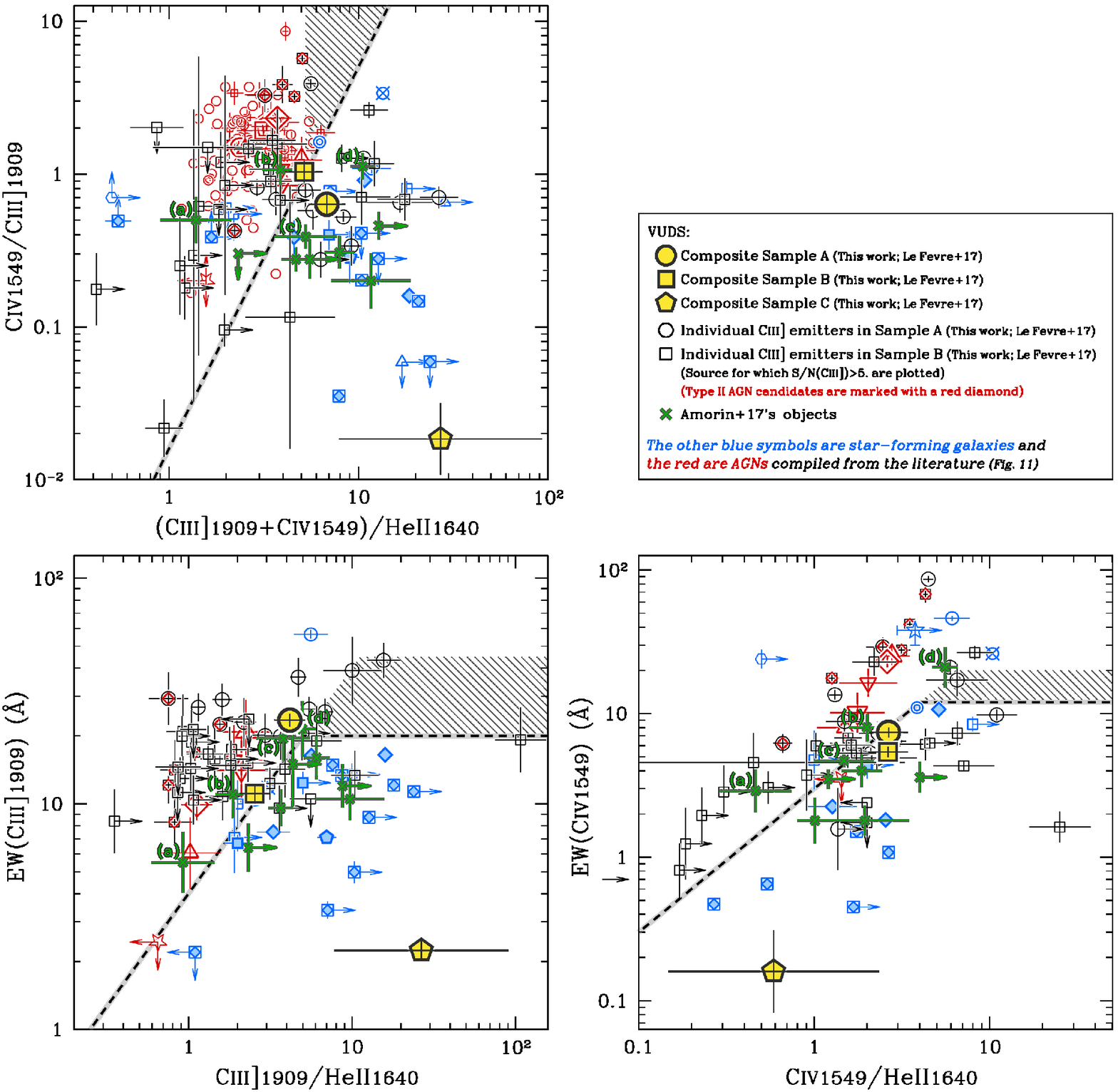}
  }
  \caption{
        UV diagnostic diagrams with observation of individual sources and stacks from VUDS.
        The large yellow-filled circle, square, and pentagon show the VUDS composites
        of the strong \CIII\ emitters (EW(\CIII)>20\,\AA, Sample A),
        the intermediate \CIII\ emitters (EW(\CIII) $=10$ -- $20$\,\AA, Sample B),
        and all star-forming galaxies identified by VUDS (Sample C), respectively 
        (\citealt{lefevre2017}; see Table \ref{tbl:samples}).
        The smaller black open circles and squares are the individual VUDS objects
        in samples A and B, respectively, 
        whose \CIII\ is identified at the $>5\sigma$ significance level \citep{lefevre2017}.
        The VUDS objects marked with a red diamond are type-II AGN candidates.
        The green crosses show metal-poor compact galaxies found by VUDS
        \citep{amorin2017}. 
        The four objects highlighted in \ref{ssec:nature_vuds_Amorin2017} are labeled 
        as (a)--(d), as in Fig. \ref{fig:Z_ew}.
        }
\label{fig:UV_vuds_sfg_agn}
\end{figure*}
%FFFFFFFFFFFFFFFFFFFFFFFFFFFFFFFFFFFFFFFFFFFFFFFFFFFFFFFFFFFFFFFFFFFFFFFFF%

%%%%%%%%%%%%%%%%%%%%%%%%%%%%%%%%%%%%%%%%%%%%%%%%%%%%
\subsection{UV diagrams to estimate ISM properties and star-formation age} 
\label{ssec:UV_ISMproperties}
%%%%%%%%%%%%%%%%%%%%%%%%%%%%%%%%%%%%%%%%%%%%%%%%%%%%

In addition to the shape of the incident radiation field, 
the UV diagrams would be helpful to infer the ISM properties 
such as metallicity and ionization parameter, especially 
for star-forming galaxies%
\footnote{
Refer to, e.g., \citet{nagao2006a} and \citet{dors2014} for AGNs.
}. 
According to the C4C3--C34 diagram, the ionization parameter
of the compiled $z=2-4$ galaxies and nearby dwarf galaxies are 
in the range from $\log U \sim -1.5$ to $-2.5$ associated with a 
metallicity of subsolar or smaller values. These are reasonably 
consistent with those estimated from the optical emission lines
\citep{christensen2012a,stark2014,berg2016}. 
The two highly ionized galaxies, Lynx arc \citep{fosbury2003}
and BX418 \citep{erb2010}, could indeed be diagnosed as showing 
a very high ionization parameter of $\log U \sim -1.0$ to $-0.5$
if their metallicities are relatively high (Z $\gtrsim 0.1\,Z_{\odot}$)
as indicated with the optical emission lines.

Once the ISM properties are well constrained, we can use the 
EWs plots to estimate the star-formation age (or \xiion), since 
EWs of UV lines are sensitive to the ratio of ionizing to non-ionizing 
UV photons
(Fig. \ref{fig:ewcs_age_rel}).
Figure \ref{fig:Z_ew} presents the distributions of EWs of \CIII\ and \CIV\
for star-forming galaxies as a function of metallicity, 
and compared to our models. 
Plotted curves are our models.
Except for some outliers with large uncertainties, 
Fig. \ref{fig:Z_ew} nicely demonstrates that 
our models give an upper-limit of EWs at a fixed metallicity.
Many of the nearby galaxies show a small EW(\CIII) of $\lesssim 5$\,\AA. 
This is because they possess an evolved stellar population with 
a low ionization parameter. 
If we adopt an empirical relation between metallicity and 
ionization parameter typically seen in the local universe; 
$\log U \sim -0.68 \times (12+\log({\rm O}/{\rm H}))+2.9$ 
\citep{kojima2017}, 
nearby galaxies in the metallicity range from 
$Z=0.2$ to $1\,Z_{\odot}$ would have an ionization parameter 
from $\log U\sim -2.5$ to $-3$, which are denoted by the 
yellow and orange curves, respectively, in Fig. \ref{fig:Z_ew}.
It is thus obvious that the $1$\,Myr models usually over-predict 
the EW(\CIII) for the given ISM condition, as expected, 
and that these galaxies experience a much longer 
star-formation of 
$\gg 1$\,Myr (cf. Fig. \ref{fig:ewcs_age_rel}).

In Fig. \ref{fig:Z_ew}, it is interesting to note that below the 
subsolar metallicity, many of the compiled galaxies at any redshift 
show an EW(\CIII) larger than $10$\,\AA. This suggests that such 
metal-poor galaxies require a hard ionizing spectrum or an efficient ionizing 
photon production achieved by a binary stellar population
(e.g., \citealt{senchyna2017}).
Moreover, the $1$\,Myr models are close to reproducing 
a number of low-$Z$ galaxies with large EWs, while 
we regard these models as providing EW upper limits 
for a given ISM condition. 
This trend indicates
that those low-$Z$ galaxies are generally a very young population
\citep{JR2016}.

Finally, we identify outlier emitters with a larger EW of \CIII\ and/or \CIV\
than predicted by our models. 
Many of them are compatible with our models within their large observational 
errors. A higher gas density than adopted in 
Fig. \ref{fig:Z_ew} (\nH\ $=10^4$ vs. $10^2$\,cm$^{-3}$) would also help 
to alleviate the apparent discrepancies
at least in the \CIII\ diagram as models would return an increase
of a factor of $1.1$--$1.2$ in EW(\CIII) at a fixed metallicity
in the subsolar metallicity regime 
(Sect.\ \ref{ssec:results_single_stellar}).
Still, a small fraction of galaxies reported in \citet{berg2016} and 
\citet{amorin2017} might show enhanced EWs of \CIII\ and \CIV.
This could be because such peculiar objects possess a higher \xiion\
parameter than predicted by the binary stellar population models.
Another possibility is that some of these objects contain an AGN.
All of the galaxies of \citet{berg2016} are confirmed to be star-forming
based on the optical BPT diagram \citep{baldwin1981}, while 
the \citet{amorin2017}'s sample remains to be checked.
This hypothesis is revisited in Sect.\ \ref{ssec:nature_vuds_Amorin2017}
for the objects of \citet{amorin2017}.
Moreover, they could have a large C$/$O abundance ratio
(Sect. \ref{ssec:results_CO}). 
In this low-$Z$ regime, the \citet{dopita2006} prescription we adopt
gives a C$/$O abundance ratio of $\log {\rm C}/{\rm O}\sim -0.8$. 
If those galaxies are more carbon-enriched, this could explain the EWs of 
\CIII\ and/or \CIV\ larger than predicted by our models. 
Indeed, higher C$/$O abundance ratios than $\log {\rm C}/{\rm O}\sim -0.8$
by $0.4-0.7$\,dex are inferred for some of these outliers
in the low-metallicity regime \citep{berg2016,amorin2017}. 
The higher C$/$O ratios could 
result in stronger EWs than predicted by the standard models 
shown in Fig. \ref{fig:Z_ew}
in the low-metallicity regime by a factor of $\sim 2.5$--$5$.
This trend could suggest an importance of the C$/$O abundance ratio 
to estimate the ISM properties and the star-formation age with the UV lines. 
(Sects.\ \ref{ssec:nature_vuds_Amorin2017} and \ref{ssec:discussion_vuds})

%ttttttttttttttttttttttttttttttttttttttttttttttttttttttttttttttttttttttttt%
\begin{table}
  \centering
  \caption{The three samples used here.}
  \label{tbl:samples}
  \renewcommand{\arraystretch}{1.2}
  \begin{tabular}{ll}
    \hline
    A &
    {\small $16$ \CIII\ emitters with EW(\CIII) $>20$\,\AA} \\
    B &  
    {\small $43$ \CIII\ emitters with EW(\CIII) $= 10-20$\,\AA} \\
    C &
    {\small $450$ star-forming galaxies} \\
     \hline
  \end{tabular}
  \renewcommand{\arraystretch}{1.0}
\end{table}
%ttttttttttttttttttttttttttttttttttttttttttttttttttttttttttttttttttttttttt%

%ttttttttttttttttttttttttttttttttttttttttttttttttttttttttttttttttttttttttt%
\begin{table*}
  \centering
  \caption{UV line EWs (rest) and fluxes of the three composite spectra.}
  \label{tbl:UV_lines_composites}
  \renewcommand{\arraystretch}{1.25}
  \begin{tabular}{@{}lcccccccccc@{}}
    \hline
     &
    \multicolumn{5}{l}{EW (\AA) - - - - - - - - - - - - - - - - - - - - - - - - - - - - - - } &
    \multicolumn{5}{l}{Flux ratio relative to \CIII\ - - - - - - - - - - - - - - - - - - - - - - - - - -} \\
     &
    % EWs
    \CIII\ &
    \CIV\ &
    \CIV$_{\rm corr}$$^{(1)}$ &
    \HeII\ &
    \HeII$_{\rm corr}$$^{(2)}$ &
    % Fluxes
    \CIII\ &
    \CIV\ &
    \CIV$_{\rm corr}$$^{(1,3)}$ &
    \HeII\ &
    \HeII$_{\rm corr}$$^{(2,3)}$ \\
    \hline
    A &
    $23.5^{+1.6}_{-1.8}$ &
    $4.4^{+0.6}_{-0.5}$ &
    $7.4^{+0.6}_{-0.5}$ &
    $4.3^{+0.7}_{-0.4}$ &
    $3.3^{+0.7}_{-0.4}$ &
    $1. $ &
    $0.31 \pm 0.03$ &
    $0.63^{+0.11}_{-0.11}$ &
    $0.27 \pm 0.03$ &
    $0.24^{+0.06}_{-0.05}$ \\
    B &
    $11.1^{+1.2}_{-1.1}$ &
    $2.4^{+0.5}_{-0.2}$ &
    $5.4^{+0.5}_{-0.2}$ &
    $3.2^{+0.5}_{-0.3}$ &
    $2.2^{+0.5}_{-0.3}$ &
    $1. $ &
    $0.37 \pm 0.02$ &
    $1.03^{+0.14}_{-0.23}$ &
    $0.49 \pm 0.03$ &
    $0.40^{+0.11}_{-0.08}$ \\
    C &
    $2.2^{+0.3}_{-0.2}$ &
    $-2.8^{+0.2}_{-0.1}$ &
    $0.2^{+0.2}_{-0.1}$ &
    $1.1^{+0.2}_{-0.1}$ &
    $0.1^{+0.2}_{-0.1}$ &
    $1. $ &
    $-0.23 \pm 0.07$ &
    $0.02^{+0.02}_{-0.01}$ &
    $0.52 \pm 0.05$ &
    $0.04^{+0.10}_{-0.08}$ \\    
     \hline
  \end{tabular}
  \renewcommand{\arraystretch}{1.0}
  \\ 
  \vspace{-3mm}
  \begin{flushleft}
        \small
        (1) \CIV\ strength corrected for stellar absorption.
        (2) \HeII\ strength corrected for stellar emission.
        (3) Corrected for reddening (see Sect.\ \ref{sec:c3emitters}).
  \end{flushleft}
\end{table*}
%ttttttttttttttttttttttttttttttttttttttttttttttttttttttttttttttttttttttttt%

%%%%%%%%%%%%%%%%%%%%%%%%%%%%%%%%%%%%%%%%%%%%%%%%%%%%
%%%%%%%%%%%%%%%%%%%%%%%%%%%%%%%%%%%%%%%%%%%%%%%%%%%%
\section{\CIII\ Emitters found by VIMOS Ultra Deep Survey}
\label{sec:c3emitters}
%%%%%%%%%%%%%%%%%%%%%%%%%%%%%%%%%%%%%%%%%%%%%%%%%%%%
%%%%%%%%%%%%%%%%%%%%%%%%%%%%%%%%%%%%%%%%%%%%%%%%%%%%

VUDS identifies the \CIII\ emission from large numbers of galaxies 
at redshifts $z=2-4$ \citep{lefevre2015}. Details of the construction of the
\CIII\ emitter sample are summarized in \citet{lefevre2017}.
The sample contains individual \CIII\ emitters with rest-frame EW(\CIII) 
above a detection limit of 3\,\AA.
Broad-line, type-I AGN are excluded from the sample.
The EW distribution has a tail toward a very high \CIII\ EW
beyond $\sim 20$\,\AA\ (see \citealt{lefevre2017}).
These strong \CIII\ emitters motivate us to create a large grid 
of photoionization models, providing a crucial test of 
what we can learn from the \CIII\ and the other UV lines that 
are available from galaxies in the early universe.

In order to address the properties of the strongly \CIII-emitting
objects we use three subsamples based on the strength 
of the \CIII\ emission (Table \ref{tbl:samples}).
Sample A contains the $16$ strongest \CIII\  emitters
whose EW(\CIII) is larger than $20$\,\AA.
The second strongest \CIII\ emitters of $43$ objects 
with EW(\CIII) $=10$--$20$\,\AA\ make up Sample B.
For reference, Sample C presents all the 
UV-continuum-selected VUDS star-forming galaxies, containing 
$450$ galaxies.

Since only broad-line type-I AGN are excluded, 
our \CIII\  emitter sample could contain type-II AGNs.
Indeed, $2$ and $4$ \CIII\ emitters in the classes of 
EW(\CIII) $>20$\,\AA\ and $10$--$20$\,\AA, 
respectively, have an X-ray counterpart based on the work of
\citet{talia2017}. 
Since one of the motivations of this paper is to understand the nature
of the strong \CIII\ emitters, we do not eliminate these likely type-II 
AGNs at this stage.
We use them subsequently to test our UV emission line diagnostics.

For a detailed analysis of the UV lines of the \CIII\ emitters
we need several emission lines to be significantly identified.
In this paper we use three UV lines: \CIII, \CIV, 
and \HeII\ recombination. 
We therefore make composite spectra in the three samples
to increase the S$/$N ratios of the important UV emission lines.
For Sample A whose sample size is quite small, we adopt 
a median stacking of the normalized individual spectra 
to get a representative composite spectrum.
For Samples B and C we simply average the normalized 
spectra. 
The composite spectra of the three samples are presented
in \citet{lefevre2017}.
The necessary emission lines are clearly identified except for 
the \CIV\ in Sample C. 
The absence of the \CIV\ emission in Sample C is due to
the strong \CIV\ absorption by the stellar population as 
indicated by the P-Cygni profile. 
To correct the \CIV\ measurements for stellar absorption, 
we use EW(\CIV) $=-3.0$\,\AA\ as obtained from the 
composite spectrum of all star-forming galaxies in VUDS
with \Lya\ in absorption \citep{lefevre2015}, 
with the idea that this would minimize the contribution from galaxies 
with some \CIV\ emission, and hence produce a \CIV\ value more 
representative of the stellar absorption value.
This value is quite close to that found by \citet{shapley2003},
but is best used for our VUDS samples.
We note that the EW correction is valid only if the \CIV\ strength
is measured in the same way, that is, over the wavelength range 
of both the \CIV\ absorption and emission components.
Another caveat is that the \HeII\ emission
is known to be a composite of the stellar and nebular emission
(e.g., \citealt{brinchmann2008,erb2010}).
Since our photoionization models can only predict the nebular emission, 
we need to subtract the stellar \HeII\ component before comparing with 
our photoionization models.
In this paper, we assume the EW of the stellar \HeII\ emission
is $1.0$\,\AA, which is predicted by models of \citet{brinchmann2008}
with metallicities below the solar value at $\sim 10$\,Myr or older.
Although the stellar component could be stronger if the binary stellar 
evolution is taken into account (e.g., \citealt{erb2010,steidel2016}), 
we cannot further constrain its strength with the currently available 
data set.

To correct for the reddening effects on the line ratios, 
we rely on the UV slope $\beta$ directly obtained from the VUDS
spectra. 
The UV slopes of $\beta =-1.68$, $-1.61$, and $-0.92$ are inferred 
from the spectra of Samples A, B, and C, respectively
\citep{lefevre2017}.
Using the SMC extinction curve that is preferred for high-$z$ galaxies 
(e.g., \citealt{reddy2017}),
these UV slopes correspond to \ebv\ $=0.08$, $0.09$, and $0.15$
for Samples A, B, and C, respectively.
We correct for the reddening of the line ratios using the \ebv\ value 
for each sample using the SMC extinction law \citep{gordon2003},
assuming that the reddening of the nebular emission and stellar 
continuum are identical.
Although we understand this assumption remains open to debate
especially at high-$z$ 
(e.g., \citealt{forsterschreiber2009,reddy2010}), 
we adopt it for the \CIII\ emitters because it is considered to be 
reasonable for actively star-forming galaxies at high-$z$ 
(e.g., \citealt{reddy2015}).
The correction for the \CIV$/$\CIII\ ratio, 
which is the most reddened ratio presented in this paper, for example, 
is a factor of $1.22$, $1.25$, and $1.45$, for Samples A, B, and C,
respectively.
Although the UV slope for Sample C is less steep than reported by 
\citet{hathi2016} with a larger sample of VUDS, 
$\beta =-1.36 \pm 0.02$,
this is likely because the \citet{hathi2016}'s sample contains 
fainter, and thus bluer galaxies. 
We do not add the errors of the $\beta$ measurements 
($\Delta\beta$=0.02 -- 0.08) as well as the difference between
Sample C and the \citet{hathi2016}'s sample
to the corrected line ratios
since they are much smaller than the errors
originating from the corrections for the stellar emission / absorption 
for the \CIV\ and \HeII\ emission.
As for the EWs, we adopt the observed values divided by $(1+z)$
as the intrinsic ones under the assumption that the reddenings of 
the nebular emission and stellar continuum are the same.

The EWs and fluxes of the UV lines for the three composite spectra are 
given in Table \ref{tbl:UV_lines_composites}, taken from \citet{lefevre2017}.
Both the observed and corrected values are listed for the \CIV\ and the \HeII\ emission.
The individual spectra of the \CIII\ emitters in Samples A and B 
are corrected in exactly the same way for each sample.

%%%%%%%%%%%%%%%%%%%%%%%%%%%%%%%%%%%%%%%%%%%%%%%%%%%%
%%%%%%%%%%%%%%%%%%%%%%%%%%%%%%%%%%%%%%%%%%%%%%%%%%%%
\section{The nature of the \CIII\ emitters from VUDS} 
\label{sec:nature_vuds}
%%%%%%%%%%%%%%%%%%%%%%%%%%%%%%%%%%%%%%%%%%%%%%%%%%%%
%%%%%%%%%%%%%%%%%%%%%%%%%%%%%%%%%%%%%%%%%%%%%%%%%%%%

We now examine the observed VUDS sources, both stacked spectra and individual sources,
in the three UV diagnostic diagrams of Fig. \ref{fig:UV_vuds_sfg_agn}, 
and we use them to infer properties of the incident radiation field and the ISM of the \CIII\ emitters
in the following subsections.

%%%%%%%%%%%%%%%%%%%%%%%%%%%%%%%%%%%%%%%%%%%%%%%%%%%%
\subsection{Stack of all VUDS star-forming galaxies at $z\sim 3$} 
\label{ssec:nature_vuds_all_sfg}
%%%%%%%%%%%%%%%%%%%%%%%%%%%%%%%%%%%%%%%%%%%%%%%%%%%%

%FFFFFFFFFFFFFFFFFFFFFFFFFFFFFFFFFFFFFFFFFFFFFFFFFFFFFFFFFFFFFFFFFFFFFFFFF%
\begin{figure*}
  \centerline{
    \includegraphics[width=0.99\textwidth]{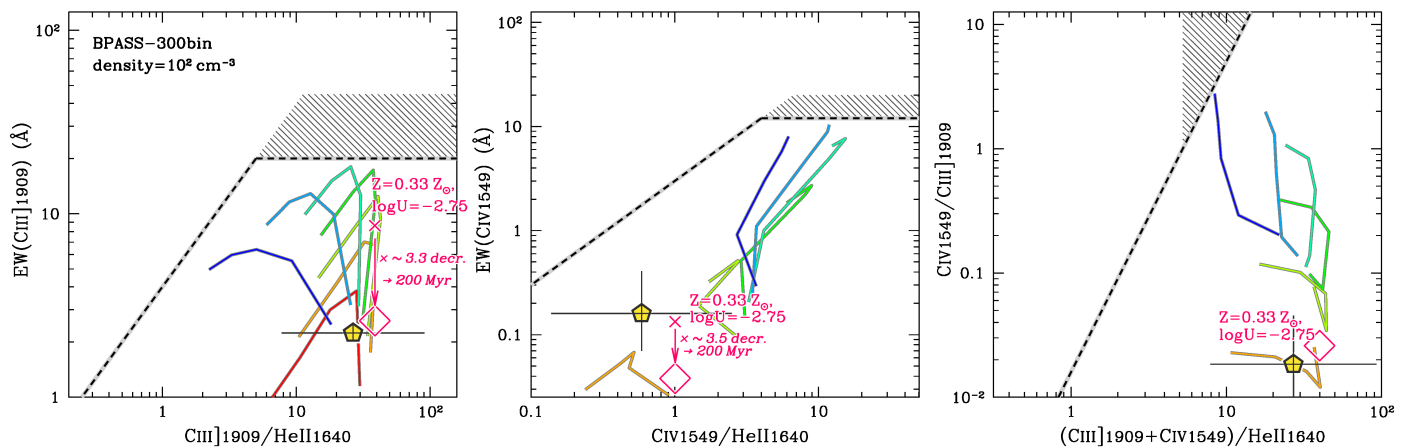}
  }
  \caption{
        Comparison with the composite of all VUDS star-forming galaxies 
        (Sample C; yellow-filled pentagon)
        and the best fit model (red open diamond).
        The model is provided by the BPASS-300bin incident radiation with 
        $Z=0.33\,Z_{\odot}$, $\log U=-2.75$, and 
        the age of $100$\,Myr. 
        In the two EW plots, a red cross presents the best-fit model at $1$\,Myr
        as derived directly from the plotted model tracks. 
        The decrease of EW by $\times 3.3$ and $\times 3.5$ corresponds to 
        the age effect for the \CIII\ and \CIV, respectively, 
        if the longer star-formation age of $200$\,Myr is adopted 
        (Fig.\ \ref{fig:ewcs_age_rel}).
        }
\label{fig:models_UVlines_bpass_compC}
\end{figure*}
%FFFFFFFFFFFFFFFFFFFFFFFFFFFFFFFFFFFFFFFFFFFFFFFFFFFFFFFFFFFFFFFFFFFFFFFFF%

The sample of all galaxies in VUDS 
(Sample C, Table \ref{tbl:samples}) includes by definition
all galaxies that are \CIII\ emitters, 
in addition to all other star-forming galaxies. 
As shown in Fig. \ref{fig:UV_vuds_sfg_agn} 
this sample (yellow-filled pentagon) falls into the star-forming 
galaxy regions in all of the three UV diagnostic diagrams, as expected.
We have confirmed that individual sources, most of which 
present only the \CIII\ line, show line ratios and EWs 
consistent with those of the composite. 
Since such individuals provide poor constraints on the properties 
based on the UV diagrams, we only adopt the composite 
of Sample C to discuss the properties in this Section.

We now examine if the UV diagrams properly characterize 
the physical properties of galaxies in Sample C
(Fig. \ref{fig:models_UVlines_bpass_compC}). 
We first look into the line ratios of the C4C3--C34 diagram 
because the EW plots depend on the age of the current star formation 
as well as the ISM condition and the incident radiation field 
(Sect.\ \ref{ssec:results_ages}).
We adopt the binary stellar population models, which is preferred 
for high-$z$ star-forming galaxies (e.g., \citealt{steidel2016}). 
Comparisons with the single stellar population models are 
presented later. 
The \CIV$/$\CIII\ ratio and the C34-index of the Sample C composite 
give constraints on the metallicity and ionization parameter
of ($Z$, $\log U$) from ($0.3\,Z_{\odot}$, $-2.7$) to ($0.5\,Z_{\odot}$, $-3$).
Next, we consider the EW plots, especially the EW(\CIII) plot. 
The composite of Sample C has large errors in the EW(\CIV) plot
due to the uncertain contribution to EW(\CIV) from stellar populations.
In the EW(\CIII) plot, the ISM properties of 
($Z$, $\log U$) $=$ ($0.3\,Z_{\odot}$, $-2.7$) -- ($0.5\,Z_{\odot}$, $-3$)
predict EWs(\CIII) larger than the observed value by a factor of $3.3$ -- $3.$,
while the predicted \CIII$/$\HeII\ ratios agree with the observed ratio.
This discrepancy is easily resolved by adopting ages of  $\sim 200$ ($50$)\,Myr
for $Z=0.3$ ($0.5$)\,$Z_{\odot}$ instead of the unrealistically young $1$\,Myr old
population adopted by default (cf.\ Fig. \ref{fig:ewcs_age_rel}).
Therefore, the UV diagnostics tell us that typical star-forming 
galaxies at $z=2-4$ present a set of physical properties of 
($Z$, $\log U$, age) $\sim$ 
($0.3\,Z_{\odot}$, $-2.7$, $200$\,Myr) -- ($0.5\,Z_{\odot}$, $-3$, $50$\,Myr).
Compared with the properties studied by other galaxy surveys
at similar redshifts
(e.g., \citealt{shapley2003,mannucci2009,onodera2016}), 
the estimated ISM properties and age are in reasonably good 
agreement. 
The best-fit model (see Fig. \ref{fig:models_UVlines_bpass_compC})
underestimates the EW(\CIV) if the age of 
$\sim 200$\,Myr is considered, albeit being consistent within
the $2\sigma$ error.
This illustrates the difficulty of using the EW(\CIV) plot alone to
estimate the physical quantities, since an accurate 
correction for stellar \CIV\ absorption is needed, 
which requires a high S$/$N and high spectral resolution,
difficult to obtain at these redshifts.

In the similar method adopted for the binary models, we derive 
the best-fit parameters using the \textsc{PopStar} single star models 
as follows: ($Z$, $\log U$, age) $\sim$ 
($0.5\,Z_{\odot}$, $-3$, $4$\,Myr) -- ($0.2\,Z_{\odot}$, $-3$, $10$\,Myr).
A notable difference when using these models compared to 
the best binary models is that they output a younger age of 
star formation of less than $10$\,Myr.
One reason is that the binary models usually predict a larger maximum EW, 
which yields a larger gap between the maximum and the observed EW.
Another is that the EW becomes weak more rapidly in the single stellar models 
if compared with the binary models (Fig. \ref{fig:ewcs_age_rel}).
These two factors result in a younger age for the current star formation
from the single stellar population models. 
It is unlikely that the population of star-forming galaxies 
at these redshifts typically shows such a young starburst age. 
Furthermore, the ionization parameter of $\log U\sim -3$
is smaller than estimated by earlier studies 
and for its relatively low-metallicity environment 
(e.g., \citealt{NO2014,onodera2016}).
The discrepancy between the predicted and observed (corrected)
EWs of \CIV\ becomes worse over the $2\sigma$ uncertainty
in the single stellar models.
Therefore, 
it is suggested that the binary stellar population
models lead to a better and more reasonable agreement
with the observation of typical star-forming galaxies at $z=2-4$.
This argument is consistent with the other individual studies
(Sect.\ \ref{ssec:UV_ISMproperties}; \citealt{steidel2016,JR2016}).

%FFFFFFFFFFFFFFFFFFFFFFFFFFFFFFFFFFFFFFFFFFFFFFFFFFFFFFFFFFFFFFFFFFFFFFFFF%
\begin{figure*}
  \centerline{
    \includegraphics[width=0.99\textwidth]{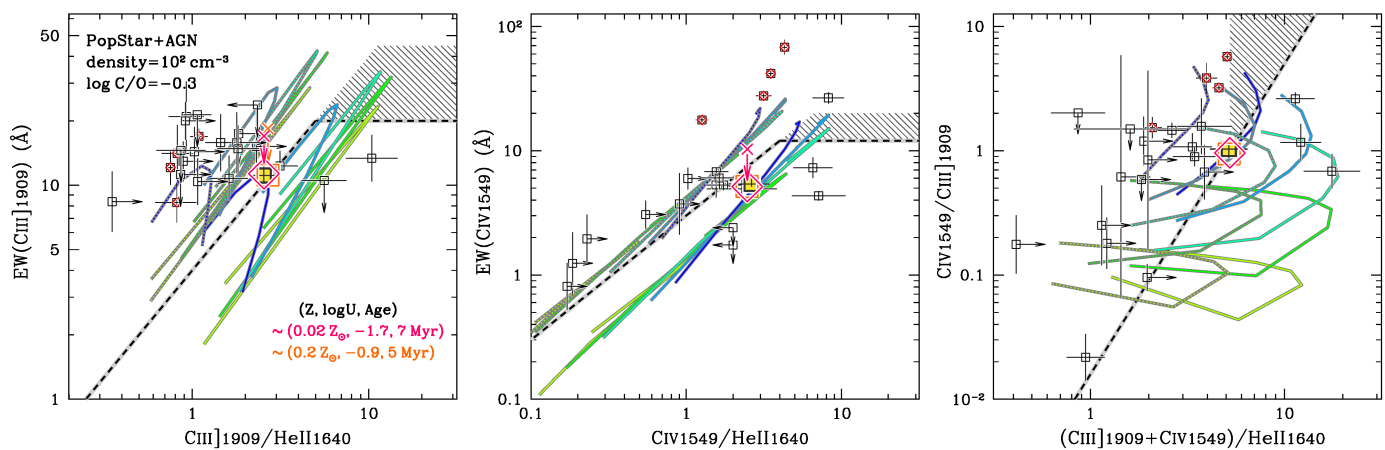}
  }
  \caption{
        Comparison with the composite of intermediately strong \CIII\ emitters
        (Sample B; yellow-filled square)
        and the two best-fit models (magenta open diamond and orange open square).
        The two best fit models are provided by \textsc{PopStar}$+$AGN models 
        with AGN contribution of $f_{\rm AGN}=0.06$ -- $0.1$, 
        fixing the C$/$O abundance ratio to the solar value
        $\log$\,C$/$O $=-0.3$.
        The magenta symbol corresponds to the low-$Z$ branch solution of $Z\sim 0.02\,Z_{\odot}$,
        and the orange to the high-$Z$ branch solution of $Z\sim 0.2\,Z_{\odot}$.
        The crosses are the same as in Fig. \ref{fig:models_UVlines_bpass_compC}.
        The individual objects in Sample B are also plotted 
        (symbols as in Fig. \ref{fig:UV_vuds_sfg_agn}).
        }
\label{fig:models_UVlines_popstar_agn_co_compB}
\end{figure*}
%FFFFFFFFFFFFFFFFFFFFFFFFFFFFFFFFFFFFFFFFFFFFFFFFFFFFFFFFFFFFFFFFFFFFFFFFF%

%%%%%%%%%%%%%%%%%%%co%%%%%%%%%%%%%%%%%%%%%%%%%%%%%%%%
\subsection{\citet{amorin2017}'s objects} 
\label{ssec:nature_vuds_Amorin2017}
%%%%%%%%%%%%%%%%%%%%%%%%%%%%%%%%%%%%%%%%%%%%%%%%%%%%

Before moving to the stronger individual \CIII-emitter samples, 
we re-examine the ten metal-poor compact galaxies found in VUDS 
at $z\sim 3$ by \citet{amorin2017}.
These objects have an intermediately strong \CIII\ EW 
ranging from $5.5$ to $21.5$\,\AA\ (median value: $11.5$\,\AA).
They are plotted with green crosses in Fig. \ref{fig:UV_vuds_sfg_agn},
corrected for the possible stellar \HeII\ emission 
in the same way as done for the other VUDS objects
(Sect.\ \ref{sec:c3emitters}).
Since \citet{amorin2017} measure the \CIV\ by fitting only the
emission component, we simply adopt their measurements as 
the nebular \CIV\ strength.
This sample is considered to be almost dust-free given a 
typical UV slope of $\beta=-2.36$.
We therefore do not correct for the reddening since the effects
on the line ratios are minimal.
Figure \ref{fig:UV_vuds_sfg_agn} demonstrates that the line emission of 
the \citet{amorin2017}'s objects 
is typically explained by star-formation as the ionizing source.
A non-detection in X-ray also disfavors bright AGNs. 
However, a relatively large scatter of the sample is also visible 
in the diagrams. Six out of the ten objects are diagnosed as being powered by 
pure star formation from all three UV diagrams. 
We now discuss the remaining $four$ sources (labeled (a)--(d) 
in Fig. \ref{fig:UV_vuds_sfg_agn}) in detail.

Among the four objects, one (ID 510583858; $z=2.4$; labeled (a)
in Fig. \ref{fig:UV_vuds_sfg_agn}) 
falls into the AGN regime in all three diagrams, irrespective of the 
corrections for the stellar \HeII\ emission. 
Based on our diagnostics, this object is considered to be powered by AGN
or mixed SF $+$ AGN with $f_{\rm AGN}\gtrsim 0.1$.
Two other objects (ID 5101421970 and 5100998761; $z=2.5$ and $2.4$; 
(b) and (c) in Fig. \ref{fig:UV_vuds_sfg_agn}, respectively) fall into the AGN 
regions, albeit close to the border. Their incident radiation field appears 
to be harder than that produced by pure star formation, and 
a SF $+$ AGN model or a soft AGN model can reproduce their 
UV lines strengths.
If true, the metallicities derived by \citet{amorin2017} for these sources 
might not be accurate.
Specifically, the object (c) with ID 5100998761 is reported
to have the lowest metallicity among the sample ($Z\sim 0.05\,Z_{\odot}$),
being an outlier in Fig. \ref{fig:Z_ew} having an EW(\CIII) too strong 
for its metallicity to be explained by our star-formation models. 
This could be because the object is (partly) powered by an AGN-like
hard ionizing source.
If we adopt a hybrid model of SF $+$ AGN of $f_{\rm AGN}=0.07$%
\footnote{
The fraction is inferred for the \CIII\ emitters in VUDS 
(below).
We simply linearly interpolate the models with the AGN 
contributions of $f_{\rm AGN}=0.03$ and $0.1$.
}
the object is estimated to have 
a four times higher metallicity of $Z\sim 0.2\,Z_{\odot}$, 
falling on a simple extrapolation of the mass-metallicity relation 
at $z\sim 2.3$ \citep{steidel2014}.
Interestingly, the other two objects (a) and (b)
(ID 510583858 and 5101421970, respectively)
are suggested to be the two most massive galaxies in the sample
($3-4\times 10^9\,M_{\odot}$)
and to fall significantly below the mass-metallicity relation \citep{amorin2017}.
In the SFR--$M_{\star}$ relation, they are almost in the star formation 
main sequence. This is in contrast to the other six pure star-forming 
metal-poor galaxies, which lie well above the star-formation main sequence. 
If we include an AGN ionizing spectrum, we find higher metallicity estimates
than \citet{amorin2017}, 
and this would possibly affect their conclusions.
If we adopt the SF$+$AGN hybrid models, the line ratios and EWs
 of object (a) can be reproduced by either a high-$Z$ ($Z= 0.5$--$1\,Z_{\odot}$) or 
a low-$Z$ ($Z\sim 0.05\,Z_{\odot}$) model with $f_{\rm AGN}=0.1$.
If the former solution is correct, this object is almost ten times more 
metal-rich than originally inferred by \citet{amorin2017}.
In the same way, the object (b) is best-fitted by a $f_{\rm AGN}=0.06$--$0.1$
model with a metallicity of $Z=0.1$--$0.2\,Z_{\odot}$, which is a factor of 
$2$--$3$ higher than the original value.
Moreover, if those objects are assumed to be completely powered by AGN 
($\alpha=-1.6$), they%
\footnote{
In this case, we use the line ratios and EWs uncorrected for the 
stellar \HeII\ emission.
} could have a metallicity of $Z=1$--$2\,Z_{\odot}$ with a very high 
ionization parameter $\log U\sim -1$ or a very low metallicity ($Z= 0.05$--$0.1$)
with $\log U \sim -2$.
Although we cannot reach a firm conclusion with the current
data set, this demonstrates the importance of the shape of the incident
radiation field when discussing ISM properties derived from the 
UV emission lines (Sect.\ \ref{ssec:discussion_vuds}).

The object ID 5100565880 ((d) in Fig. \ref{fig:UV_vuds_sfg_agn}), 
showing the highest EWs of \CIII\ and \CIV\ ($\sim 20$\,\AA) among 
the sample of \citet{amorin2017} deserves special mention; 
also since it resembles many objects from
the high-EW sample (A) discussed later.
From the high EWs this source is classified as AGN, whereas 
it falls in the star-formation regime in the C4C3-C34 diagram.
A hint to understanding this puzzling object is provided by the C$/$O
abundance ratio. This object shows a weak \OIII$\lambda 1665$ 
doublet with respect to the strong carbon emission lines. 
\citet{amorin2017} thus derive an exceptionally high C$/$O abundance 
ratio, $\log {\rm C}/{\rm O}=-0.38\pm 0.1$, for its low metallicity of 
$Z\sim0.08\,Z_{\odot}$. 
This C$/$O ratio is close to the solar abundance ratio 
($\log {\rm C}/{\rm O}=-0.26$), a factor of $\sim 2.5$ higher than 
predicted by the \citet{dopita2006}'s formula for its metallicity.
The high  C$/$O abundance for its metallicity explains the ``boost'' of the 
\CIII\ and \CIV\ EWs of this object, and hence its unusual location 
in Fig.\ \ref{fig:Z_ew} as well as Fig.\ \ref{fig:UV_vuds_sfg_agn}, 
in accordance with the fact that the EWs scale approximately linearly with 
C$/$O (Sect.\ \ref{ssec:results_CO}).
This also suggests that the classifications using the EW plots
could diagnose a galaxy as AGN if it possesses an enhanced 
C$/$O abundance ratio.

In summary, \citet{amorin2017}'s sample is mainly explained by 
star-formation as the ionizing source with a low metallicity.
However, it could contain three objects that are, in part, powered 
by an AGN-like hard ionizing spectrum, and their metallicities could
be higher than originally estimated if the harder ionizing spectrum
is taken into account. 
Moreover, one source from \citet{amorin2017} shows a very high 
C$/$O abundance ratio, and presents boosted EWs of \CIII\ and \CIV\ 
for its ISM condition.

%%%%%%%%%%%%%%%%%%%%%%%%%%%%%%%%%%%%%%%%%%%%%%%%%%%%
\subsection{Intermediate \CIII\ emitters with EW(\CIII) $=10$--$20$\,\AA} 
\label{ssec:nature_vuds_mediumC3}
%%%%%%%%%%%%%%%%%%%%%%%%%%%%%%%%%%%%%%%%%%%%%%%%%%%%

Next, we examine the properties of the objects in Sample B, 
composed of intermediately strong \CIII\ emitters with EW(\CIII)=$10$--$20$\,\AA,
using the diagnostics shown in Fig.\ \ref{fig:models_UVlines_popstar_agn_co_compB}.

% -% -% -% -% -% -% -% -% -% -% -% -% -% -% -% -% -% -% -% -% -% -% -% -% -% -% -% -% -% -% 
\subsubsection{Stars versus active galactic nuclei} 
\label{sssec:nature_vuds_mediumC3_fraction_SF_AGN}
% -% -% -% -% -% -% -% -% -% -% -% -% -% -% -% -% -% -% -% -% -% -% -% -% -% -% -% -% -% -% 

As seen from Fig. \ref{fig:models_UVlines_popstar_agn_co_compB}
(cf.\ also Fig.\ \ref{fig:UV_vuds_sfg_agn}), the composite of 
Sample B (yellow-filled square) falls on the border 
between the star-forming galaxy and the AGN regions. 
Individual objects in this class (black open squares) are distributed 
widely on the UV diagnostics.
These trends suggest this galaxy population has an incident radiation field 
that is not fully explained by either a pure star-formation or an AGN.
Based on our UV diagnostics, $\sim 28$\,\%\ of the sample is suggested
to be likely star-forming galaxies, and $\sim 34$\,\%\ are likely AGNs
or objects reproduced by a SF $+$ AGN model.
The rest are uncertain due to a weak constraint on the \HeII\ and/or \CIV\ strengths.

Sample B includes four X-ray-identified type-II AGN candidates
(Sect.\ \ref{sec:c3emitters}).
They are marked with a red diamond in Fig. 
\ref{fig:models_UVlines_popstar_agn_co_compB},
all being classified as AGNs based on the UV diagrams.
This further confirms that these classification diagrams can be 
reliably used to separate star formation from AGN.

In the following, we examine how different models can explain 
the properties of the composite Sample B.

% -% -% -% -% -% -% -% -% -% -% -% -% -% -% -% -% -% -% -% -% -% -% -% -% -% -% -% -% -% -% 
\subsubsection{Mixed SF + AGN models} 
\label{sssec:nature_vuds_mediumC3_SF+AGN}
% -% -% -% -% -% -% -% -% -% -% -% -% -% -% -% -% -% -% -% -% -% -% -% -% -% -% -% -% -% -% 

To infer the typical properties of Sample B, 
we reproduce the line ratios and EWs of the composite of Sample B 
with our models. 
Since the composite falls on the border between the star-forming galaxy 
and the AGN regions in Fig. \ref{fig:UV_vuds_sfg_agn}, 
one simple interpretation of this population 
is that the shape of the incident radiation field is 
a combination of stellar population and AGN.
Indeed, the measurements are reproduced by a hybrid model 
with an AGN contribution of $f_{\rm AGN}\sim 0.07$, 
ISM properties of ($Z$, $\log U$) $=$ ($0.2\,Z_{\odot}$, $-1.5$),
and the current star-formation age of $\sim 3$\,Myr.
We note that the AGN component becomes significant in the high-energy regime 
of $E\gtrsim40$\,eV. 
Adding the hard ionizing AGN spectrum to the stellar light implies
that galaxies with a relatively strong \CIII\ emission 
typically require a harder ionizing spectrum than produced by conventional stellar 
population synthesis codes.
The binary stellar evolution might not be sufficient to fully reproduce these strong 
\CIII\ emitters.
If the SF $+$ AGN hybrid model is correct, a concern remains, however, 
as to whether such a high ionization parameter ($\log U = -1.5$) is compatible with 
a relatively metal-enriched gaseous condition ($Z=0.2\,Z_{\odot}$). 
Previous studies have suggested that galaxies with $Z=0.2\,Z_{\odot}$
typically have an ionization parameter of $\log U = -2.5$ 
irrespective of redshift (e.g., \citealt{onodera2016,kojima2017}). 
If the best-fit parameters are correct, Sample B on average would have 
an ionization parameter higher than expected from the metallicity by nearly 
one order of magnitude.
The presence of AGN could contribute to increase the 
ionization parameter for its metallicity. 
Other studies may be needed to further corroborate this result and 
interpretation. 
Another possible issue of the model is an underestimation of the flux of 
the \OIII$\lambda 1665$ doublet, which is not explicitly 
visible in the UV diagrams but is significantly detected in the composite spectrum
\citep{lefevre2017}. 
This indicates that the assumed C$/$O ratio, given the measured oxygen abundance,
may not be correct for the composite of Sample B, as it is for the default models
(Sect.\ \ref{ssec:results_CO}).
This issue is revisited in Sect.\ \ref{sssec:nature_vuds_mediumC3_highCO}.

% -% -% -% -% -% -% -% -% -% -% -% -% -% -% -% -% -% -% -% -% -% -% -% -% -% -% -% -% -% -% 
\subsubsection{Pure AGN models} 
\label{sssec:nature_vuds_mediumC3_AGN}
% -% -% -% -% -% -% -% -% -% -% -% -% -% -% -% -% -% -% -% -% -% -% -% -% -% -% -% -% -% -% 

If a pure AGN is assumed, a caution is that we should not correct for the 
stellar \CIV\ absorption and the \HeII\ emission for the observed line strength 
(Sect.\ \ref{sec:c3emitters}).
If we assume the observed \CIV\ and \HeII\ emission are entirely of nebular origin,
the composite of Sample B
would marginally fall in the AGN regime in the C4C3--C34 and EW(\CIII) plots
and become compatible with a faint ($\alpha=-2.0$) AGN model 
with ISM properties of ($Z$, $\log U$) $=$ ($0.2\,Z_{\odot}$, $-2$).
However, the best-fit model predicts an EW(\CIV) much larger than
observed.
Given the caution in Sect.\ \ref{ssec:AGNs} that EWs of our AGN models
can be larger if the attenuation of UV-continuum by torus is 
more properly considered, the disagreement of the EW(\CIV) (and possibly
that of EW(\CIII)) would become much larger. 
Moreover, the composite of Sample B presents a \CIV\ emission with 
a weak P-Cygni profile. We thus consider that Sample B should be dominated 
by a stellar population on average.

% -% -% -% -% -% -% -% -% -% -% -% -% -% -% -% -% -% -% -% -% -% -% -% -% -% -% -% -% -% -% 
\subsubsection{Pure SF models} 
\label{sssec:nature_vuds_mediumC3_SF}
% -% -% -% -% -% -% -% -% -% -% -% -% -% -% -% -% -% -% -% -% -% -% -% -% -% -% -% -% -% -% 

Allowing for uncertainties in the corrections for the \HeII\ and the \CIV\ lines, 
there remains another possibility that the composite of 
Sample B could be explained by a pure star-formation model 
if the correction applied to \HeII\ for the stellar contribution is too small.
If the observed \HeII\ emission is dominated by stellar emission ($\sim 90$\,\%), 
Sample B could be reproduced by
a binary stellar population with the ISM properties 
of ($Z, \log U$) $\sim$ ($0.1\,Z_{\odot}$, $-1.5$) -- ($0.15\,Z_{\odot}$, $-1.0$)
and with a starburst age of less than a few Myr.
Under this assumption, the stellar \HeII\ emission
would be $\sim 3$\,\AA. 
However such a large EW(\HeII) is 
only expected from a stellar population with metallicities
of $Z>1\,Z_{\odot}$ and ages of $\gtrsim 10$\,Myr
\citep{brinchmann2008};
both of which are inconsistent with our estimations based on the 
UV diagrams. 
Therefore, we conclude that an ionizing field coming only from 
star formation is unlikely to explain the properties of this 
galaxy sample.

% -% -% -% -% -% -% -% -% -% -% -% -% -% -% -% -% -% -% -% -% -% -% -% -% -% -% -% -% -% -% 
\subsubsection{Blackbody SEDs}  
\label{sssec:nature_vuds_mediumC3_BB}
% -% -% -% -% -% -% -% -% -% -% -% -% -% -% -% -% -% -% -% -% -% -% -% -% -% -% -% -% -% -% 

 We explore blackbody spectra as an alternative explanation.
Blackbodies with a high temperature 
could produce enough high-energy ionizing photons to 
explain the relatively strong \HeII\ emission and the large EWs
of \CIII\ and \CIV\ seen in Sample B.
Indeed, a model with ISM properties of 
($Z, \log U$) = ($\sim 0.2\,Z_{\odot}$, $-0.7$) 
using a temperature of $T\sim 6\times 10^{4}$\,K
could be consistent with the UV lines and EWs.
However, as mentioned earlier, blackbody spectra may be unrealistic
and hence predict large EWs due to their very high \xiion\ parameter,
$\log$\,\xiion$/$\ergsHz\ $=26.2$ for $T\sim 6\times 10^{4}$\,K,
which is much higher than typically found in high-$z$ galaxies 
($\log$\,\xiion$/$\ergsHz\ $=25.3\pm 0.3$; \citealt{bouwens2016}).
This fit shows that Sample B
could be explained with the properties above 
if such an efficient ionizing photon production is achieved.
We note that values as high as $\log$\,\xiion$/$\ergsHz\ $\sim 26$ can be 
reached in the BPASS 300bin models if the lowest metallicity 
($Z=0.05\,Z_{\odot}$) and the youngest starburst age ($1$\,Myr) are adopted.
In this case one would need to invoke a significantly lower
stellar metallicity compared to the gas phase value to mimic
the blackbody case (cf. \citealt{steidel2016}).
Furthermore, we have indicated in Sect.\ \ref{ssec:results_popIII}
that a popIII-like stellar spectrum is approximated by a high-temperature
blackbody. 
It could thus be claimed that Sample B has an extremely young
stellar population, although it is unclear if such a population 
is associated with a relatively enriched ($Z\sim 0.2\,Z_{\odot}$) 
ISM condition.

%FFFFFFFFFFFFFFFFFFFFFFFFFFFFFFFFFFFFFFFFFFFFFFFFFFFFFFFFFFFFFFFFFFFFFFFFF%
\begin{figure*}
  \centerline{
    \includegraphics[width=0.99\textwidth]{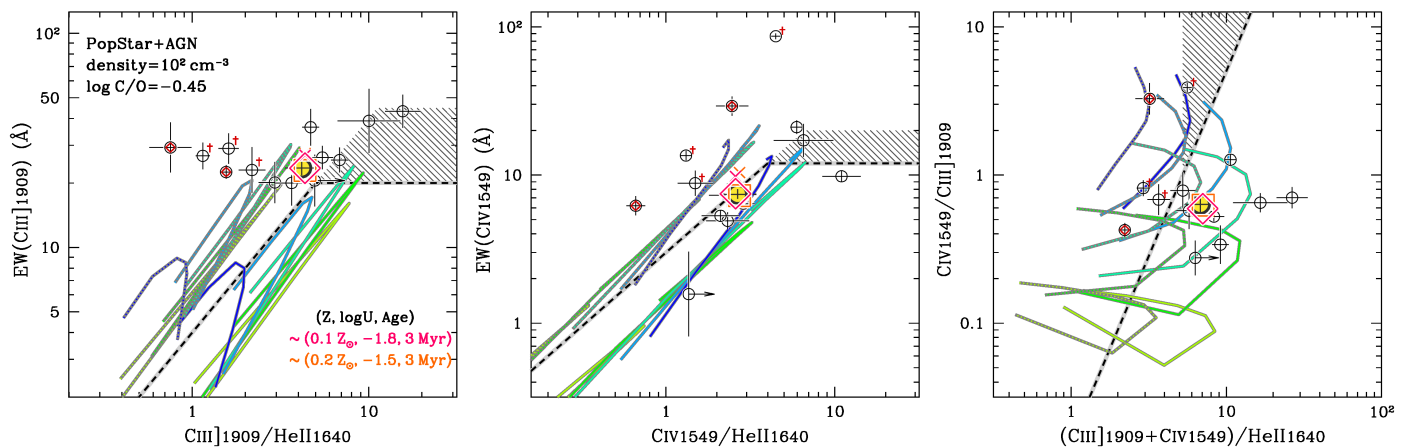}
  }
  \caption{
        Comparison with the composite of strong \CIII\ emitters
        (Sample A; yellow-filled circle)
        and the two best fit models 
        (red open diamond and orange open square).
        The models are provided by the 
        \textsc{PopStar}$+$AGN model 
        with the AGN contribution of $f_{\rm AGN}=0.08$,
        fixing the C$/$O abundance ratio to
        $\log$ C$/$O $=-0.45$.
        The magenta and orange symbols show the low-$Z$ and high-$Z$
        branch solution, respectively,
        as shown in the legend. 
        The crosses are the same as in Fig. \ref{fig:models_UVlines_bpass_compC}.
        The individual objects in Sample A are also plotted 
        (same symbols as Fig. \ref{fig:UV_vuds_sfg_agn}).
        Objects that are diagnosed to be AGN-powered with 
        the UV diagrams but not X-ray detected
        are marked with a red dagger. 
        }
\label{fig:models_UVlines_popstar_agn_co_compA}
\end{figure*}
%FFFFFFFFFFFFFFFFFFFFFFFFFFFFFFFFFFFFFFFFFFFFFFFFFFFFFFFFFFFFFFFFFFFFFFFFF%

% -% -% -% -% -% -% -% -% -% -% -% -% -% -% -% -% -% -% -% -% -% -% -% -% -% -% -% -% -% -% 
\subsubsection{High C$/$O abundance} 
\label{sssec:nature_vuds_mediumC3_highCO}
% -% -% -% -% -% -% -% -% -% -% -% -% -% -% -% -% -% -% -% -% -% -% -% -% -% -% -% -% -% -% 

One uncertainty in the above model fits is the C$/$O abundance ratio.
As mentioned in Sect. \ref{ssec:results_CO}, 
the \CIII\ and \CIV\ EWs would be larger than predicted by the standard models
if the sources turn out to be more carbon-enriched for their oxygen abundance. 
This could also affect the estimates of the ISM conditions.
In order to assess this possibility, we use the method of 
\citet{PA2017} to estimate the C$/$O ratio from 
the (\CIII$+$\CIV)$/$\OIII\ ratio%
\footnote{See Eq.\ (7) of \citet{PA2017}.
This equation could overestimate the C$/$O abundance ratio for 
a metal-rich ($Z\gtrsim 0.5\,Z_{\odot}$), low-ionization parameter 
($\log U\lesssim -3$) galaxy.
},  for the composite of Sample B. 
We find a high ratio $\log$\,(\CIII$+$\CIV)$/$O\,{\sc iii}] $\sim 1$, 
which corresponds to an abundance ratio $\log$\,C$/$O $\sim -0.3$.
which is usually found in chemically evolved galaxies 
($Z\sim 1\,Z_{\odot}$).
Therefore, the high C$/$O abundance ratio indicates that the sources 
of Sample B should have a solar metallicity or 
a higher carbon-to-oxygen abundance ratio than expected 
for its oxygen abundance.

To be more quantitative, we adopt the models with a 
fixed C$/$O abundance ratio of $\log$\,C$/$O $\sim -0.3$
instead of the lower values as detailed in Section 3.3
from the prescription of \citet{dopita2006}.
Using models with the high C$/$O abundance ratio, 
we find that the composite of 
Sample B is nicely explained by hybrid models 
combining star formation and AGN.
The best-fit models include two different sets of ISM properties,
one is a low-$Z$ solution; 
($Z$, $\log U$, age, $f_{\rm AGN}$) $=$ 
($0.02\,Z_{\odot}$, $-1.7$, $6$\,Myr, $0.085$) --
($0.025\,Z_{\odot}$, $-1.65$, $8$\,Myr, $0.1$),
and another is a high-$Z$ one; 
($Z$, $\log U$, age, $f_{\rm AGN}$) $=$ 
($0.2\,Z_{\odot}$, $-1.0$, $8$\,Myr, $0.1$) --
($0.25\,Z_{\odot}$, $-0.75$, $3$\,Myr, $0.07$).
Figure \ref{fig:models_UVlines_popstar_agn_co_compB} shows 
the two best-fit models.
We have also checked with our models that the 
(\CIII$+$\CIV)$/$O\,{\sc iii}] ratio becomes as high as 
observed if the high C$/$O ratio is adopted.
Pure star formation would predict overly high carbon-to-helium
line ratios, and adding an AGN is necessary to lower
the ratios as seen in the composite spectrum.
Although we do not have any definitive clue to determine
which metallicity solutions are preferred, we consider 
the low-$Z$ solution of $Z\sim 0.02\,Z_{\odot}$ 
more likely since it leads to a better agreement with the 
known $Z$--$\log U$ relation. 
If the best-fit model is correct, the C$/$O abundance ratio
is $\sim 3.5$ times higher than predicted by the relation of 
\citet{dopita2006} for its low metallicity.
The physical cause for the high C$/$O abundance remains to be 
determined, and is briefly discussed below 
in Sect.\ \ref{ssec:discussion_origin_highCO}.

% -% -% -% -% -% -% -% -% -% -% -% -% -% -% -% -% -% -% -% -% -% -% -% -% -% -% -% -% -% -% 
\subsubsection{Summary for intermediate-strength emitters EW(\CIII) $=10-20$\,\AA\ (Sample B)} 
\label{sssec:nature_vuds_mediumC3_summary}
% -% -% -% -% -% -% -% -% -% -% -% -% -% -% -% -% -% -% -% -% -% -% -% -% -% -% -% -% -% -% 

In summary, the most likely interpretation of the composite of Sample B 
calls for a star-forming galaxy with a small contribution ($\sim 7-10$\,\%)
of an AGN-like hard ionizing radiation field,
having an overabundance of carbon for its metal-poor 
condition ($Z=0.02\,Z_{\odot}$).
However, the standard prescription of C$/$O ratio as a function of 
metallicity could also reproduce the UV lines' fluxes and EWs. 
The latter case requires an ionizing spectrum as hard, 
and an ionization parameter as high, as 
in the former case, but has a metallicity higher by an order of magnitude
($Z\sim 0.2\,Z_{\odot}$).
To reach a more definitive conclusion, we need further observations,
especially of the rest frame-optical nebular emission lines, 
to confirm the ISM properties.
The possible but more hypothetical blackbody 
model can be tested with the \xiion\ parameter, which will be directly 
estimated with the Hydrogen Balmer line luminosity 
as described in Sect.\ \ref{sssec:nature_vuds_strongC3_BB}.

The above discussion is based on the composite of Sample B.
As we note early in this Section, the sample contains a variety 
of objects with at least $\sim 28$\,\%\ nebular emission 
produced by a stellar ionizing spectrum. 
With our models, the likely star-forming galaxy population 
in this class is typically reproduced by a low-metallicity 
($Z=0.05$ to $0.2\,Z_{\odot}$, or could be smaller), 
high-ionization-parameter ($\log U=-1$ to $-2$) model
including binary stars. 
The age of the current star formation is predicted
to be approximately $30$\,Myr or shorter, and the \xiion\ parameter
to be $\log$\,\xiion$/$\ergsHz\ $\sim 25.55$ or 
as high as $25.7$.
This population is thus likely equivalent to the low-$Z$ 
galaxy population selected by the \CIII$+$\OIII\ emission
in VUDS \citep{amorin2017}.
Hence the sample could contain a variety of objects, including 
low-metallicity star-forming galaxies, pure AGNs, and 
objects with a hard ionizing spectrum reproduced by 
a mix of stars and AGN.

%ttttttttttttttttttttttttttttttttttttttttttttttttttttttttttttttttttttttttt%
\begin{table*}
  \centering
  \caption{Summary of the inferred properties of the VUDS \CIII\ emitters.}
  \label{tbl:properties_composites}
  \renewcommand{\arraystretch}{1.25}
  \begin{tabular}{@{}cclccccc@{}}
    \hline
    Sample &
    Percentile &
    Input radiation &
    $Z$ ($Z_{\odot}$)&
    $\log U$ &
    $\log$\,C$/$O &
    Age (Myr)&
    $\log$\,\xiion\ (\ergsHz) \\
    \hline
    A &
    $\sim 30$\,\% &
     pure AGN &
     -- &
     -- &
     -- & 
     -- &
     -- \\
     &
    $\sim 70$\,\% &
     {\sc PopStar}$+$AGN &
     $0.1$ $\cdots$ $0.2$ &
     $-1.75$ $\cdots$ $-1.5$ &
     $-0.45$ $\cdots$ $-0.3$ & 
     $\sim 3$ &
     $25.7$ $\cdots$ $25.75$ \\
      &
      (On average )&
     {\tiny ($f_{\rm AGN}=0.08^{+0.02}_{-0.05}$)} &
      &
      &
      &
      &
      \\
      \\
    B &
    $\gtrsim 34$\,\% &
     pure AGN &
     -- &
     -- &
     -- & 
     -- &
     -- \\
     &
    $\gtrsim 28$\,\% &
     BPASS 300bin &
     $0.05$ $\cdots$ $0.2$ &
     $-2$ $\cdots$ $-1$ &
     $-0.8$ $\cdots$ $-0.6$ & 
     $\lesssim 30$ &
     $25.55$ $\cdots$ $25.7$ \\
      &
     On average &
     {\sc PopStar}$+$AGN&
     $0.02$ $\cdots$ $0.025$ &
     $-1.7$ $\cdots$ $-1.65$ &
     $-0.3$ & 
     $6$ $\cdots$ $8$ &
     $25.6$ $\cdots$ $25.7$ \\
      &
      &
     {\tiny ($f_{\rm AGN}=0.07$ -- $0.1$)} &
     {\it or} \,\, $0.2$ $\cdots$ $0.25$ &
     $-1.0$ $\cdots$ $-0.75$ &
     $-0.3$ & 
     $3$ $\cdots$ $8$ &
     $25.55$ $\cdots$ $25.7$ \\
     \\
    C &
     On average &
     BPASS 300bin &
     $0.3$ $\cdots$ $0.5$ &
     $-3.0$ $\cdots$ $-2.7$ &
     $-0.6$ $\cdots$ $-0.4$ &
     $50$ $\cdots$ $200$ &
     $25.3$ $\cdots$ $25.4$ \\

     \hline
  \end{tabular}
\end{table*}
%ttttttttttttttttttttttttttttttttttttttttttttttttttttttttttttttttttttttttt%

% -% -% -% -% -% -% -% -% -% -% -% -% -% -% -% -% -% -% -% -% -% -% -% -% -% -% -% -% -% -% 
\subsection{Strongest \CIII\ emitters with EW(\CIII) $>20$\,\AA} 
\label{ssec:nature_vuds_strongC3}
% -% -% -% -% -% -% -% -% -% -% -% -% -% -% -% -% -% -% -% -% -% -% -% -% -% -% -% -% -% -%

Finally, we discuss the properties of Sample A
whose \CIII\ equivalent widths are larger than $20$\,\AA, and 
whose line properties are shown in 
Figs. \ref{fig:UV_vuds_sfg_agn} and 
\ref{fig:models_UVlines_popstar_agn_co_compA}.

% -% -% -% -% -% -% -% -% -% -% -% -% -% -% -% -% -% -% -% -% -% -% -% -% -% -% -% -% -% -% 
\subsubsection{Stars versus active galactic nuclei} 
\label{sssec:nature_vuds_strongC3_fraction_SF_AGN}
% -% -% -% -% -% -% -% -% -% -% -% -% -% -% -% -% -% -% -% -% -% -% -% -% -% -% -% -% -% -% 

As shown in Fig. \ref{fig:UV_vuds_sfg_agn} (or also Fig. \ref{fig:models_UVlines_popstar_agn_co_compA}) 
these sources occupy diverse areas in
our SF--AGN diagnostic diagrams.
Five sources, representing $\sim 30$\,\%\ of sample A, are consistently found in 
the AGN area of the three diagrams,
suggesting a clear AGN classification.
Of these five sources, two turn out to be detected in X-rays and identified as 
type-II AGN by \cite{talia2017}.
The other three are highlighted with a red dagger in Fig. \ref{fig:models_UVlines_popstar_agn_co_compA}.
Attention is now focused on the remaining  $\sim 70$\,\%\ of sample A
that are more puzzling. 
In the C4C3--C34 diagram on Fig. \ref{fig:models_UVlines_popstar_agn_co_compA}, 
all of them (i.e., black circles without a red diamond or a red dagger) 
fall in the star-formation region. 
On the other hand, all show an EW of \CIII\ that 
exceeds the limit found by standard stellar population models,
as well as an EW(\CIV) exhibiting a similar behavior. 
The apparent inconsistency suggests that this small galaxy population  
has unusual properties regarding their incident radiation field 
and/or ISM condition.

We now explore the \cloudy\ models with the several different 
incident radiation fields as presented in Sect.\ \ref{sec:modelling} 
to constrain the nature of this sample. 
For this purpose we first use the median composite of the sample 
to investigate the detailed properties.

% -% -% -% -% -% -% -% -% -% -% -% -% -% -% -% -% -% -% -% -% -% -% -% -% -% -% -% -% -% -% 
\subsubsection{Blackbody models} 
\label{sssec:nature_vuds_strongC3_BB}
% -% -% -% -% -% -% -% -% -% -% -% -% -% -% -% -% -% -% -% -% -% -% -% -% -% -% -% -% -% -% 

Normal star-forming galaxies cannot have such a large EW(\CIII)
even with the binary stellar populations at the youngest starburst age 
of $1$\,Myr. 
Pure AGNs are unlikely scenarios since the relatively weak \HeII\ emission
to the other metal lines is incompatible with any AGN model. 
Through the analyses using SF$+$AGN, blackbody, and PopIII star
as an incident radiation field, we find only one explanation 
for Sample A: the SED of a high-temperature blackbody.
The blackbody model with $T_{\rm eff}\sim 6.5\times 10^4$\,K 
can reproduce the EWs and the flux line ratios with 
an ISM condition of 
($Z, \log U$, \nH) = ($\sim 0.1\,Z_{\odot}$, $-1.6$, $10^4\,{\rm cm}^{-3}$).
Again, as discussed in Sect.\ \ref{sssec:nature_vuds_mediumC3_BB}, 
if the blackbody model were to give the best-fit to the UV spectrum, 
the objects in this sample would require a very high \xiion\ parameter 
($\log$\,\xiion$/$\ergsHz\ $=26.2$) plus a hard ionizing spectrum.
However, it remains unclear if such an 
efficient ionizing photon production is achievable 
with stellar populations. 
In the study of \citet{bouwens2016} a small number of galaxies 
is suggested to have a \xiion\ parameter  
as high as $\log$\,\xiion$/$\ergsHz\ $\sim 25.8$ for the bluest galaxies. 
Even \Lya\ emitters that are considered to present a strong 
\CIII\ emission (cf. \citealt{stark2014,rigby2015}) possess 
a \xiion\ parameter reaching at maximum $\log$\,\xiion$/$\ergsHz\ $=25.8$
at $z\sim 3$ \citep{nakajima2016}.
Although a higher \xiion\ would be expected for these galaxies 
if their escape fraction of ionizing photons was non-zero, 
potential local counterparts of high-$z$ galaxies, so called
green pea galaxies, show a relatively modest \xiion\
parameter ($\log$\,\xiion$/$\ergsHz\ $\sim 25.0$ -- $25.6$) 
after correction for the amount of escaped ionizing photons 
\citep{schaerer2016}.
In conclusion, sources with a very high ionizing photon production 
per UV luminosity ($\log$\,\xiion$/$\ergsHz\ $> 26$), corresponding to the 
blackbody spectrum which can explain Sample A, are not known 
so far and are difficult to justify theoretically.

% -% -% -% -% -% -% -% -% -% -% -% -% -% -% -% -% -% -% -% -% -% -% -% -% -% -% -% -% -% -%
\subsubsection{High C/O abundance} 
\label{sssec:nature_vuds_strongC3_highCO}
% -% -% -% -% -% -% -% -% -% -% -% -% -% -% -% -% -% -% -% -% -% -% -% -% -% -% -% -% -% -%

An alternative explanation to reconcile the high EWs
of \CIII\ and \CIV\ -- indicative of AGNs -- and the position in the
C4C3--C34 diagram, which indicates stellar photoionization, is 
to invoke a high C$/$O abundance, following the case of the individual
source ID 5100565880 and possibly the composite of Sample B
discussed earlier
(Sects.\ \ref{ssec:nature_vuds_Amorin2017} and \ref{sssec:nature_vuds_mediumC3_highCO}).
Using the (\CIII$+$\CIV)$/$O\,{\sc iii}] ratio and the method of \citet{PA2017}
we estimate a typical C$/$O ratio of this class of $\log$\,C$/$O $=-0.45$. 
With the relatively high C$/$O ratio we re-analyze the composite 
through the several incident radiation fields, finding that the composite 
is reproduced by SF $+$ AGN models.
The best-fit model has the ISM properties of 
($Z$, $\log U$, age) $=$ 
($0.1\,Z_{\odot}$, $-1.75$, $3$\,Myr) -- 
($0.2\,Z_{\odot}$, $-1.5$, $3$\,Myr)
with a fraction of AGN-like hard ionizing radiation field of 
$f_{\rm AGN}\sim 0.08$. 
If the best fit model shown in 
Fig. \ref{fig:models_UVlines_popstar_agn_co_compA} 
is correct, the C$/$O ratio would be elevated 
by $0.2$ -- $0.3$\,dex compared with the standard models.

% -% -% -% -% -% -% -% -% -% -% -% -% -% -% -% -% -% -% -% -% -% -% -% -% -% -% -% -% -% -% 
\subsubsection{The nature of the strongest emitters in Sample A} 
\label{sssec:nature_vuds_strongC3_summary}
% -% -% -% -% -% -% -% -% -% -% -% -% -% -% -% -% -% -% -% -% -% -% -% -% -% -% -% -% -% -% 

The individual sources included in Sample A are also shown in 
Fig. \ref{fig:models_UVlines_popstar_agn_co_compA}. 
As mentioned earlier (Sect. \ref{sssec:nature_vuds_strongC3_fraction_SF_AGN}), 
$\sim 30$\,\% of this sample are clearly identified as AGN. 
Interestingly, the other dominant population, which is represented 
by the median composite, shows only relatively little variety.
Based on our method using the C4C3--C34 diagram, 
all of them are diagnosed as being star-forming galaxies,
while their large EWs of \CIII\ and \CIV\ are incompatible 
with any standard model of star-formation.
The most likely scenario for this puzzling but interesting population 
is that it is made of star-forming galaxies 
with a small contribution 
from an AGN-like hard ionizing radiation field. 
We derive the properties for the individual sources 
in the same manner as for the composite, finding that 
they require a radiation field made up of a mixture of a young stellar population
plus an AGN component with $f_{\rm AGN} = 3$ -- $10$\,\%\ 
and possess a moderately high C$/$O ratio from 
$\log$ C$/$O $=-0.45$ to $-0.3$ for their subsolar metal 
condition ($Z=0.1$ -- $0.2\,Z_{\odot}$).
These properties are well represented by the composite. 
A contribution from an AGN with a hard ionizing radiation field is necessary
even for the objects that are not likely AGNs in the C4C3--C34 diagram.

%%%%%%%%%%%%%%%%%%%%%%%%%%%%%%%%%%%%%%%%%%%%%%%%%%%%
%%%%%%%%%%%%%%%%%%%%%%%%%%%%%%%%%%%%%%%%%%%%%%%%%%%%
\section{Discussion} 
\label{sec:discussion}
%%%%%%%%%%%%%%%%%%%%%%%%%%%%%%%%%%%%%%%%%%%%%%%%%%%%
%%%%%%%%%%%%%%%%%%%%%%%%%%%%%%%%%%%%%%%%%%%%%%%%%%%%

We have shown in Sect.\ \ref{sec:results_UV_diagrams} that 
the UV diagnostics proposed in Sect.\ \ref{sec:results_model}  
can distinguish AGN from star-formation-dominated galaxies, and that they can be used to constrain
ISM properties of galaxies.
We now discuss the properties of the \CIII\ emitters from VUDS
derived in Sect.\ \ref{sec:nature_vuds}, compare them with 
other studies at high redshift, and point out some implications.

%%%%%%%%%%%%%%%%%%%%%%%%%%%%%%%%%%%%%%%%%%%%%%%%%%%%
\subsection{The nature and physical properties of the \CIII\ emitters in VUDS sample of star-forming galaxies}
\label{ssec:discussion_vuds}
%%%%%%%%%%%%%%%%%%%%%%%%%%%%%%%%%%%%%%%%%%%%%%%%%%%%

The main radiation field and ISM properties derived from the comparison of 
the UV emission lines with our grid of photoionization models are given in 
Table \ref{tbl:properties_composites} for the three samples of \CIII\ emitters
defined above (cf.\ Table \ref{tbl:samples}).

The composite spectrum of $450$ star-forming 
galaxies at $z \sim 3$, whose average \CIII\ EW is 
approximately equal to 2 \AA,  
is well described by stellar photoionization, as expected. 
We also suggest that adding binary stars \citep{stanway2015} is favored,
as already pointed out by other studies 
(e.g., \citealt{steidel2016,JR2016}).
The typical values inferred for the metallicity, ionization parameter,
and age of the ionizing stellar population are
($Z$, $\log U$, age) from ($0.3\,Z_{\odot}$, $-2.7$, $200$\,Myr) to 
($0.5\,Z_{\odot}$, $-3$, $50$\,Myr).
This is in agreement with earlier work 
(e.g., \citealt{steidel2014,onodera2016} and references therein).
Using EW(\CIII) and the inferred ISM properties, we constrain 
the average efficiency of the ionizing photon production of this sample 
as $\log$\,\xiion$/$\ergsHz\ $=25.3$ -- $25.4$.
The \xiion\ is higher than the canonical value usually assumed by 
typical models of reionization ($25.2$; e.g., \citealt{robertson2013}),
but comparable to those of (bluer) Lyman break galaxy populations
at $z=4-5$ \citep{bouwens2016}. 
A revised relationship between UV slope and dust attenuation
could elevate the canonical \xiion\ to reconcile the difference 
\citep{reddy2017}.

Among the strong and intermediate \CIII\ emitters 
(Samples A and B), 
approximately $30$\,\%\ are likely to be dominated by AGNs
based on our UV diagnostics. 
Their average UV spectra indicate that the combination of a young stellar 
population and an AGN-like hard component is required 
for these strong \CIII\ emitters. 
Furthermore, to fully explain the large EWs of \CIII\ (and \CIV)
but faint \OIII$\lambda 1665$ emission, we suggest that 
the population typically has a high C$/$O abundance ratio
($\log$\,C$/$O = $-0.45$ and $-0.3$ for samples A and B,
respectively).
With this assumption and a hard stellar $+$ AGN SED, 
we are able to reproduce the average UV observations. 
The inferred ISM properties are 
($Z$, $\log U$) $=$ ($0.02\,Z_{\odot}$, $-1.7$) or ($0.2\,Z_{\odot}$, $-1.0$)
for sample B, and 
($0.1$ -- $0.2\,Z_{\odot}$, $-1.75$ to $-1.5$) for sample A. 
Since the EW of \CIII\ peaks for metallicities of $Z=0.1-0.2\,Z_{\odot}$, 
the selection of objects by strong \CIII\ emission should result in
a population with a subsolar metallicity (with a boost of C$/$O,
whose  origin is discussed  in Sect.\ \ref{ssec:discussion_origin_highCO}). 
In this sense, Sample B covering weaker \CIII\  emitters could contain 
galaxies that are less chemically enriched than Sample A.
The large EW(\CIII)s indicate a young starburst age of $\sim 3$ -- $8$\,Myr
and an efficient ionizing photon production of 
$\log$\,\xiion$/$\ergsHz\ $=25.55$ -- $25.75$ for Samples A and B,
as high as those observed in the bluest population of continuum-selected 
galaxies \citep{bouwens2016}, LAEs \citep{nakajima2016}, and galaxies
at $z>6$ (\citealt{stark2017}; Sect.\ \ref{ssec:discussion_comparison}).
As shown in \citet{lefevre2017}, galaxies in the A and B samples overall 
show a smaller SFR than expected for their stellar mass compared to 
the main sequence of star-forming galaxies. This could be interpreted 
as a sign of feedback from the AGN component which suppresses 
surrounding star-formation \citep{lefevre2017}, 
further reinforcing the analysis presented here.

The presence of an AGN-like hard radiation field  can change the ISM properties
derived from the UV high-ionization lines.
Reexamining the VUDS sample of \CIII$+$\OIII\ emitters presented 
by \citet{amorin2017}, we find  
that three out of the ten \CIII\ emitters could be partly powered by AGN.
The stellar $+$ AGN hybrid models can reproduce their UV line spectra
with metallicities by a factor of $2$--$10$ higher than originally inferred,
although we cannot fully rule out the low-metallicity solution for one of the
three objects.
If such high metallicities are considered, these objects would lie on 
(a simple extrapolation of) the mass-metallicity relation previously reported 
at similar redshifts ($z\sim 2.3$; e.g., \citealt{steidel2014}).
Although we require the observations of rest-frame optical emission lines to draw 
a firm conclusion, our study shows that a careful investigation of the 
powering source is essential for understanding the evolution of the 
ISM gaseous properties, particularly when only the UV high-ionization 
lines (and complicated \Lya) are available.

%%%%%%%%%%%%%%%%%%%%%%%%%%%%%%%%%%%%%%%%%%%%%%%%%%%%
\subsection{On the origin of high C$/$O abundance ratios}
\label{ssec:discussion_origin_highCO}
%%%%%%%%%%%%%%%%%%%%%%%%%%%%%%%%%%%%%%%%%%%%%%%%%%%%

If the high C$/$O abundance inferred for the intermediate and strong 
\CIII\ emitters is correct as reported in Sects.\ \ref{sssec:nature_vuds_mediumC3_highCO}
and \ref{sssec:nature_vuds_strongC3_highCO}, 
what is its physical origin?
Generally, solar C$/$O ratios are observed (e.g., \citealt{dopita2006}) and
predicted by chemical evolution models (e.g., \citealt{mattsson2010}) 
at high metallicities ($Z \ga 1$ \zsun). 
The trend of the increasing C$/$O with metallicity at \Oabundance\ $\ga 8$ 
is usually explained with a delayed contribution to carbon enrichment from 
low- and intermediate-mass (i.e., longer-lived) stars along with metal-rich (PopI) 
massive stars.
At very low metallicities, fast rotating stars are also predicted to produce 
yields with a high C$/$O ratio
\citep[cf.][]{chiappini2006}.
However, this behavior is only expected for metallicities \Oabundance\
$< 7$, which is significantly lower than the estimated metallicities
of the star-forming galaxies at $z=2-4$ in our sample.
Asymptotic giant branch (AGB) stars of intermediate mass, which could
contribute to chemical enrichment on the relatively short timescales
appropriate for these high-redshift sources, do not produce Carbon,
since they are predicted to undergo hot bottom burning, which in fact
destroys this element \citep{marigo2001}.
Only under special circumstances
may this phase be avoided and high C$/$O ratios be reached
\citep{marigo2007}.
From these available nucleosynthesis predictions for single stars it
appears therefore difficult to explain a solar or super-solar C$/$O abundance ratio 
at the subsolar metallicities in the galaxies we are observing.

Alternatively, 
because both carbon and nitrogen are secondary nucleosynthesis 
elements at high metallicity, these apparent over-abundances at 
low metallicity (i.e., low O$/$H ratio) could be due to a massive inflow of 
pristine gas into the system. This event would lower the O$/$H ratio for 
a chemically evolved system but not alter the C$/$O and N$/$O ratios,
resulting in apparent over-abundances of secondary nucleosynthetic 
elements of nitrogen and carbon for its metallicity.
A similar idea has been suggested for explaining a nitrogen excess 
as the origin of the offset seen in the BPT diagram in high-$z$ star-forming 
galaxies (e.g., \citealt{sanders2016} and references therein)
and their low-$z$ analogs (e.g., \citealt{amorin2010}).
If this scenario were correct, the VUDS intermediate and strong \CIII-emitter 
samples would contain galaxies that are experiencing such a massive 
inflow and undergoing a burst of star formation. 
Moreover, incomplete metal mixing in the system might leave 
some pockets of pristine gas in which PopIII stars could form.
The PopIII stellar population, comprising a small fraction of 
the total star-formation activity, could explain narrow \HeII-line emitters (\citealt{cassata2013}; see also \citealt{schaerer2003,raiter2010})
as well as the hard ionizing spectra as suggested for the 
VUDS intermediate and strong \CIII-emitter samples.

Furthermore, a significant population of carbon-sequence W-R (WC) stars
could eject carbon into the ISM, leading to the observed enhancement
of the C$/$O abundance.
This was also initially introduced to explain a possible cause of the nitrogen
enhancement in galaxies with nitrogen-sequence W-R (WN) stars 
(e.g., \citealt{masters2014} and references therein).
However, we consider this scenario unlikely, because 
(i) the fraction of massive stars entering the W-R phase becomes small 
in a low-metallicity regime,
and because 
(ii) the W-R population is dominated by WN stars in a metal-poor 
ISM environment (e.g., \citealt{crowther2007}).

We can test these scenarios by examining the C$/$N abundance ratio 
as a function of metallicity, for example.
A further discussion of the possible variation of the C$/$O
abundance ratio in high-$z$ objects will be presented 
elsewhere using our VUDS sample (Amor\'{i}n et al. in prep. 2017).
In any case, we emphasize that the objects potentially possessing a very high 
C$/$O ratio are quite rare in the large VUDS sample, 
thus only rarely requiring such extraordinary explanations.

%%%%%%%%%%%%%%%%%%%%%%%%%%%%%%%%%%%%%%%%%%%%%%%%%%%%
\subsection{Comparison with other high-redshift sources}
\label{ssec:discussion_comparison}
%%%%%%%%%%%%%%%%%%%%%%%%%%%%%%%%%%%%%%%%%%%%%%%%%%%%

% -% -% -% -% -% -% -% -% -% -% -% -% -% -% -% -% -% -% -% -% -% -% -% -% -% -% -% -% -% -%
\subsubsection{The Lynx  arc at $z=3.4$}
\label{sssec:discussion_comparison_Lynx}
% -% -% -% -% -% -% -% -% -% -% -% -% -% -% -% -% -% -% -% -% -% -% -% -% -% -% -% -% -% -%

Interestingly, we note that the Lynx arc at $z=3.4$ 
\citep{fosbury2003,villar-martin2004}
presents similar UV spectroscopic properties to the 
strongest \CIII-emitter sample from VUDS, with very large EWs of 
\CIII\ and \CIV\ ($\sim 50-60$\,\AA)%
\footnote{
We calculate the EWs by using the given fluxes and the continuum flux 
densities, which are estimated from the SED plot \citep{fosbury2003}.
}
. However, line ratios place this object in the star-formation regime on the C4C3-C34 
diagram (blue circle in Figs. \ref{fig:UV_SFGsAGNs_both} and \ref{fig:UV_vuds_sfg_agn}). 
Its rest-frame optical nebular spectrum suggests an ISM condition of 
a high-ionization parameter ($\log U \simeq -1$) and 
a low metallicity ($Z\sim 0.1\,Z_{\odot}$). 
The presence of an AGN is not required according to 
\citet{fosbury2003} and \citet{villar-martin2004},
although \citet{binette2003} show that AGN models with an absorbed 
power-law distribution are able to reproduce the UV-to-optical line spectrum.
We thus need a relatively high C$/$O abundance ratio to explain 
the large EWs of \CIII\ and \CIV\ for this object. 
Indeed,  $\log({\rm C}/{\rm O})=-0.37$ is inferred using the method of 
\citet{PA2017}.
Furthermore, a high N$/$O abundance ratio is suggested for this object
by a factor of $\sim 2$--$3$ compared with the solar abundance
\citep{villar-martin2004}.

% -% -% -% -% -% -% -% -% -% -% -% -% -% -% -% -% -% -% -% -% -% -% -% -% -% -% -% -% -% -%
\subsubsection{Recent \CIII\ detections at high redshift}
\label{sssec:discussion_comparison_otherC3}
% -% -% -% -% -% -% -% -% -% -% -% -% -% -% -% -% -% -% -% -% -% -% -% -% -% -% -% -% -% -%

\cite{stark2014} have reported  \CIII\ emission in 16 lensed $z \sim 1.5-2$ galaxies. 
Five of them also show \CIV, which is weaker than \CIII, as we would expect. 
When possible these sources have been included in the relevant Figures 
discussed above.
For three sources with multiple line detections, \cite{stark2014} determine metallicities 
from their complex photoionization modeling including a Bayesian analysis, 
finding \Oabundance $\sim 7.8$ with relatively large uncertainties 
(see Fig.\ \ref{fig:Z_ew}).
The comparison with our models shows that the observed line strengths of their 
\CIII\ emitters can be reproduced with the ionization field of star-forming galaxies, 
either from  \textsc{PopStar} or BPASS models. 
This is in agreement with the conclusions of \cite{stark2014}.

A more peculiar source, named A383-5.2 with strong \CIII\ emission 
(EW(\CIII) $=22.5 \pm 7.5$ \AA), was found at $z=6.03$ by \cite{stark2015_c3}. 
Fitting only one UV line plus the broad-band SED of this galaxy, \cite{stark2015_c3} 
determine a very low-metallicity \Oabundance $\sim 7.3\pm0.2$, a high ionization 
parameter, and a relatively high C$/$O abundance for its low metallicity. 
In addition, \citet{stark2017} recently report another \CIII\ detection with 
a large EW of $22\pm 2$\,\AA\ from a luminous high-$z$ source, EGS-zs8-1, 
at $z=7.73$.
The ISM properties including the C$/$O ratio are inferred to be quite
similar to those of A383-5.2.
Whether these explanations are unique remains, however, questionable. 
If the metallicity is correct, it places these sources in a very similar location 
as some sources of \cite{amorin2017}, as seen from Fig.\ \ref{fig:Z_ew}. 
From this consideration and from the resemblance of these sources to 
those from our intermediate and strong \CIII-emitter sample, we suspect that 
these high-$z$ sources could also be enriched in Carbon. 
More data are needed to better constrain these sources.

% -% -% -% -% -% -% -% -% -% -% -% -% -% -% -% -% -% -% -% -% -% -% -% -% -% -% -% -% -% -%
\subsubsection{\CIV\ emitters at $z \sim 3$--$7$}
\label{sssec:discussion_comparison_otherC4}
% -% -% -% -% -% -% -% -% -% -% -% -% -% -% -% -% -% -% -% -% -% -% -% -% -% -% -% -% -% -%

Recent spectroscopic observations of galaxies at redshift $z>6$ 
have identified strong \CIV\ emission with rest-frame EW $\sim 20$\,\AA\ or larger 
(\citealt{stark2015_c4,schmidt2017,mainali2017}; Fig. \ref{fig:UV_vuds_sfg_agn}).
If these objects are powered by star formation, their high \CIV\ EWs 
cannot be reproduced by conventional binary or single
stellar populations, as demonstrated above. 
This suggests that these early galaxies possess a much harder ionizing spectrum, 
a more efficient ionizing photon production (i.e., a higher \xiion\ parameter), 
and/or an elevated C$/$O abundance ratio. 
Although no evidence for AGN has so far been found for these objects, 
we cannot fully rule out an AGN-powering radiation field 
with a weak upper-limit on the \HeII\ emission.
For example, \citet{schmidt2017} discuss the ionizing source of 
a $z=6.1$ object using the AGN models of \citet{feltre2016} and 
galaxy models of \citet{gutkin2016} on the line ratios diagrams 
of \CIV$/$\CIII\ and \CIV$/$\HeII, concluding that the object does not 
favor an AGN. We think the situation is unclear. It is true that their emission
line limit of \CIV$/$\HeII\ $\gtrsim 0.5$ $(2\sigma)$ is consistent
with many of the star-forming galaxy models and with only a small number 
of AGN models. However, most of the observed AGNs are reproduced 
by the ``minority'' of the AGN models, with line ratios of 
\CIV$/$\HeII\ $\sim 1$--$3$ 
(see top-right panel of Fig. \ref{fig:UV_SFGsAGNs_both}; 
\citealt{nagao2006a,dors2014,hainline2011}). 
Since the AGN models of \citet{feltre2016} are based on a large 
grid of parameter sets, some of which might be unrealistic, 
 comparing an observation with the 
abundance of the generated models to judge its ionizing source could be misleading.
For this particular example of a $z=6.1$ object, \citet{mainali2017}
provide a stronger constraint on the \CIV$/$\HeII\ ratio of $\gtrsim 7.6$ $(2\sigma)$, 
leading to the conclusion that the $z=6.1$ source is powered by 
star formation.

A lensed \Lya\ emitter at $z=7.045$ showing strong \CIV\ emission 
with EW(\CIV) $\sim$ 40 \AA\ was discovered by \cite{stark2015_c4}. 
Again, using the models mentioned above, they derive a very low metallicity 
(\Oabundance $\sim 7.0 \pm 0.3$) and a high ionizing photon flux.
The assumed or derived C$/$O abundance is not mentioned.
According to  \cite{stark2015_c4} this source can be explained by stellar 
photoionization or by an AGN. From our modeling we do not see how such 
a high \CIV\ EW could be reproduced without a significant C$/$O 
overabundance, that is,\ close to solar value, in case of stellar photoionization. 
Alternatively, we agree that an AGN-like hard spectrum could explain the currently 
available observations. 
In fact the properties of this source resemble those of some of our strongest \CIII\ emitters (Sample A), whose nature, including an AGN-powered source of ionizing 
photons, was discussed above.

%FFFFFFFFFFFFFFFFFFFFFFFFFFFFFFFFFFFFFFFFFFFFFFFFFFFFFFFFFFFFFFFFFFFFFFFFF%
\begin{figure*}
  \centerline{
    \includegraphics[width=0.9\textwidth]{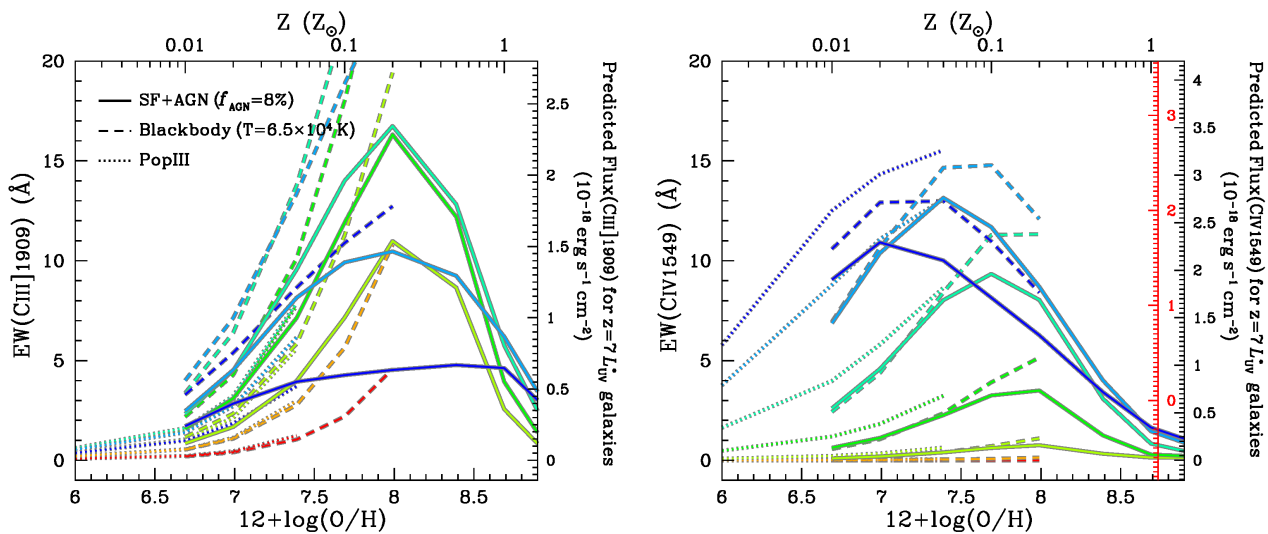}
  }
  \caption{
        As in Fig. \ref{fig:Z_ew}, but with models of
        mixed SF$+$AGN ($f_{\rm AGN}=8$\,\%, solid),
        blackbody ($T=6.5\times 10^4$\,K below $Z=0.2\,Z_{\odot}$, dashed), 
        and 
        PopIII star (below $Z=0.05\,Z_{\odot}$, dotted).
        In each plot, the right ordinate presents a flux predicted 
        for $L^{\star}_{\rm UV}$ galaxies at $z=7$ with an EW 
        corresponding to the left ordinate. 
        For the \CIV\ emission, the red right ordinate is added 
        which shows a predicted flux if the stellar \CIV\ absorption
        exists with EW$_{\rm abs}$(\CIV) $=-3$\,\AA.
        }
\label{fig:Z_ew_predict}
\end{figure*}
%FFFFFFFFFFFFFFFFFFFFFFFFFFFFFFFFFFFFFFFFFFFFFFFFFFFFFFFFFFFFFFFFFFFFFFFFF%

Since \CIV\ line photons can be trapped in the ionized nebular regions
by resonant scattering, they would be preferentially absorbed and 
weakened if internal dust existed. 
Strong \CIV\ emission is thus likely to be observed from young, 
dust-free sources. 
Indeed, at $z=3-5$, low-mass ($\lesssim 10^7\,M_{\odot}$) young galaxies 
and super star clusters with a blue UV continuum (UV slope $\beta<-2$)
are observed using strong-lensing effect to present strong \CIV\ emission of 
EW $\gtrsim 10$\,\AA\ 
\citep{vanzella2016,vanzella2017,smit2017}.
The \CIV\ emitters found at higher-$z$ \citep{stark2015_c4,mainali2017}
would be a similarly dust-free population if the main ionizing sources were 
stars.

%%%%%%%%%%%%%%%%%%%%%%%%%%%%%%%%%%%%%%%%%%%%%%%%%%%%
\subsection{Implications}
\label{ssec:discussion_implications}
%%%%%%%%%%%%%%%%%%%%%%%%%%%%%%%%%%%%%%%%%%%%%%%%%%%%

Finally, we discuss the possibility of using the \CIII\ and/or \CIV\  as a probe
of galaxies in the early universe. 
First, we revisit the claimed positive correlation between EW(\Lya) and EW(\CIII)
\citep{shapley2003,stark2014,rigby2015}.
\citet{rigby2015} suggest that the correlation is dominated by the strongest 
emitters with EW(\Lya) $\gtrsim 50$\,\AA\ and EW(\CIII) $\gtrsim 5$\,\AA. 
The correlation looks weaker in the range below EW(\Lya) $\sim 50$\,\AA.
Our models corroborate these trends. 
One of the key quantities that govern the correlation is the 
UV-continuum level, or the star-formation age.
We demonstrated in Sect.\ \ref{ssec:UV_ISMproperties}
that strong \CIII\ emitters require a very young starburst age
to have a large EW(\CIII). In the early phase of the current 
starburst, galaxies should have a large EW(\Lya) as well. 
If galaxies are more evolved, that is, if the metallicity and age become increased,
both of the EWs of \Lya\ and \CIII\ become smaller. 
In addition, in the evolved system, radiative transfer effects on \Lya\
complicate the \Lya\ visibility. In the correlation of \citet{rigby2015},
there are several galaxies that present a modest \CIII\ emission 
(EW $\lesssim 5$\,\AA) while no \Lya\ in emission. This would indicate that
these galaxies have more dust extinction, which preferentially absorbs the 
resonantly-scattered \Lya\ photons due to their relatively long mean free paths. Alternatively, a scatter in the C$/$O abundance ratio
as a function of metallicity could add another uncertainty 
(cf. Sect.\ \ref{ssec:UV_ISMproperties}). 
Such complexities would result in the weaker correlation in the range of 
EW(\Lya) $\lesssim 20$\,\AA\ as suggested by \citet{rigby2015}.

We can also speculate on the relationship between EWs of \Lya\ and \CIII\ 
for much stronger emitters, that is, in the range of EW(\Lya) $\gtrsim 100$\,\AA.
If we assume such a strong \Lya\ is emitted by a very young galaxy 
with intense bursts of star formation, a key property is the hardness of 
the incident radiation field to further enhance the EW(\Lya). Based on 
our models that assume the gas-phase and stellar metallicities are 
the same, a more metal-poor galaxy tends to have a harder ionizing 
spectrum and a higher \xiion. 
Therefore, the EW(\Lya) is considered to monotonically increase with 
decreasing metallicity. 
On the other hand, as we present in Fig. \ref{fig:Z_ew}, the EW(\CIII) 
becomes smaller with decreasing metallicity below $Z=0.1-0.2\,Z_{\odot}$
due to a smaller amount of carbon.
We thus predict that the positive correlation between EW(\Lya) and 
EW(\CIII) becomes weaker in the range of EW(\Lya) $\gtrsim 100$\,\AA,
and that EW(\CIII) could even start to drop with EW(\Lya).
Indeed, 
\citet{lefevre2017} use a large sample of \CIII\ emitters to find
that these objects are not always strong \Lya\ emitters, 
and vice versa.
Although their plot might be contaminated by \CIII\ emitters powered
by a hidden AGN, the positive correlation is suggested to not be 
as strong as previously thought.
More \CIII\ observations for strong \Lya\ emitters will clarify that trend.

Figure \ref{fig:Z_ew_predict} presents our model predictions of the EWs of 
\CIII\ and \CIV\ as a function of metallicity if the radiation field is given by
the mix of SF $+$ AGN ($f_{\rm AGN}=0.08$), by a blackbody ($T=6.5\times 10^4$\,K),
or by a PopIII star. The former two are favored to reproduce our strong and 
intermediately strong \CIII\ emitters.
Here we adopt the ``standard'' assumption of the C$/$O abundance ratio
\citep{dopita2006}. All of these models are considered to be at a very young 
star-formation age: $1$\,Myr for the SF $+$ AGN models, and zero-age for 
the blackbody and PopIII star models. 
Compared with the BPASS binary-star models in Fig. \ref{fig:Z_ew}, 
the SF $+$ AGN models present a similar trend of EWs with metallicity.
The SF $+$ AGN models have a peak of EW(\CIII) at $Z\sim 0.2\,Z_{\odot}$,
predicting a lower EW with higher and lower metallicities. EW(\CIV) is 
predicted to have a peak at a somewhat lower metallicity of $Z= 0.02-0.1\,Z_{\odot}$.
These trends are well reproduced and extrapolated to a lower metallicity regime 
with the blackbody and PopIII star models. 
Therefore, it is likely that these models enable us to make a firm prediction of the 
EWs of \CIII\ and \CIV\ in the low-metallicity regime, where high-$z$ galaxies 
in the reionization era should lie.

In order to evaluate the detectability of the carbon lines from early galaxies,
we estimate typical fluxes of these lines from galaxies at $z\sim 7$ for reference. 
An $L^{\star}_{\rm UV}$ galaxy at $z\sim 7$ has an absolute UV magnitude of 
$M_{\rm UV}\sim -20.7$ \citep{bouwens2015}.
Using this UV continuum level and assuming a flat $f_{\nu}$ in the FUV range, 
we derive fluxes of \CIII\ and \CIV\ predicted for $L^{\star}_{\rm UV}$ galaxies at $z=7$
as shown in the right ordinate of each panel in Fig. \ref{fig:Z_ew_predict}.
One suggestion we obtain is that at a low metallicity regime of $Z\lesssim 0.05\,Z_{\odot}$, 
the \CIV\ emission looks stronger, and therefore appears to be a more prominent probe 
than the \CIII\ emission if a high-ionization parameter is associated with the low 
metallicity ($\log U\gtrsim -1.5$).
Such low metallicity is considered to be likely, since brighter (and more evolved) 
galaxies at $z=6-8$ are suggested to have $Z\sim 0.1\,Z_{\odot}$ 
\citep{stark2017}.
We thus focus on the metallicity range of $Z=0.01-0.1\,Z_{\odot}$ here. 
If the metallicity is $Z=0.1\,Z_{\odot}$, both of the EWs could be as large as 
$\sim 15$\,\AA, which corresponds to 
$\sim 2\times 10^{-18}$ and $3\times 10^{-18}$ erg\,s$^{-1}$\,cm$^{-2}$ in fluxes
of \CIII\ and \CIV, respectively. These are obtainable with currently available 
ground-based instruments such as Keck/MOSFIRE 
within a reasonable amount of time 
of a few hours on-source integration, 
given the typical sensitivity provided by 
\citet{steidel2014}.
If a lower metallicity of $Z=0.01\,Z_{\odot}$ is considered, the maximum EW of 
\CIII\ drops to $\sim 3-4$\,\AA, while that of EW(\CIV) remains as large as 
$\sim 12$\,\AA.
In this case, the predicted flux of \CIII\ and \CIV\ is $\sim 0.5\times 10^{-18}$ 
and $2.5\times 10^{-18}$ erg\,s$^{-1}$\,cm$^{-2}$, respectively. 
Therefore, for such a metal-poor galaxy, \CIV\ is considered to provide 
a better probe for its identification. 
Although the observed \CIV\ emission might become weaker if the stellar 
absorption exists, our calculation suggests that the stellar absorption of 
EW$_{\rm abs}$(\CIV) $=-3$\,\AA, which we adopt for the VUDS sample, 
does not significantly decrease the visibility of the \CIV\ emission 
(cf. red ordinate in the right panel of Fig. \ref{fig:Z_ew_predict}).

However, we need to acknowledge some caveats. 
One is that these calculations are for $L^{\star}_{\rm UV}$ galaxies at $z=7$. 
The predicted fluxes would be decreased by $\sim \times 1/10$ for 
sub-$L^{\star}_{\rm UV}$ galaxies.
Another difficulty is that we cannot simply state that brighter galaxies 
present stronger carbon emission lines, because the EWs presented in 
Fig. \ref{fig:Z_ew_predict} assume a very young starburst age.
If brighter galaxies were more evolved, 
their EWs would become lower as shown in 
Fig. \ref{fig:ewcs_age_rel}.
Indeed, \citet{shibuya2017} argue a tentative%
\footnote{
Since the anti-correlation is dominated by strong \CIV\ emitters
with EW $>20$\,\AA, which our standard models cannot explain,
it is not obvious whether the correlation is universally correct or not.
} anti-correlation between 
EW(\CIV) and UV-continuum luminosity.
This suggests that a less luminous galaxy tends to be younger, 
less chemically enriched
(and hence to have a higher ionization parameter), 
and have a harder ionizing spectrum.
Finally, the absolute strengths of the Carbon emission would become
stronger if the C$/$O ratio was enhanced.
An analysis of multiple UV lines such as the one we perform with the VUDS \CIII\ emitters,
ideally including the \OIII$\lambda\lambda 1660,1666$ doublet, 
is preferred for resolving the degeneracy.
Such studies should be possible for high-redshift ($z\gtrsim 7$) galaxies
from the ground with NIR spectrographs and 
using the upcoming James Webb Space Telescope (JWST).

%%%%%%%%%%%%%%%%%%%%%%%%%%%%%%%%%%%%%%%%%%%%%%%%%%%%
%%%%%%%%%%%%%%%%%%%%%%%%%%%%%%%%%%%%%%%%%%%%%%%%%%%%
\section{Summary and conclusions} 
\label{sec:conclusions}
%%%%%%%%%%%%%%%%%%%%%%%%%%%%%%%%%%%%%%%%%%%%%%%%%%%%
%%%%%%%%%%%%%%%%%%%%%%%%%%%%%%%%%%%%%%%%%%%%%%%%%%%%

In order to interpret the observed UV spectra of distant galaxies, 
we present the theoretical predictions 
for the behavior of the UV emission, 
in particular of \CIII$\lambda 1909$, \CIV$\lambda 1549$, and 
\HeII$\lambda 1640$ lines,
from extensive grids of photoionization models.
The models use several incident radiation fields 
(stellar population, AGNs, a mixture of stars and AGNs, and so on) 
and cover a wide range of ISM properties. 
We have tested these models on existing sources with 
known ionization field and/or ISM properties. 
Our main results concerning the UV emission line predictions
are summarized as follows.
\begin{itemize}
        \setlength{\itemsep}{5pt}
        \setlength{\parskip}{0pt}
        \item From the photoionization models 
                we propose new spectral UV line diagnostics 
                using rest-frame EWs of 
                \CIII\ and \CIV\ and the line ratios of \CIII, 
                \CIV, and \HeII. 
                Extending earlier predictions for UV lines 
                \citep[e.g.,][]{feltre2016,gutkin2016,JR2016},
                we demonstrate that the UV diagrams of
                \CIV$/$\CIII\ versus (\CIII$+$\CIV)$/$\HeII,
                EW(\CIII) versus \CIII$/$\HeII, and  
                EW(\CIV) versus \CIV$/$\HeII\ 
                are useful 
                for classifying the nature of the ionizing radiation field 
                of galaxies
                and for constraining their ISM properties. 
        \item For star-forming galaxies, our models predict that \CIII\ EW 
                peaks at subsolar metallicities ($\sim 0.1-0.2$ \zsun),
                whereas \CIV\ EW becomes maximal at even lower metallicities.
                Standard models show that sources with 
                EW(\CIII)$\ga 20$ \AA\ and EW(\CIV)$\ga 12$ \AA\ 
                cannot be explained by purely stellar photoionization: 
                an enhanced carbon abundance (with C$/$O up to solar) or 
                an AGN-like ionizing spectrum is needed to explain such sources.
\end{itemize}
%
%
%\noindent
Subsequently, we used these models and UV diagnostic diagrams 
to analyze new spectroscopic observations from the VIMOS Ultra Deep 
Survey (VUDS). 
\citet{lefevre2017} identify $450$ \CIII-emitting galaxies at $z=2-4$ in VUDS.
We  binned the sources in  three categories, according to 
strong (Sample A) and intermediate (Sample B) EW(\CIII), and a reference sample,
Sample (C), of all star-forming galaxies. 
The average properties of the three samples are the following: 
\begin{itemize}
        \setlength{\itemsep}{5pt}
        \setlength{\parskip}{0pt}
        \item {\bf All star-forming galaxies} 
                (Sample C; $N=450$ sources; EW(\CIII) $\sim 2$\,\AA): 
                We find that they typically require a stellar photoionization 
                including binary stars.
                Using the incident radiation field generated by BPASS 
                \citep{stanway2015}, 
                the ISM properties of metallicity ($Z$) and ionization parameter ($U$) 
                are inferred to be ($Z$, $\log U$) from ($0.3\,Z_{\odot}$, $-2.7$) to 
                ($0.5\,Z_{\odot}$, $-3$), and the age of the current star formation
                to be from $200$\,Myr to $50$\,Myr. 
                The ionizing photon production efficiency parameterized by \xiion\
                is calculated to be $\log$\,\xiion$/$\ergsHz\ $= 25.3$ -- $25.4$.
        \item {\bf Intermediate \CIII\ emitters}
                (Sample B; $N=43$; EW(\CIII) $=10$--$20$\,\AA): 
                We suggest the sample contains a variety of objects, 
                at least $\sim 28$\,\%\ of which are metal-poor 
                ($Z\lesssim 0.05$ -- $0.2\,Z_{\odot}$) star-forming galaxies
                and $\sim 34$\,\%\ are pure AGNs. 
                A pure stellar photoionization is insufficient to explain their 
                average UV observations,
                irrespective of the inclusion of binary stars.
                They are best described by an SED including a mix 
                of stellar $+$ AGN photoionization with a contribution
                of $f_{\rm AGN}\sim 7$ -- $10$\,\%\ of the ionizing photons 
                from the AGN.
                Furthermore, to fully explain the large EWs of \CIII\ and \CIV\
                but faint \OIII$\lambda 1665$ emission, we indicate 
                that the population typically requires a high C$/$O abundance ratio 
                ($\log$\,C$/$O $=-0.3$).
                With these conditions, the ISM properties are best inferred as 
                ($Z$, $\log U$) $\sim$ ($0.02\,Z_{\odot}$, $-1.7$) or 
                ($0.2\,Z_{\odot}$, $-0.9$).
                A young starburst age of $\sim 3-8$\,Myr and 
                an efficient ionizing photon production of 
                $\log$\,\xiion$/$\ergsHz\ $\sim 25.55-25.7$ are derived.
        \item {\bf Strong \CIII\ emitters} 
                (Sample A; $N=16$; EW(\CIII) $>20$\,\AA): 
                With the UV diagrams we classify $\sim 30$\,\%\ of the sample 
                as AGNs.
                We find that stellar photoionization is clearly insufficient 
                to explain the UV observation of the average and the other 
                $\sim 70$\,\%\ dominant population.
                A radiation field consisting of a mixture of a young stellar population
                ($\log$\,\xiion$/$\ergsHz\ $= 25.7$ -- $25.75$) plus an AGN component
                with $f_{\rm AGN}\sim 8^{+2}_{-5}$\,\%\
                is necessary. 
                Furthermore, an enhanced C$/$O ratio ($\log$\,C$/$O $=-0.45$) 
                is needed for ISM properties of 
                ($Z$, $\log U$) from ($0.1\,Z_{\odot}$, $-1.75$) to 
                ($0.2\,Z_{\odot}$, $-1.5$).
\end{itemize}
%
%
%\noindent
The analyses of the VUDS \CIII\ emitters have provided 
the following discussion for the high-$z$ star-forming population: 
\begin{itemize}
        \setlength{\itemsep}{5pt}
        \setlength{\parskip}{0pt}
        \item Since the EW of \CIII\ peaks for metallicities of 
                $Z=0.1-0.2\,Z_{\odot}$,
                the selection of objects by strong \CIII\ emission should 
                result in a population with a subsolar metallicity.
                A young starburst age of $\lesssim 10$\,Myr
                and a high ionizing photon production efficiency of
                $\log$\,\xiion$/$\ergsHz\ $\sim 25.55-25.75$
                are required for the strong \CIII\ emission of 
                EW(\CIII) $\gtrsim 10$\,\AA.
        \item The possible addition of AGN to the ionizing spectrum 
                indicates that the shape of stellar ionizing spectrum
                in the very high-energy regime might not be sufficiently 
                reproduced by conventional stellar population synthesis 
                codes.
                If the inclusion of an AGN-like hard radiation field is correct
                for the strong \CIII\ emitters, this could alter the 
                ISM properties that are estimated with the UV high ionization 
                lines alone under the assumption of a pure stellar photoionization.
        \item Using the ionizing spectrum as inferred 
                for the VUDS intermediate and strong \CIII\ emitters, 
                we offer a prospect of the visibility of the \CIII\ and \CIV\ emission
                from galaxies in the reionization epoch of $z\sim 7$.
                We predict that both \CIII\ and \CIV\ could be equally as strong as 
                EW $\sim 15$\,\AA\
                and detectable
                with currently available instruments 
                from $z\sim 7$ $L^{\star}_{\rm UV}$ galaxies if the metallicity is 
                subsolar. 
                If these galaxies are more metal-poor by an order of magnitude, 
                the maximum EW of \CIII\ would drop to $\sim 4$\,\AA,
                while that of \CIV\ would remain as large as $\sim 12$\,\AA.
                In a metal-poor environment, \CIV\ would provide a better probe 
                of objects in the early universe.
\end{itemize}
%
%
%\noindent
This paper presents a useful means to identify the main source of ionizing photons 
in distant galaxies based on UV rest-frame emission lines. This will be further 
improved, including the addition of a full suite of emission lines over the rest-frame UV$+$optical 
wavelength range. 
The methods presented in this paper provide useful analysis and classification tools 
to study the properties of galaxies with the upcoming JWST, with future extended 
large spectroscopic surveys such as Subaru/PFS \citep{takada2014} or
VLT/MOONS \citep{cirasuolo2014}, and $30$\,m-class telescopes.

%%%%%%%%%%%%%%%%%%%%%%%%%%%%%%%%%%%%%%%%%%%%%%%%%%%%
%%%%%%%%%%%%%%%%%%%%%%%%%%%%%%%%%%%%%%%%%%%%%%%%%%%%
\begin{acknowledgements}
This work is supported by funding from the European Research Council 
Advanced Grant ERC--2010--AdG--268107--EARLY and by
INAF Grants PRIN 2010, PRIN 2012 and PICS 2013. 
K.N. acknowledges the JSPS Overseas Research Fellowships, and 
R.A. acknowledges support from the ERC Advanced Grant 695671 `QUENCH'. 
This work is based on data products made available at the CESAM data center, 
Laboratoire d'Astrophysique de Marseille.\end{acknowledgements}
%%%%%%%%%%%%%%%%%%%%%%%%%%%%%%%%%%%%%%%%%%%%%%%%%%%%
%%%%%%%%%%%%%%%%%%%%%%%%%%%%%%%%%%%%%%%%%%%%%%%%%%%%

%RRRRRRRRRRRRRRRRRRRRRRRRRRRRRRRRRRRRRRRRRRRRRRRRRRRRRRRRRRRRRR%

%RRRRRRRRRRRRRRRRRRRRRRRRRRRRRRRRRRRRRRRRRRRRRRRRRRRRRRRRRRRRRR%


\begin{thebibliography}{}
\addcontentsline{toc}{chapter}{\bibname}
\expandafter\ifx\csname natexlab\endcsname\relax\def\natexlab#1{#1}\fi
\bibitem[\protect\citeauthoryear{Allen et al.}{2008}]{allen2008} Allen, M.~G., Groves, B.~A., Dopita, M.~A., Sutherland, R.~S., \& Kewley, L.~J. 2008, \apjs, 178, 20
Amor\'{i}n, R.~O., P\'{e}rez-Montero, E., \& V\'{i}lchez, J.~M. 2010, \apjl, 715, L128
\bibitem[\protect\citeauthoryear{Amor\'{i}n et al.}{2010}]{amorin2010} Amor\'{i}n, R.~O., P\'{e}rez-Montero, E., \& V\'{i}lchez, J.~M. 2010, \apjl, 715, L128
\bibitem[\protect\citeauthoryear{Amor\'{i}n et al.}{2017}]{amorin2017} Amor\'{i}n, et al. 2017, NatAs, 1, 0052
\bibitem[\protect\citeauthoryear{Asplund et al.}{2009}]{asplund2009} Asplund, M., Grevesse, N., Sauval, A.~J., \& Scott, P. 2009, ARA\&A, 47, 481
\bibitem[\protect\citeauthoryear{Baldwin et al.}{1981}]{baldwin1981} Baldwin, J.~A., Phillips, M.~M., \& Terlevich, R. 1981, \pasp, 93, 5
\bibitem[\protect\citeauthoryear{Bayliss et al.}{2014}]{bayliss2014} Bayliss, M.~B., et al. 2014, \apj, 790, 144
\bibitem[\protect\citeauthoryear{Berg et al.}{2016}]{berg2016} Berg, D.~A., Skillman, E.~D., Henry, R.~B.~C., Erb, D.~K., \& Carigi, L. 2016, \apj, 827, 126
\bibitem[\protect\citeauthoryear{Binette et al.}{2003}]{binette2003} Binette, L., Groves, B., Villar-Mart\'{i}n, M., Fosbury, R.~A.~E., \& Axon, D.~J. 2003, \aap, 405, 975
\bibitem[\protect\citeauthoryear{Bouwens et al.}{2015}]{bouwens2015} Bouwens, R.~J., et al. 2015, \apj, 803, 34
\bibitem[\protect\citeauthoryear{Bouwens et al.}{2016}]{bouwens2016} Bouwens, R.~J., et al. 2016, \apj, 831, 176
\bibitem[\protect\citeauthoryear{Brinchmann et al.}{2004}]{brinchmann2004} Brinchmann, J., Charlot, S., White, S.~D.~M., et al. 2004, \mnras, 351, 1151
\bibitem[\protect\citeauthoryear{Brinchmann et al.}{2008}]{brinchmann2008} Brinchmann, J., Pettini, M., \& Charlot, S. 2008, \mnras, 385, 769
\bibitem[\protect\citeauthoryear{Cassata et al.}{2013}]{cassata2013} Cassata, P., et al. 2013, \aap, 556, A68
\bibitem[\protect\citeauthoryear{Cassata et al.}{2015}]{cassata2015} Cassata, P., et al. 2015, \aap, 573, A24
\bibitem[\protect\citeauthoryear{Chabrier}{2003}]{chabrier2003} Chabrier, G. 2003, \pasp, 115, 763
\bibitem[\protect\citeauthoryear{Chiappini}{2006}]{chiappini2006} Chiappini, C., Hirschi, R., Meynet, G., Ekstr\"{o}m, S., Maeder, A., \& Matteucci, F. 2006, \aap, 449, 27
\bibitem[\protect\citeauthoryear{Christensen et al.}{2012}]{christensen2012a} Christensen, L., et al. 2012, \mnras, 427, 1953
\bibitem[\protect\citeauthoryear{Cirasuolo et al.}{2014}]{cirasuolo2014} Cirasuolo, M., et al. 2014, Proc. SPIE, 9147, 91470N
\bibitem[\protect\citeauthoryear{Crowther}{2007}]{crowther2007} Crowther, P.~A. 2007, ARA\&A, 45, 177
Cresci, G., Mannucci, F., Sommariva, V., Maiolino, R., Marconi, A., Brusa, M., 2012, \mnras, 421, 262
\bibitem[\protect\citeauthoryear{Crowther}{2016}]{crowther2016} Crowther, P.~A., et al. 2016, \mnras, 458, 624
\bibitem[\protect\citeauthoryear{de Barros et al.}{2016}]{debarros2016} de Barros, S., et al. 2016, \aap, 585, A51
\bibitem[\protect\citeauthoryear{Dopita et al.}{2006}]{dopita2006} Dopita, M.~A., et al. 2006, \apjs, 167, 177
\bibitem[\protect\citeauthoryear{Dors et al.}{2014}]{dors2014} Dors, O.~L., Jr, Cardaci, M.~V., H\"{a}gele, G.~F., \& Krabbe, \^{A}.~C. 2014, \mnras, 443, 1291 
\bibitem[\protect\citeauthoryear{Elvis et al.}{1994}]{elvis1994} Elvis, M., et al. 1993, \apjs, 95, 1
\bibitem[\protect\citeauthoryear{Elvis et al.}{2002}]{elvis2002} Elvis, M., Risaliti, G., \& Zamorani, G. 2002, \apjl, 565, L75
\bibitem[\protect\citeauthoryear{Erb et al.}{2010}]{erb2010} Erb, D.~K., Pettini, M., Shapley, A.~E., Steidel, C.~C., Law, D.~R., Reddy, N.~A., 2010, \apj, 719, 1168
\bibitem[\protect\citeauthoryear{Fan et al.}{2006}]{fan2006} Fan, X., Carilli, C.~L., \& Keating, B. 2006, ARA\&A, 44, 415
\bibitem[\protect\citeauthoryear{Feltre et al.}{2016}]{feltre2016} Feltre, A., Charlot, S., \& Gutkin, J. 2016, \mnras, 456, 3354
\bibitem[\protect\citeauthoryear{Ferland et al.}{1998}]{ferland1998} Ferland, G.~J., et al. 1998, \pasp, 110, 761
\bibitem[\protect\citeauthoryear{Ferland et al.}{2013}]{ferland2013} Ferland, G.~J., et al. 2013, RMxAA, 49, 137
\bibitem[\protect\citeauthoryear{Fosbury et al.}{2003}]{fosbury2003} Fosbury, R.~A.~E., et al. 2003, \apj, 596, 797
\bibitem[\protect\citeauthoryear{F\"{o}rster Schreiber et al.}{2009}]{forsterschreiber2009} F\"{o}rster Schreiber et al. 2009, \apj, 706, 1364
\bibitem[\protect\citeauthoryear{Francis et al.}{1993}]{francis1993} Francis, P.~J., Hooper, E.~J., \& Impey, C.~D. 1993, \aj, 106, 417
\bibitem[\protect\citeauthoryear{Garnett et al.}{1995}]{garnett1995} Garnett, D.~R., et al. 1995, \apj, 443, 64
\bibitem[\protect\citeauthoryear{Gordon et al.}{2003}]{gordon2003} Gordon, K.~D., Clayton, G.~C., Misselt, K.~A., Landolt, A.~U., \& Wolff, M.~J. 2003, \apj, 594, 279
\bibitem[\protect\citeauthoryear{Gutkin et al.}{2016}]{gutkin2016} Gutkin, J., Charlot, S., \& Bruzual, G. 2016, \mnras, 462, 1757
\bibitem[\protect\citeauthoryear{Hainline et al.}{2011}]{hainline2011} Hainline, K.~N., Shapley, A.~E., Greene, J.~E., \& Steidel, C.~C. 2011, \apj, 733, 31
\bibitem[\protect\citeauthoryear{Hathi et al.}{2016}]{hathi2016} Hathi, N.~P., et al. 2016, \aap, 588, A26
\bibitem[\protect\citeauthoryear{Heckman et al.}{2011}]{heckman2011} Heckman, T.~M., et al. 2011, \apj, 730, 5
\bibitem[\protect\citeauthoryear{Jaskot \& Oey}{2013}]{JO2013} Jaskot, A.~E. \& Oey, M.~S. 2013, \apj, 766, 91
\bibitem[\protect\citeauthoryear{Jaskot \& Ravindranath}{2016}]{JR2016} Jaskot, A.~E. \& Ravindranath, S. 2016, \apj, 833, 136
\bibitem[\protect\citeauthoryear{Kauffmann et al.}{2003}]{kauffmann2003_agn} Kauffmann, et al. 2003, \mnras, 346, 1055
\bibitem[\protect\citeauthoryear{Kennicutt}{1998}]{kennicutt1998} Kennicutt, R.~C., Jr., 1998, ARA\&A, 36, 189
\bibitem[\protect\citeauthoryear{Kewley et al.}{2001}]{kewley2001} Kewley, L.~J., Dopita, M.~A., Sutherland, R.~S., Heisler, C.~A., \& Trevena, J. 2001, \apj, 556, 121
\bibitem[\protect\citeauthoryear{Kewley \& Dopita}{2002}]{KD2002} Kewley, L.~J. \& Dopita, M.~A. 2002, \apjs, 142, 35
\bibitem[\protect\citeauthoryear{Kewley et al.}{2006}]{kewley2006} Kewley, L.~J., Groves, B., Kauffmann, G., \& Heckman, T. 2006, \mnras, 372, 961
\bibitem[\protect\citeauthoryear{Kewley et al.}{2013}]{kewley2013_theory} Kewley, L.~J., et al. 2013, \apj, 774, 100
\bibitem[\protect\citeauthoryear{Kojima et al.}{2017}]{kojima2017} Kojima, T., Ouchi, M., Nakajima, K., et al. 2017, \pasj\ in press, arXiv e-prints, arXiv:1605.03436
\bibitem[\protect\citeauthoryear{Konno et al.}{2014}]{konno2014} Konno, A., Ouchi, M., Ono, M., et al. 2014, \apj, 797, 16
\bibitem[\protect\citeauthoryear{Kroupa}{2001}]{kroupa2001} Kroupa, P. 2001, \mnras, 322, 231
\bibitem[\protect\citeauthoryear{Laporte et al.}{2017}]{laporte2017} Laporte, N., Nakajima, K., Ellis, R.~S., Zitrin, A., Stark, D.~P., Mainali, R., \& Roberts-Borsani, G. 2017, arXiv e-prints, arXiv:1708.05173
\bibitem[\protect\citeauthoryear{Le F\`evre et al.}{2015}]{lefevre2015} Le F\`evre, O., et al. 2015, \aap, 576, A79
\bibitem[\protect\citeauthoryear{Le F\`evre et al.}{2017}]{lefevre2017} Le F\`evre, O., et al. 2017, arXiv e-prints, arXiv:1710.10715
\bibitem[\protect\citeauthoryear{L\'{o}pez-S\'{a}nchez et al.}{2012}]{lopez-sanchez2012} L\'{o}pez-S\'{a}nchez, \'{A}.~R., Dopita, M.~A., Kewley, L.~J., Zahid, H.~J., Nicholls, D.~C., \& Scharw\"{a}chter, J. 2012, \mnras, 426, 2630 
\bibitem[\protect\citeauthoryear{Mainali et al.}{2017}]{mainali2017} Mainali, R., et al., 2017, \apjl, 836, L14
\bibitem[\protect\citeauthoryear{Masters et al.}{2014}]{masters2014} Masters, D., et al. 2014, \apj, 785, 153
\bibitem[\protect\citeauthoryear{Mattsson}{2010}]{mattsson2010} Mattsson, L. 2010, \aap, 515, A68
\bibitem[\protect\citeauthoryear{Mannucci et al.}{2009}]{mannucci2009} Mannucci, F., et al. 2009, \mnras, 398, 1915
\bibitem[\protect\citeauthoryear{Marigo}{2001}]{marigo2001} Marigo, P. 2001, \aap, 370, 194
\bibitem[\protect\citeauthoryear{Marigo}{2007}]{marigo2007} Marigo, P. 2007, \aap, 467, 1139
\bibitem[\protect\citeauthoryear{Moll\'{a} et al.}{2009}]{molla2009} Moll\'{a}, M., Garc\'{i}a-Vargas, M.~L., \& Bressan, A. 2009, \mnras, 398, 451
\bibitem[\protect\citeauthoryear{Nagao et al.}{2006}]{nagao2006a} Nagao, T., Maiolino, \& R., Marconi, A. 2006, \aap, 447, 863
\bibitem[\protect\citeauthoryear{Nagao et al.}{2011}]{nagao2011} Nagao, T., Maiolino, R., Marconi, A., \& Matsuhara, H. 2011, \aap, 526, 149
\bibitem[\protect\citeauthoryear{Nakajima \& Ouchi}{2014}]{NO2014} Nakajima, K. \& Ouchi, M. 2014, \mnras, 442, 900
\bibitem[\protect\citeauthoryear{Nakajima et al.}{2016}]{nakajima2016} Nakajima, K., Ellis, R.~S., I. Iwata, et al. 2016, \apjl, 831, L9
\bibitem[\protect\citeauthoryear{Onodera et al.}{2016}]{onodera2016} Onodera M., et al. 2016, \apj, 822, 42 
\bibitem[\protect\citeauthoryear{Pagel et al.}{1979}]{pagel1979} Pagel, B.~E.~J., Edmunds, M.~G., Blackwell, D.~E., Chun, M.~S., \& Smith, G. 1979, \mnras, 189, 95
\bibitem[\protect\citeauthoryear{Pentericci et al.}{2014}]{pentericci2014} Pentericci, L., et al. 2014, \apj, 793, 113 
\bibitem[\protect\citeauthoryear{Pettini \& Pagel}{2004}]{PP2004} Pettini, M. \& Pagel, B.~E.~J. 2004, \mnras, 348, L59
\bibitem[\protect\citeauthoryear{P\`erez-Montero \& Amor\'in}{2017}]{PA2017} P\`erez-Montero, E. \& Amor\'in, R. 2017, \mnras, 467, 1287
\bibitem[\protect\citeauthoryear{Raiter et al.}{2010}]{raiter2010} Raiter, A., Schaerer, D., \& Fosbury, R.~A.~E. 2010, \aap, 523, 64
\bibitem[\protect\citeauthoryear{Ranalli et al.}{2003}]{ranalli2003} Ranalli, P., Comastri, A., \& Setti, G. 2003, 
\aap, 399, 39
\bibitem[\protect\citeauthoryear{Reddy et al.}{2010}]{reddy2010} Reddy, N.~A., et al. 2010, \apj, 712, 1070
\bibitem[\protect\citeauthoryear{Reddy et al.}{2015}]{reddy2015} Reddy, N.~A., et al. 2015, \apj, 806, 259
\bibitem[\protect\citeauthoryear{Reddy et al.}{2017}]{reddy2017} Reddy, N.~A., et al. 2017, arXiv e-prints, arXiv:1705.09302
\bibitem[\protect\citeauthoryear{Rigby et al.}{2015}]{rigby2015} Rigby, J.~R., et al. 2015, \apjl, 814, L6
\bibitem[\protect\citeauthoryear{Robertson et al.}{2013}]{robertson2013} Robertson, B.~E., et al. 2013, \apj, 768, 71 
\bibitem[\protect\citeauthoryear{Salpeter}{1955}]{salpeter1955} Salpeter, E.~E. 1955, \apj, 121, 161
\bibitem[\protect\citeauthoryear{Sanders et al.}{2016}]{sanders2016} Sanders, R.~L., et al. 2016, \apj, 816, 23
\bibitem[\protect\citeauthoryear{Schaerer}{2003}]{schaerer2003} Schaerer, D. 2003, \aap, 397, 527
\bibitem[\protect\citeauthoryear{Schaerer et al.}{2016}]{schaerer2016} Schaerer, D., et al. 2016, \aap, 591, L8
\bibitem[\protect\citeauthoryear{Schenker et al.}{2014}]{schenker2014} Schenker, M.~A., Ellis, R.~S., Konidaris, N.~P., \& Stark, D.~P. 2014, \apj, 795, 20
\bibitem[\protect\citeauthoryear{Schmidt et al.}{2017}]{schmidt2017} Schmidt, K.~B., et al. 2017, \apj, 839, 17
\bibitem[\protect\citeauthoryear{Senchyna et al.}{2017}]{senchyna2017} Senchyna, P., et al. 2017, arXiv e-prints, arXiv:1706.000881
\bibitem[\protect\citeauthoryear{Shapley et al.}{2003}]{shapley2003} Shapley, A.~E., Steidel, C.~C., Pettini, M., \& Adelberger, K.~L. 2003, \apj, 588, 65
\bibitem[\protect\citeauthoryear{Shibuya et al.}{2017}]{shibuya2017} Shibuya, T., et al. 2017, arXiv e-prints, arXiv:1705.00733
\bibitem[\protect\citeauthoryear{Smit et al.}{2017}]{smit2017} Smit, R., Swinbank, A,~M., Massey, R., Richard, J., Smail, I., \& Kneib, J.~-P. 2017, \mnras, 467, 3306
\bibitem[\protect\citeauthoryear{Stanway et al.}{2015}]{stanway2015} Stanway, E.~R., Eldridge, J.~J. \& Becker, G.~D. 2015, \mnras, 456, 485
\bibitem[\protect\citeauthoryear{Stanway}{2017}]{stanway2017} Stanway, E.~R. 2017, the Proceedings of IAU Symposium 329: ``The Lives and Death Throws of Massive Stars'', arXiv e-prints, arXiv:1702.07303
\bibitem[\protect\citeauthoryear{Stark et al.}{2011}]{stark2011} Stark, D.~P., Ellis, R.~S., \& Ouchi, M. 2011, \apj, 728, L2
\bibitem[\protect\citeauthoryear{Stark et al.}{2014}]{stark2014} Stark, D.~P., et al. 2014, \mnras, 445, 3200
\bibitem[\protect\citeauthoryear{Stark et al.}{2015a}]{stark2015_c3} Stark, D.~P., et al. 2015a, \mnras, 450, 1846
\bibitem[\protect\citeauthoryear{Stark et al.}{2015b}]{stark2015_c4} Stark, D.~P., et al. 2015b, \mnras, 454, 1393
\bibitem[\protect\citeauthoryear{Stark et al.}{2017}]{stark2017} Stark, D.~P., et al. 2017, \mnras, 464, 469
\bibitem[\protect\citeauthoryear{Stasi\'{n}ska et al.}{2015}]{stasinska2015} Stasi\'{n}ska, G., Izotov, Y., Morisset, C., \& Guseva, N. 2015, \aap, 576, 83
\bibitem[\protect\citeauthoryear{Steidel et al.}{2002}]{steidel2002} Steidel, C.~C., et al., 2002, \apj, 576, 653
\bibitem[\protect\citeauthoryear{Steidel et al.}{2014}]{steidel2014} Steidel, C.~C., Rudie, G.~C., Strom, A.~L., et al., 2014, \apj, 795, 165
\bibitem[\protect\citeauthoryear{Steidel et al.}{2016}]{steidel2016} Steidel, C.~C., Strom, A.~L., Pettini, M., Rudie, G.~C., Reddy, N.~A., \& Trainor, R. 2016, \apj, 826, 159
\bibitem[\protect\citeauthoryear{Storchi-Bergmann et al.}{1994}]{storchi-bergmann1994} Storchi-Bergmann, T., Calzetti, D., \& Kinney, A.~L. 1994, \apj, 429, 572
\bibitem[\protect\citeauthoryear{Stroe et al.}{2017}]{stroe2017} Stroe, A., Sobral, D., Matthee, J., Calhau, J., \& Oteo, I. 2017, \mnras, 471, 2558
\bibitem[\protect\citeauthoryear{Takada et al.}{2014}]{takada2014} Takada, M., et al. 2014, \pasj, 66, R1
\bibitem[\protect\citeauthoryear{Talia et al.}{2017}]{talia2017} Talia, M., et al. 2017, \mnras, 471, 4527
\bibitem[\protect\citeauthoryear{Trainor et al.}{2016}]{trainor2016} Trainor, R.~F., et al. 2016, \apj, 832, 171
\bibitem[\protect\citeauthoryear{van Hoof et al.}{2004}]{vanhoof2004} van Hoof, P.~A.~M., Weingartner, J.~C., Martin, P.~G., Volk, K., \& Ferland, G.~J. 2004, \mnras, 350, 1330
\bibitem[\protect\citeauthoryear{Vanzella et al.}{2009}]{vanzella2009} Vanzella, E., et al. 2009, \apj, 695, 1163
\bibitem[\protect\citeauthoryear{Vanzella et al.}{2016}]{vanzella2016} Vanzella, E., et al. 2016, \apjl, 821, L27
\bibitem[\protect\citeauthoryear{Vanzella et al.}{2017}]{vanzella2017} Vanzella, E., et al. 2017, \apj, 842, 47
\bibitem[\protect\citeauthoryear{Villar-Mart\'{i}n et al.}{1997}]{villar-martin1997} Villar-Mart\'{i}n, M., Tadhunter, C., \& Clark, N. 1997, \aap, 323, 21
\bibitem[\protect\citeauthoryear{Villar-Mart\'{i}n et al.}{2004}]{villar-martin2004} Villar-Mart\'{i}n, M., Cervi\~{n}o, M., \& Delgado, G. 2004, \mnras, 355, 1132
\bibitem[\protect\citeauthoryear{Zamorani et al.}{1981}]{zamorani1981} Zamorani, G., Henry, J.~P., Maccacaro, T., et al. 1981, \apj, 245, 357

\end{thebibliography}
\end{document}